\documentclass[12pt]{article}

\usepackage{amsmath,amsfonts,epsfig}
\usepackage{times}
\usepackage{bm}
\usepackage{natbib}
\usepackage[usenames]{color}
\usepackage{xcolor}
\usepackage{url}

\usepackage{subfigure}

\usepackage{verbatim}

\usepackage[plain,noend]{algorithm2e}

\def\N{{\mbox{N}}}
\def\Be{{\mbox{Be}}}
\def\E{{E}}
\def\logit{{\mbox{logit}}}

\newtheorem{lemma}{Lemma}
\newtheorem{theorem}{Theorem}

\newtheorem{proposition}{Proposition}
\newtheorem{remark}{Remark}

\begin{document}




\title{In Search of Lost (Mixing) Time: Adaptive Markov chain Monte Carlo schemes for Bayesian variable selection with very large $p$}
\author{J. E. Griffin, K. {\L}atuszy\'nski and M. F. J. Steel\thanks{Jim Griffin, Department of Statistical Science, University College London,  WC1E 6BT, U.K.
(Email: j.griffin@ucl.ac.uk), Krys {\L}atuszy\'nski, Department of Statistics, University of Warwick, Coventry, CV4 7AL, U.K. (Email: K.G.Latuszynski@warwick.ac.uk) and Mark Steel, Department of Statistics, University of Warwick, Coventry, CV4 7AL, U.K. (Email: m.steel@warwick.ac.uk). }}

\maketitle

\abstract{
The availability of data sets with large numbers of variables is rapidly increasing. The effective application of Bayesian variable selection methods for regression with these data sets has proved difficult since available Markov chain Monte Carlo methods do not perform well in typical problem sizes of interest. The current paper proposes new adaptive Markov chain Monte Carlo algorithms to address this shortcoming. The adaptive design of these algorithms exploits the observation that in large $p$ small $n$ settings, the majority of the $p$ variables will be approximately uncorrelated a posteriori. The algorithms adaptively build suitable non-local proposals that result in moves with squared jumping distance significantly larger than standard methods. Their performance is studied empirically in high-dimensional problems (with both simulated and actual data) and speedups of up to 4 orders of magnitude are observed. The proposed algorithms are easily implementable on multi-core architectures and are well suited for parallel tempering or sequential Monte Carlo implementations.

\noindent {\bf Keywords}:
variable selection; spike-and-slab priors; high-dimensional data; large $p$, small $n$ problems; linear regression: expected squared jumping distance; optimal scaling}

\section{Introduction}

The availability of large data sets has led to an increasing interest in variable selection methods applied to regression models with many potential variables but few observations ({\it large} $p$, {\it  small} $n$ problems). Frequentist approaches have mainly concentrated on providing point estimates under assumptions of sparsity using penalized maximum likelihood procedures \citep{HaTiWa15}.
However, some recent work has considered constructing confidence intervals and taking into account model uncertainty
\citep{ ShSa13, DeBuMeMe15}. 
Bayesian approaches to variable selection are an attractive and natural alternative and lead to a posterior distribution on all possible models which can be used to address model uncertainty for variable selection and prediction.  A growing literature provides a theoretical basis for good properties of the posterior distribution in large $p$ problems \citep[see {\it e.g.}][]{CaSHVDV15, johnson2012bayesian}.

The posterior probabilities of all models can usually only be calculated or approximated if $p$ is smaller than 30.
If $p$ is larger,
 Markov chain Monte Carlo methods are typically used to sample from the posterior distribution
 \citep{george1997approaches, o2009review,  ClGhLi11}.
 \cite{GDMB13} discuss the
 benefits of such methods. The most widely used Markov chain Monte Carlo algorithm in this context is
the Metropolis-Hastings sampler where new models are proposed using  add-delete-swap samplers \citep{BVF98, Chip01}. For example, this approach is used by \cite{NiWaJo16} in a binary regression model with a non-local prior for the regression coefficients on a data set with 7129 genes. Some supporting theoretical understanding of convergence is available for the  add-delete-swap samplers, {\it e.g.}~conditions for rapid mixing in linear regression model have been derived by \cite{YaWaJo16}. Others have considered  more targeted  moves in model space. For example, \cite{TitsiasYau} introduce the Hamming Ball sampler which more carefully selects the proposed model in a Metropolis-Hastings sampler (in a similar way to shotgun variable selection, \cite{HaDoWe07}) and
\cite{ScCh11} develop a sequential Monte Carlo method that uses a sequence of  annealed posterior distributions.
However, the mixing of these methods is often thought to be poor, when applied to data sets with thousands of potential covariates. As an alternative, several authors use more general shrinkage priors and develop suitable  MCMC algorithms for high-dimensional problems \citep[see {\it e.g.}][]{ABACBM16}. Nonlocal priors \citep{johnson2012bayesian} are adopted in \cite{Shin_etal_18}, who use screening for high dimensions. \cite{RobertsZanella_19} combine Markov chain Monte Carlo and importance sampling ideas in their tempered Gibbs sampler.



The challenge of performing Markov chain Monte Carlo for Bayesian variable selection in high dimensions has lead to several developments sacrificing exact posterior exploration. For example, 
 \cite{LiSoYu13} used the stochastic approximation Monte Carlo algorithm \citep{LiLiCa07} to efficiently explore model space. In another direction, variable selection can be performed as a post-processing step after fitting a model including all variables \citep[see {\it e.g.}][]{BoRe12, HaCa15}. Several authors develop algorithms that focus on high posterior probability models. In particular \cite{rovckova2014emvs} propose a deterministic expectation-maximisation based algorithm for identifying posterior modes, while \cite{papaspiliopoulos2016scalable} develop an exact deterministic algorithm to find the most probable model of any given size in block-diagonal design models.

Alternatively, Markov chain Monte Carlo methods for variable selection can be tailored to the data to allow faster convergence and mixing
using the adaptive Markov chain Monte Carlo framework \citep[see {\it e.g.}][Section 2.4, and references therein]{MR3360496}.
Several strategies have been developed in the literature for both the Metropolis-type algorithms \citep{lamnisos12, JiSchmidler} and Gibbs samplers \citep{nottkohn05, richardson2010bayesian}.
 Our proposal is a Metropolis-Hastings kernel that learns the relative importance of the variables, unlike \cite{JiSchmidler} who use an independent proposal, and unlike \cite{lamnisos12} who only tune the number of variables proposed to be changed.
This leads to substantially more efficient algorithms than commonly-used methods in high-dimensional settings and for which the computational cost of one step scales linearly with $p$.
The design of these algorithms utilizes the observation that in large $p$, small $n$ settings posterior correlations will be negligible for the vast majority of the $p$ inclusion indicators. The algorithms adaptively build suitable non-local Metropolis-Hastings type proposals that result in moves with expected squared jumping distance \citep{MR1425429} significantly larger than standard methods. In idealized examples the limiting versions of our adaptive algorithms converge in $\mathcal{O}(1)$ and result in super-efficient sampling. They outperform independent
 sampling in terms of the expected squared jump distance and also in the sense of the central limit theorem asymptotic variance. This is in contrast to the behaviour of optimal local random walk Metropolis algorithms that on analogous idealized targets need at least $\mathcal{O}(p)$ samples to converge \citep{RGG97}.  The performance of our algorithms is studied empirically in realistic high-dimensional problems for both synthetic and real data. In particular, in Section \ref{sec_sim_data},
for a well studied synthetic data example, speedups of up to 4 orders of magnitude are observed compared to standard algorithms.
Moreover, in Section \ref{sec_tecator}, we show the efficiency of the method in the presence of multicollinearity on a real data example with $p=100$ variables, and in Section \ref{sec_PCR}, we present real data gene expression examples with $p=22\ 576$ and with $p=79\ 748$, 
and reliably  estimate the posterior inclusion probabilities for all variables. All proofs are grouped in  the Supplementary Material.

\section{Design of the Adaptive Samplers}

\subsection{The Setting} \label{sec:setting}

Our approach  is applicable to general regression settings but we will focus on normal linear regression models. This will allow for clean efficiency comparisons independent of model-specific sampling details (e.g.~of a reversible jump implementation). We define
$\gamma=(\gamma_1,\dots,\gamma_p) \in \Gamma = \{0,1\}^p$ to be a vector of
 indicator variables with  $\gamma_i=1$ if the $i$-th variable is included in the model and
 $p_{\gamma}=\sum_{j=1}^p \gamma_j$. We
 consider the model specification
\begin{equation}
y=\alpha {\bf 1}_n + X_{\gamma}\beta_{\gamma} + e,\qquad e\sim\N(0,\sigma^2 I_n)
\label{linreg}
\end{equation}
where $y$ is an $(n\times 1)$-dimensional vector of responses, ${\bf a}_q$ represents a $q$-dimensional column vector with entries $a$,
 and  $X_\gamma$ is
a $(n\times p_\gamma)$-dimensional data matrix
formed using the included variables.
We consider Bayesian variable selection and, for clarity of exposition and validity of comparisons, we will assume the commonly used prior structure
\begin{equation}p(\alpha,\sigma^2,\beta_{\gamma},\gamma)\propto \sigma^{-2}p(\beta_{\gamma}\mid\sigma^2, \gamma)p(\gamma)\label{prior}
\end{equation}
with $\beta_{\gamma}  \mid  \sigma^2,\gamma  \sim  \N(0,\sigma^2 V_{\gamma})$, and $p(\gamma)  =  h^{p_{\gamma}}(1-h)^{p-p_{\gamma}}$.
The hyperparameter $0<h<1$ is the prior probability that a particular variable is included in the model and $V_{\gamma}$ is often chosen as proportional to $(X_\gamma^T X_\gamma)^{-1}$ (a $g$-prior) or to the identity matrix (implying conditional prior independence between the regression coefficients). For both priors, the marginal likelihood $p(y\mid \gamma)$ can be calculated analytically. 
The prior can be further extended with hyperpriors, for example, assuming that $h\sim\Be(a,b)$.

We will consider sampling from the target distribution $\pi_p(\gamma)=p(\gamma\vert y)$ using a non-symmetric Metropolis-Hastings kernel. Let the probability of proposing to move from model $\gamma$ to $\gamma'$ be
\begin{equation}
q_{\eta}(\gamma,\gamma')=\prod_{j=1}^p q_{\eta,j}(\gamma_j,\gamma'_j)
\label{gen_prop}
\end{equation}
where $\eta = (A,D) = (A_1,\dots, A_p, D_1, \dots, D_p)$,
$q_{\eta,j}(\gamma_j=0, \gamma_j'=1)=A_j$ and
 $q_{\eta,j}(\gamma_j=1,\gamma_j'=0)=D_j$.
 The proposal can be quickly sampled, the parametrisation allows
optimisation of the expected squared jumping distance,  and multiple variables can be added to or deleted from the model in one iteration.
The proposed model is accepted using the standard Metropolis-Hastings acceptance probability
\begin{equation}
a_{\eta}(\gamma,\gamma') = \min\left\{1,\frac{\pi_p(\gamma') q_{\eta}(\gamma',\gamma)}{\pi_p(\gamma) q_{\eta}(\gamma,\gamma')}\right\}.
\label{MH_accept}
\end{equation}


\subsection{In Search of Lost Mixing Time: Optimising the Sampler} \label{sec:optimising}

The transition kernel in (\ref{gen_prop}) is highly parameterised with $2p$ parameters and these will be tuned using adaptive Markov chain Monte Carlo methods
\citep[see {\it e.g.}][]{andrieuthoms08, roberts2009examples, MR3360496}. These methods allow the tuning of parameters on the fly to improve mixing using some computationally accessible performance criterion whilst maintaining the ergodicity of the chain. Suppose that $\mu_p$ is a $p$-dimensional probability density function which has the form $\mu_p = \prod_{j=1}^p f$.
A commonly used result is that the optimal scale of a random walk proposal for $\mu_p$ leads to a mean acceptance rate of $0.234$ as $p\rightarrow\infty$  for some smooth enough $f$.
The underlying analysis also implies that the optimised random walk Metropolis  will converge to stationarity in $\mathcal{O}(p)$ steps.
This is a useful guide even in moderate dimensions and well beyond the restrictive independent, identically distributed product form assumption of \cite{RGG97}. \cite{lamnisos12} show that this rule can be effectively used to tune a Metropolis-Hastings sampler for Bayesian variable selection.
However, other results suggest that other optimal scaling rules could work well in Bayesian variable selection problems.
Firstly,  \cite{MR3025684} established,
under additional regularity conditions,
that if $f$
is discontinuous, the optimal mean acceptance rate for a Metropolis-Hastings random walk is $e^{-2}\approx 0.1353$ and the chain mixes
in $\mathcal{O}(p^2)$ steps, an order of magnitude slower than with smooth target densities $f$. Rather surprisingly, \cite{neal2016optimal} show that the optimally tuned independence sampler in this settings recovers the $O(p)$ mixing and acceptance rate of $0.234$
 without any additional smoothness conditions.
Secondly, \cite{MR1613256}
considered optimal scaling of the random walk Metropolis-Hastings algorithm on $\Gamma=\{0,1\}^p$ for the product measures
\[\mu_p(\gamma_1, \dots, \gamma_p) =   s^{p_\gamma}(1-s)^{p- p_\gamma},\qquad \gamma=(\gamma_1,\dots,\gamma_p)\in\Gamma, \quad 0<s<1.
\]
If $s$ is close to $1/2$, the optimal $\mathcal{O}(p)$ mixing rate occurs as $p$ tends to infinity if the mean acceptance rate is 0.234. If $s\rightarrow 0$ as $p\rightarrow \infty$,
the numerical results of Section 3 in \cite{MR1613256}  indicate that the optimally tuned random walk Metropolis proposes to change two $\gamma_j$'s at a time but that
 the acceptance rate deteriorates to zero resulting in the chain not moving. This suggests the actual mixing in this regime is slower than the $\mathcal{O}(p)$ observed for smooth continuous densities.

In Bayesian variable selection, it is natural to assume that the variables differ in posterior inclusion probabilities and so we consider target densities that have the form
\begin{equation}\label{our-target}
\pi_p(\gamma) = \prod_{j=1}^{p}\pi_j^{\gamma_j}(1-\pi_j)^{1-\gamma_j},\qquad \gamma\in\Gamma
\end{equation}
where $0<\pi_j<1$ for $j=1,\dots,p$.
Consider  the non-symmetric Metropolis-Hastings algorithm with the product form proposal $q_{\eta}(\gamma, \gamma')$ given by \eqref{gen_prop} targeting the posterior distribution given by~\eqref{our-target}.
%
Note that  $\alpha_{\eta}(\cdot, \cdot) \equiv 1$ for any choice of $\eta = (A,D)$ satisfying \begin{equation}\label{accept1}
\frac{A_j}{D_j} = \frac{\pi_j}{1-\pi_j}, \quad \textrm{for every } j.\end{equation} To discuss optimal choices of $\eta,$ we consider several commonly used criteria for Markov chains with stationary distribution $\pi$ and transition kernel $P$ on a finite discrete state space $\Gamma$.  The \emph{mixing time} of a Markov chain
 \citep{roberts2004general}
is
$
\rho := \min\{t: \max_{\gamma \in \Gamma} \|P^t(\gamma, \cdot) - \pi(\cdot)\|_{TV} < 1/2\}
$ where $\|\cdot\|_{TV}$ is the total variational norm.
 If  $\Gamma = \{0,1\}^p$,
it is natural to define the \emph{expected squared jumping distance}  \citep{MR1425429} as
$\E_{\pi}\left[\sum_{j=1}^p \vert\gamma^{(0)}_j - \gamma^{(1)}_j\vert^2\right]$ where $\gamma^{(0)}$ and $\gamma^{(1)}$
are two consecutive values in a Markov chain trajectory, which
  is the average number of variables changed in one iteration.
  Suppose that the Markov chain is ergodic, then, for any function $f: \Gamma \to \mathbb{R}$,
$
\frac{1}{\sqrt{n}}\sum_{k=0}^{n-1} f(\gamma^{(k)}) \stackrel {D}{\rightarrow} N(E_{\pi}f, \sigma^2_{P,f}),
$
where the constant $\sigma^2_{P,f}$ depends on the transition kernel $P$ and function $f$.
 Consider transition kernels $P_1$ and $P_2$. If $\sigma^2_{P_1,f} \leq \sigma^2_{P_2,f} $  for every $f,$ then $P_1$ dominates $P_2$ in \emph{Peskun ordering} \citep{MR0362823}. If $P_1$ dominates all other kernels from a given class, then $P_1$ is optimal in this class with respect to Peskun ordering. Apart from toy examples, Peskun ordering can be rarely established without further restrictions. Hence, for the variable selection problem, where posterior inclusion probabilities are often of interest, we consider Peskun ordering for the class $\mathbb{L}(\Gamma)$  of linear combinations of univariate functions,
\begin{equation} \label{lin_fun}\mathbb{L}(\Gamma) := \left\{f:\Gamma \to \mathbb{R}: f(\gamma) = a_0 + \sum_{j=1}^p a_j f_j(\gamma_j) \right\}.
\end{equation}

We consider two proposals which satisfy \eqref{accept1}. The \emph{independent proposal} for which $A_j=1-D_j=\pi_j $ and
the \emph{random walk proposal} for which $A_j = \min\{1, \frac{\pi_j}{1-\pi_j}\}$ and $D_j =  \min\{1, \frac{1- \pi_j}{\pi_j}\}$. The following proposition shows that the random walk proposal has more desirable properties.

\begin{proposition} \label{prop_AD_choices} Consider the class of Metropolis-Hastings algorithms with target distribution given by~\eqref{our-target}  and proposal $q_{\eta}(\gamma, \gamma')$ given by~\eqref{gen_prop} with the independent or random walk proposal. Let $Var_{\pi}f $ be the stationary variance of $f$ under $\pi_p(\gamma)$ and
$\pi^{(j)} := \{1-\pi_j, \pi_j\}$. 
Then,
\vspace*{-8pt}
\begin{enumerate}
\item[(i)] the independent proposal leads to
\begin{itemize}
\item[(a)] independent sampling and optimal mixing time $\rho = 1;$
\item[(b)] the  expected squared jumping distance is $E_{\pi}[\Delta^2] = 2\sum_{j=1}^p \pi_j(1-\pi_j)$;
\item[(c)] the asymptotic variances is $\sigma_{P,f}^2 = Var_{\pi}f$ for arbitrary $f$ and  $\sigma_{P,f}^2 = Var_{\pi}f = \sum_{j=1}^p a_j^2 Var_{\pi^{(j)}}f_j$  for $f \in \mathbb{L}(\Gamma)$;
\end{itemize}

\item[(ii)] the random walk proposal leads to
\begin{itemize}
\item[(a)] the expected squared jumping distance is
$E_{\pi}[\Delta^2] = 2\sum_{j=1}^p\min\{1-\pi_j, \pi_j\}$, which is maximal;
\item[(b)]
 the asymptotic variance is $\sigma_{P,f}^2 = \sum_{j=1}^p \big(2 \max \{1-\pi_j, \pi_j\} -1 \big)a_j^2 Var_{\pi^{(j)}}f_j$
  for $f \in \mathbb{L}(\Gamma)$
and it is
optimal with respect to the Peskun ordering for the class of linear functions~$\mathbb{L}(\Gamma)$ defined in \eqref{lin_fun}.\end{itemize}
\end{enumerate}
\end{proposition}


\begin{remark}
The differences of the expected squared jumping distance and  asymptotic variance for the two proposals is largest when $\pi_j$ is close to $1/2$.
\end{remark}

\begin{remark}
In discrete spaces, \cite{ScCh11} argue that the mutation rate
\begin{equation*}
\bar{a}_M=\int \mathbb{I}(\gamma\neq \gamma')a_\eta(\gamma,\gamma')q_{\eta}(\gamma,\gamma')\pi(\gamma) d\gamma'\,d\gamma,
\end{equation*}
which excludes moves which do not change the model, is more appropriate than  average acceptance rate.
%
The mutation rate is
$
\bar{a}_M
=1 - \prod_{j=1}^p [
(1 - \pi_j)^2 + \pi_j^2
]$
with independent sampling and  is
$
\bar{a}_M
=1 - \prod_{j=1}^p \vert 2\pi_j-1\vert$
with the random walk proposal.
Therefore, the random walk proposal always leads to a higher mutation rate.
\end{remark}

These results suggest that the random walk proposal should be preferred to the independent proposal when designing a Metropolis-Hastings sampler for idealised posteriors of the form in \eqref{our-target}.
In practice, the posterior distribution will not have a product form but can anything be said about its form when $p$ is large? The following result sheds some light on this issue. We define $\mbox{BF}_j(\gamma_{-j})$ to be the Bayes factor of including the $j$-th variable given the values of $\gamma_{-j}=(\gamma_1,\dots,\gamma_{j-1},\gamma_{j+1},\dots,\gamma_p)$ and denote by $\gamma_0$ the vector $\gamma$ without $\gamma_j$ and $\gamma_k$.
\begin{proposition}
Let
$a=\frac{\mbox{BF}_j(\gamma_k=1,\gamma_0)}
 {\mbox{BF}_j(\gamma_k=0,\gamma_0)}$.
If (i) $a \rightarrow 1$ or
(ii) $a\rightarrow A<\infty$ and $\mbox{BF}_j(\gamma_k=0,\gamma_{0})
h\rightarrow 0$ then
$
p(\gamma_j=1\vert \gamma_k=1,\gamma_0)
\rightarrow p(\gamma_j=1\vert \gamma_k=0,\gamma_0)$.
\end{proposition}

\begin{figure}[!h]
\begin{center}
\includegraphics[scale=1, trim=0mm 0mm 130mm 220mm, clip]{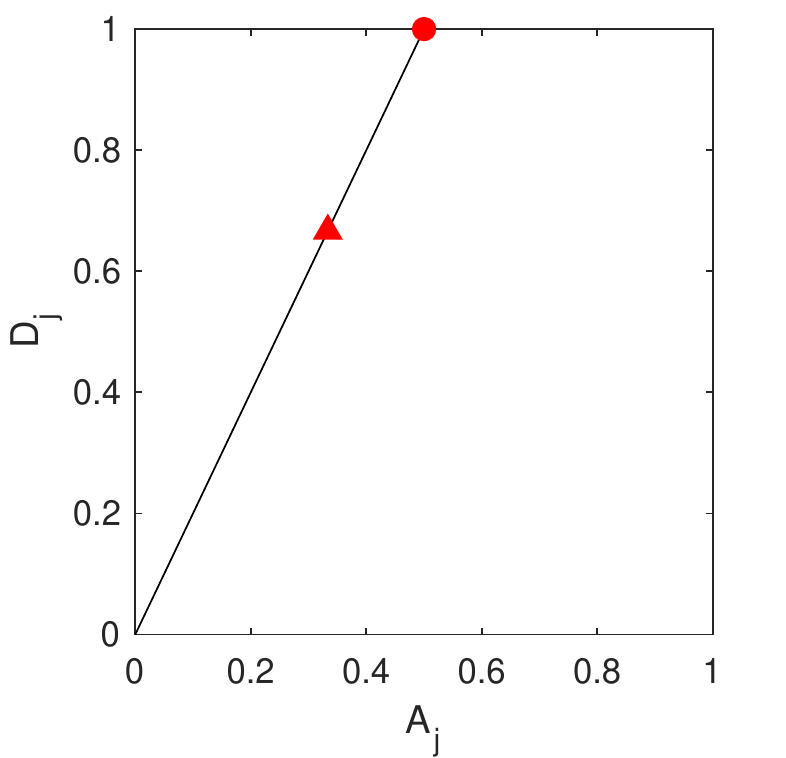}
\end{center}
\caption{\label{fig_segment} \small The solid black segment presents $A_j$'s and $D_j$'s corresponding to different choices of  $\zeta_j \in [0,1]$ in \eqref{individually_scaled} with $\pi_j=1/3$. Any point in the segment results in acceptance probability~$=1$ for the idealized target~\eqref{our-target}. The iid sampling (i), marked with a triangle, is a shrunk version of the superefficient sampling~(ii), marked with a bullet. }
\end{figure}

This result gives condition under which $\gamma_j$ and $\gamma_k$ are approximately independent.  Condition (ii) is interesting in large $p$ settings: $\gamma_j$ and $\gamma_k$ are approximately independent if $p$ is large (and so $h$ is small) and $\mbox{BF}_j(\gamma_k=0,\gamma_{0})$ is not large, {\it i.e.} the evidence in favour of including $\gamma_j$ is not large. This will be the case for all variables apart from the most important.
Although this result provides some reassurance, there will be some posterior correlation in many problems and the random walk proposal may propose to change too many variables leading to low acceptance rates. This can be addressed by using a scaled proposal  of the form \begin{equation}\label{individually_scaled}
A_j = \zeta_j \min\left\{1, \frac{\pi_j}{1-\pi_j}\right\}, \qquad D_j = \zeta_j \min\left\{1, \frac{1- \pi_j}{\pi_j}\right\}.\end{equation}
The family of these proposals for $\zeta_j \in [0,1]$ form a line segment for $(A_j, D_j)$ between $(0, 0)$ and $\left(\min\left\{1, \frac{\pi_j}{1-\pi_j}\right\}, \min\left\{1, \frac{1- \pi_j}{\pi_j}\right\}\right)$, illustrated in Figure~\ref{fig_segment} (Supplementary Material \ref{add_fig}). The independent proposal corresponds to the point on this line where $\zeta_j=\max\{\pi_j, 1-\pi_j\}$.

In the next section, we devise adaptive MCMC algorithms to tune proposals of the form \eqref{gen_prop} so that
$A_j$'s and $D_j$'s lie approximately on this line.
Larger values of $\zeta_j$ tend to lead to larger jumps whereas smaller values of $\zeta_j$ tend to increase  acceptance. These algorithms aim to find a point which balances this trade-off.
We define two strategies for adapting $\eta$: \emph{Exploratory Individual Adaptation} and \emph{Adaptively Scaled Individual Adaptation}.

\cite{craiu2009learn} showed empirically that running multiple independent Markov chains with the same adaptive parameters
 improves the rate of convergence of adaptive algorithms towards their target acceptance rate in the context of the classical adaptive Metropolis algorithm of \cite{haario2001adaptive} (see also Bornn et al. 2013)
\nocite{BJDD12}. Therefore, we consider
 algorithms with different numbers of independent parallel chains (but the same parameters of the proposal) and refer to this as { multiple chain acceleration}.
To avoid the algorithms becoming trapped in well separated modes, we also consider parallel tempering versions of the algorithms, following \cite{MiMoVi12} as explained in Supplementary Material \ref{PT}.

At this point, it is helpful to define some notation.
Let $\eta^{(i)}=(A^{(i)},D^{(i)})$ and
$\gamma^{(i)}$ be the values of $\eta$ and $\gamma$ at the start of the $i$-th iteration, and $\gamma'$
be the subsequently proposed value. Let $
a_i = a_{\eta^{(i)}}(\gamma^{(i)}, \gamma')$ be the acceptance probability at the $i$-th iteration.
We define for $j=1,\dots,p$,
 \[
\gamma^{A\,(i)}_j=
\left\{\begin{array}{ll}
1\mbox{ if }\gamma'_j\neq \gamma_j^{(i)}\mbox{ and }\gamma_j^{(i)}=0\\
0\mbox{ otherwise}
\end{array}\right.,\qquad
\gamma^{D\,(i)}_j=
\left\{\begin{array}{ll}
1\mbox{ if }\gamma'_j\neq \gamma_j^{(i)}\mbox{ and }\gamma_j^{(i)}=1\\
0\mbox{ otherwise}
\end{array}\right.
\]
and the map $\logit_{\epsilon}: (\epsilon,1-\epsilon)\rightarrow \mathbb{R}$ by
$
\logit_{\epsilon}(x) = \log(x-\epsilon) - \log(1 - x - \epsilon),
$
where $0\leq\epsilon\leq1/2$. This reduces to the usual logit transform if $\epsilon=0$.

\subsection{Remembrance of Things Past: Exploratory Individual Adaptation}

 The first adaptive strategy is a general purpose method that we term {\it Exploratory Individual Adaptation} (EIA). It aims to find pairs $(A_j, D_j)$ on the line segment defined by \eqref{accept1}
which lead to good mixing. Proposals with larger values of $A_j$ and $D_j$ will tend to propose more changes to the included variables but will also tend to reduce the average acceptance probability or mutation rate. The method introduces two tuning parameter $\tau_L$ and $\tau_U$.
  There are three types of updates for $A^{(i)}$ and $D^{(i)}$ which move towards the correct ratio $A_j/D_j$ and then along the segment (note that the slope of the segment is not known in practice, as it depends on $\pi_j$). Unless otherwise stated, $A^{(i+1)}_j=A^{(i)}_j$ and $D^{(i+1)}_j=D^{(i)}_j$:
\begin{enumerate}
\item Both the \emph{expansion step} and the \emph{shrinkage step} change $A^{(i+1)}_j$ and $D^{(i+1)}_j$ for $j$ in
$\gamma^{A(i)}$ and $\gamma^{D(i)}$ to adjust the average squared jumping distance whilst maintaining that
$A^{(i+1)}_j / D^{(i+1)}_j \approx A^{(i)}_j / D^{(i)}_j$.
The expansion step is used if a promising move is proposed (if $a_i>\tau_U$) and sets $A^{(i+1)}_j$ and $D^{(i+1)}_j$ larger than $A^{(i)}_j$ and $D^{(i)}_j$ respectively. Similarly, the shrinkage step is used if an unpromising move has been proposed (if $a_i<\tau_L$) and $A^{(i+1)}_j$ and $D^{(i+1)}_j$ are set smaller than $A^{(i)}_j$ and $D^{(i)}_j$.
\item The \emph{correction step} aims to  increase the average acceptance rate by correcting the ratio between $A$'s and $D$'s. If $\tau_L<a_i<\tau_U$, we set $A^{(i+1)}_j>A^{(i)}_j$ and $D^{(i+1)}_j<D^{(i}_j$
 if $\gamma^{D(i)}_j=1$ and $A^{(i+1)}_j<A^{(i)}_j$
 and
 $D^{(i+1)}_j>D^{(i)}_j$ if $\gamma^{A(i)}_j=1$.\end{enumerate}

\begin{figure}[!h]
\begin{center}
\begin{tabular}{cc}
\includegraphics[scale=0.37, clip, trim=0mm 0mm 0mm 0mm]{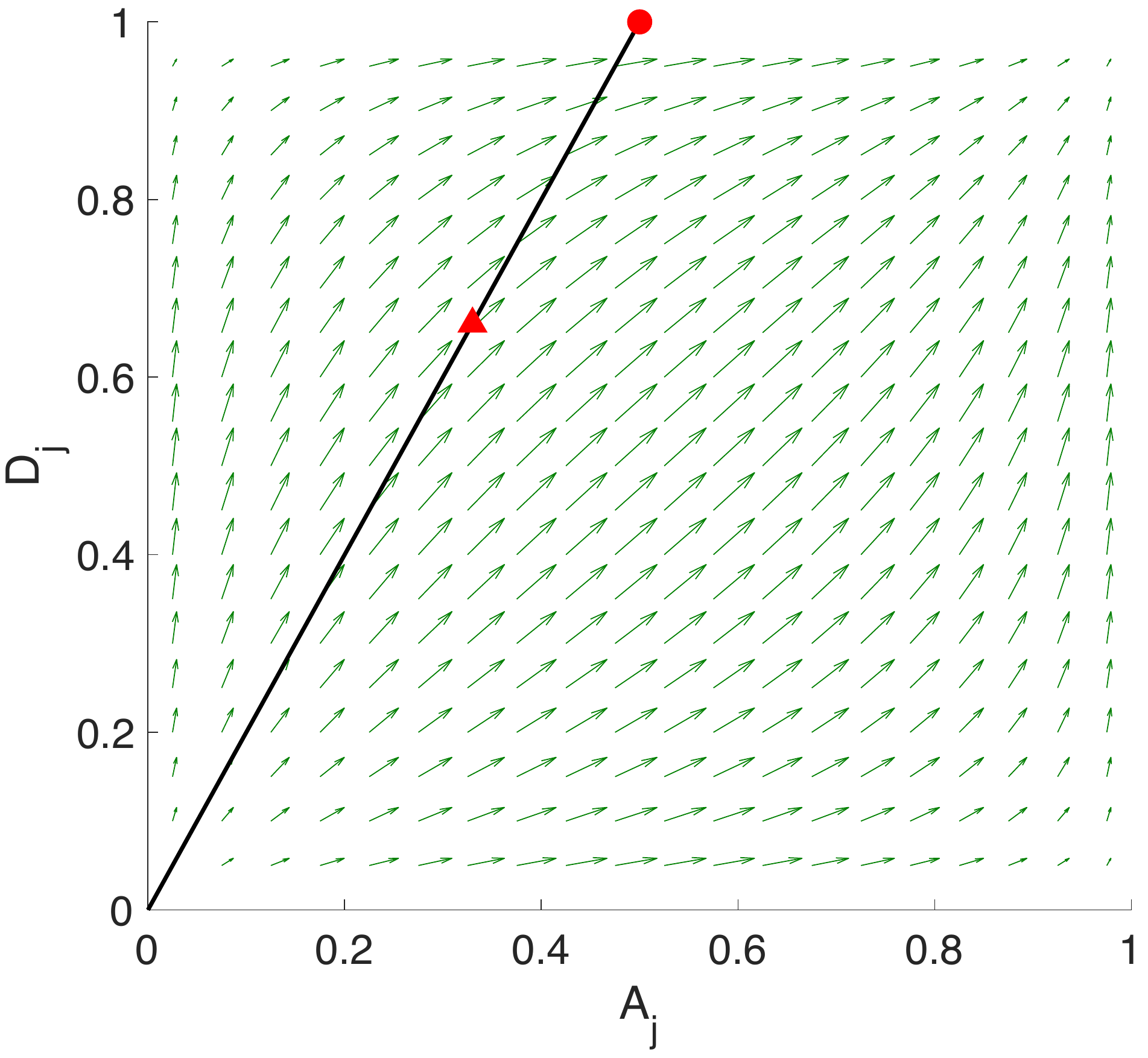} &
\includegraphics[scale=0.37, clip, trim=0mm 0mm 0mm 0mm]{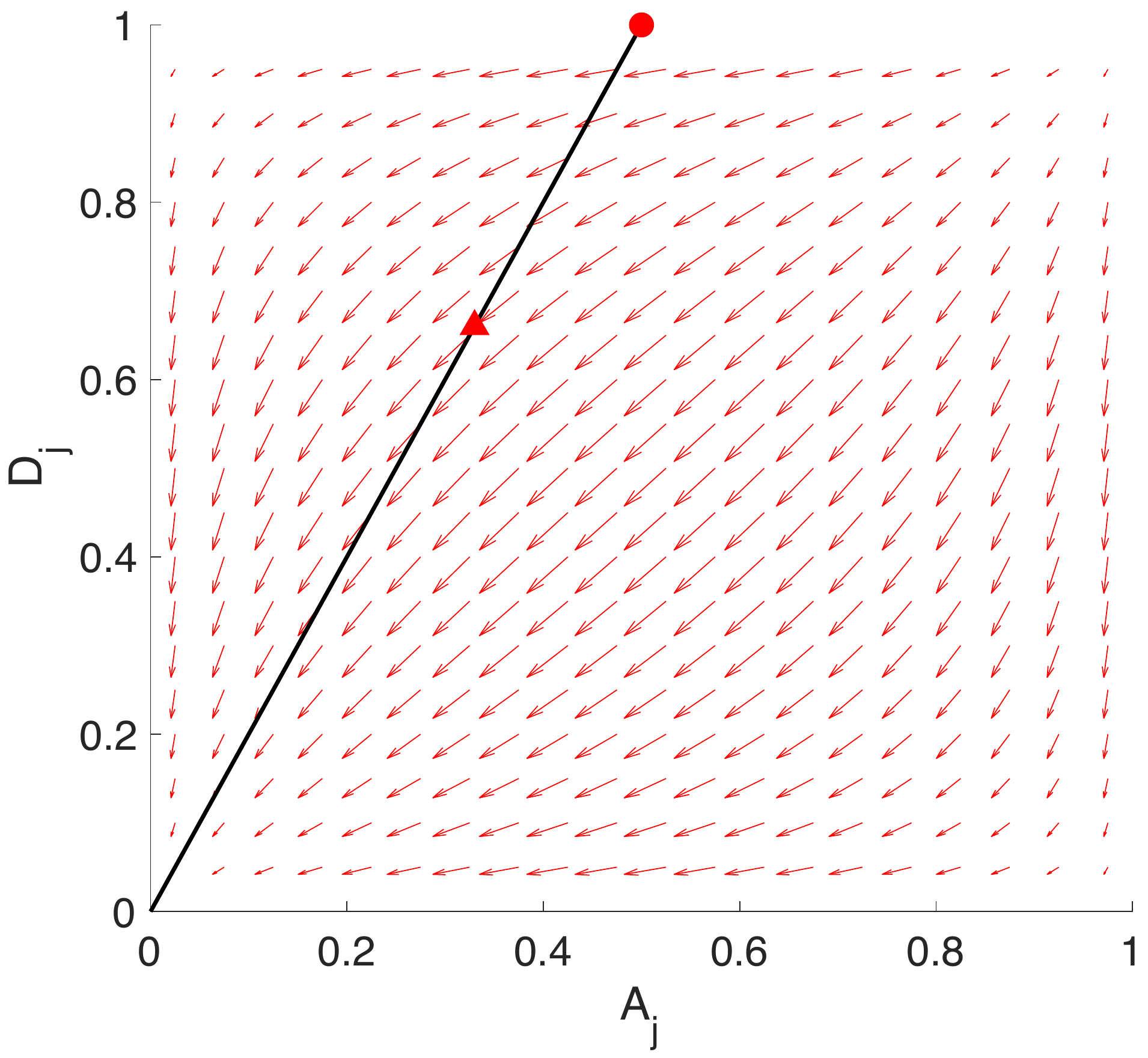}\\
Gradient field of the \emph{expansion step}  & Gradient field of the \emph{shrinkage step} \\
for $\gamma^{A(i)}$ and $\gamma^{D(i)}$ & for $\gamma^{A(i)}$ and $\gamma^{D(i)}$\\
\includegraphics[scale=0.37, clip, trim=0mm 0mm 0mm 0mm]{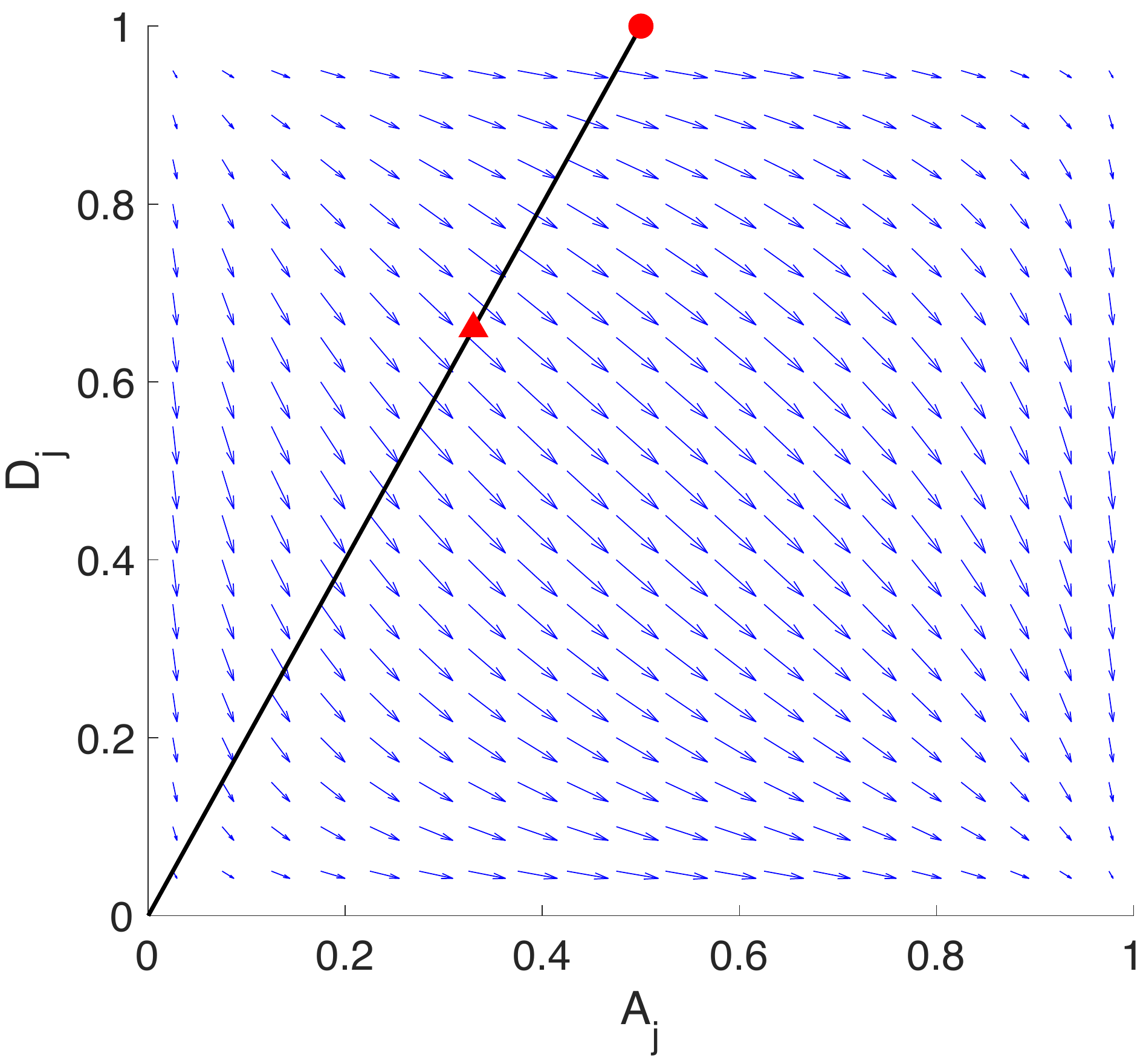} &
\includegraphics[scale=0.37, clip, trim=0mm 0mm 0mm 0mm]{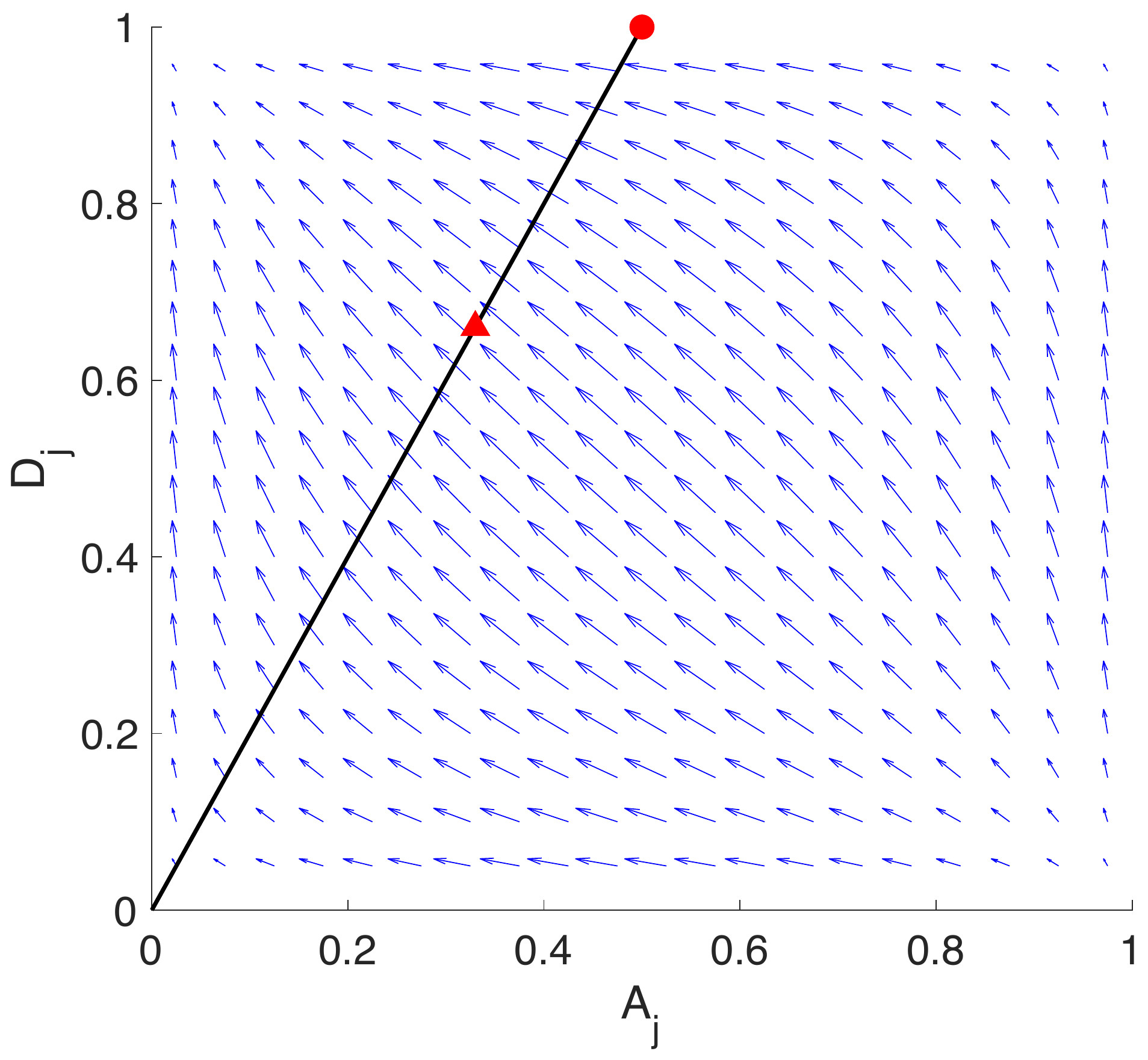}\\
Gradient field of the \emph{correction step}  &
Gradient field of the \emph{correction step}\\
for  $\gamma^{D(i)}$ &  for  $\gamma^{A(i)}$
\end{tabular}
\end{center}
\caption{\label{gradient} \small Gradient fields guiding parameter updates of the exploratory individual adaptation algorithm towards and along the segment.}
\end{figure}

 The gradient fields of these updates are shown in Figure \ref{gradient} (Supplementary Material \ref{add_fig}). These three moves can be combined into the following adaptation of  $A^{(i)}$ and $D^{(i)}$
\begin{eqnarray}
\logit_{\epsilon}A^{(i+1)}_j& = &\logit_{\epsilon}A^{(i)}_j
+
\phi_i \left(
\gamma^{A(i)}_j d_i(\tau_U) +\, \gamma^{D(i)}_j
d_i(\tau_L)
- \gamma^{A(i)}_j (1-d_i(\tau_U))\right), \label{eqn:adap_A}
\\
\logit_{\epsilon}D^{(i+1)}_j & = &\logit_{\epsilon}D^{(i)}_j
+
\phi_i \left(
\gamma^{D(i)}_j d_i(\tau_U)
 +\, \gamma^{A(i)}_j
 d_i(\tau_L)
- \gamma^{D(i)}_j (1 - d_i(\tau_U))\right), \label{eqn:adap_B}
\end{eqnarray}
for $j=1\dots,p$ where  $d_i(\tau) =
 \mathbb{I}{\big\{a_i\geq \tau\big\}}$ and
$\phi_i=O(i^{-\lambda})$ for some constant $1/2<\lambda\leq 1$. The transformation implies that $\epsilon<A_j^{(i)}<1-\epsilon$ and
$\epsilon<D_j^{(i)}<1-\epsilon$ and we assume that $0<\epsilon<1/2$. It also implies diminishing adaptation (essentially since the derivative of the inverse logit is bounded, see Lemma \ref{lemma:diminishing_adaptation}). Based on several simulation studies, we suggest to take $\tau_L = 0.01$ and $\tau_U = 0.1$. As discussed in Section \ref{sec:optimising}, targeting a low acceptance rate is often beneficial in irregular cases, so we expect this choice to be robust in real data applications. In all our simulations with this parameter setting, the resulting mean acceptance rate was between $0.15$ and $0.35$, i.e. in the high efficiency region identified in \cite{RGG97}. We also suggest the initial choice of parameters such that  $A_j^{(1)}/D_j^{(1)} \approx h/(1-h)$ as this summarises the prior information on $\pi_j / (1-\pi_j)$, and in particular $D_j^{(1)}\equiv 1$ and  $A_j^{(1)}\equiv h$ often works well. 
The parameter $\epsilon$ controls the minimum and maximum values of $A_i$ and $D_i$. In the large $p$ setting, $A_i\approx\epsilon$ for unimportant variables and the expected number of those unimportant variables  proposed to be included at each iteration will be approximately $p\epsilon$ (since the number of excluded, unimportant variables will be close to $p$). This expected value can be controlled by choosing $\epsilon=0.1/p$.
The EIA algorithm is described in Algorithm~\ref{explore_IA} and we indicate its transition kernel at time $i$ as $P_{\eta^{(i)}}^{\textrm{EIA}}$. 

\begin{algorithm}[!h]
\label{explore_IA}
\caption{Exploratory Individual Adaptation (EIA)}
\vspace*{-12pt}
\begin{tabbing}
\enspace for $i=1$ to $i=M$\\
\qquad sample $\gamma' \sim q_{\eta^{(i)}}(\gamma^{(i)}, \cdot) \;$  \textrm{and} $\; U \sim U(0,1);$\\
\qquad if {$\;U < a_{\eta^{(i)}}(\gamma^{(i)}, \gamma')\;$}  then\   $\gamma^{(i+1)}:=\gamma'$,     else
 $ \gamma^{(i+1)}:=\gamma^{(i)}$ \\
\qquad \textrm{update $A^{(i+1)}$ using (\ref{eqn:adap_A}) and
$D^{(i+1)}$ using (\ref{eqn:adap_B})}\\
\enspace endfor
\end{tabbing}
\end{algorithm}
\vspace*{-12pt}

\subsection{Remembrance of Things Past: Adaptively Scaled Individual Adaptation}

Algorithm~\ref{explore_IA} learns two parameters $A_j^{(i)}$ and $D_j^{(i)}$ for each variable and can be slow to converge to optimal values if $p$ is large. Alternatively, we could learn $\pi_1,\dots,\pi_p$ from the chain to approximate the slope of the line defined by \eqref{accept1}
 and use the proposal \eqref{individually_scaled} with the same scale parameter for all variables. We term this approach the {\it Adaptively Scaled Individual Adaptation} (ASI) proposal. In particular, we use \begin{equation} A_j^{(i)}=\zeta^{(i)}\min\left\{1,\hat\pi^{(i)}_j/\left(1-\hat\pi^{(i)}_j\right)\right\} \quad \mbox{and} \quad D_j^{(i)}=\zeta^{(i)}\min\left\{1,\left(1-\hat\pi^{(i)}_j\right)/\hat\pi^{(i)}_j\right\}, \label{ASI_update} \end{equation}
for $j=1,\dots,p$ where $0<\zeta^{(i)}<1$ is a tuning parameter
and  $\hat\pi^{(i)}_j$ is a Rao-Blackwellised estimate of the posterior inclusion probability of variable
 $j$ at the  $i$-th iteration.
  Like
\cite{GhCl11}, we work with the Rao-Blackwellised estimate conditional on the model, marginalizing out $\alpha$, $\beta_{\gamma}$ and $\sigma^2$,
in contrast to
\cite{GuSt11} who condition on the model parameters. We assume that $V_{\gamma} = g I_{p_{\gamma}}$, where $I_q$ is the $q\times q$ identity matrix.
After $N$ posterior samples, $\gamma^{(1)},\dots,\gamma^{(N)}$,
the Rao-Blackwellised estimate of $\pi_j=p(\gamma_j=1\vert y)$ is
\begin{equation}\label{eq:RB}
\hat\pi_j = \frac{1}{N}
 \sum_{k=1}^N
 \frac{\tilde{h}_j^{(k)}\,\mbox{BF}_j\left(\gamma_{-j}^{(k)}\right)}{1-\tilde{h}_j^{(k)}+\tilde{h}_j^{(k)}\,\mbox{BF}_j\left(\gamma_{-j}^{(k)}\right)}
\end{equation}
where
$\tilde{h}_j^{(k)} = h$ if $h$ is fixed or $\tilde{h}_j^{(k)} = \frac{\#\gamma^{(k)}_{-j}+1+a}{p+a+b}$ if $h\sim\Be(a,b)$.
Let $Z_{\gamma} = [{\bf 1}_n \ X_{\gamma}]$,
$\Lambda_{\gamma} = \left(
\begin{array} {cc}
0 & {\bf 0}_{p_{\gamma}}^T\\
{\bf 0}_{p_{\gamma}} & V^{-1}_{\gamma}
\end{array}
\right)$,
$F = (Z_{\gamma}^TZ_{\gamma} + \Lambda_{\gamma})^{-1}$
and $A = y^Ty - y^T Z_{\gamma} F Z_{\gamma}^T y$.
If $\gamma_j=0$,  
\[
\mbox{BF}_j(\gamma_{-j})={d_j^{\uparrow}}^{-1/2}g^{-1/2}
\left(
\frac{A-\frac{1}{d_j^\uparrow}
(y^T x_j-y^T Z_{\gamma} F Z_{\gamma}^T x_j)^2}{A}
\right)^{-n/2}
\]
with $
d_j^\uparrow = x_j^T x_j + g^{-1} - (x_j^T Z_{\gamma})F(Z_{\gamma}^T x_j)$.
If $\gamma_j=1$, we define $z_j$ to be ordered position of the included variables
($z_j=1$ if $j$ is the first included variable, etc.), then
\[
\mbox{BF}_{j}(\gamma_{-j})={d_j^{\downarrow}}^{-1/2}g^{-1/2}
\left(
\frac{A}{A +
{d_j^\downarrow}
(y^T Z_{\gamma}
F_{\cdot, z_j+1})^2}
\right)^{-n/2}
\]
where $
d_j^\downarrow = 1/F_{z_j+1,z_j+1}$.  
These results allow the contribution to the Rao-Blackwellised estimates for all values of $j$ to be calculated in $O(p)$ operations at each iteration
 if the values of $F$  and $A$ (which are needed for calculating the marginal likelihood) are stored. Derivations 
are provided in Supplementary Material \ref{SM:RB}.
The value of $\zeta^{(i)}$ is updated using
\begin{equation}
\label{eqn:adap_C}
\logit_{\epsilon} \zeta^{(i+1)} = \logit_{\epsilon} \zeta^{(i)} + \phi_i (a_i - \tau),
\end{equation}
where $\tau$ is a targeted acceptance rate. We use $\epsilon=0.1/p$ as in Algorithm 1. We shall see (in Lemma \ref{lemma:diminishing_adaptation}) that ASI also satisfies diminishing adaptation by verifying that the Rao-Blackwellised estimate in \eqref{eq:RB} evolves at the rate $1/i$ and reiterating the argument about inverse logit derivatives.
To avoid proposing to change no variable with high probability, we set $\zeta^{(i+1)} = 1/\Delta^{(i+1)}$ if $\zeta^{(i+1)}\Delta^{(i+1)}<1$ where $\Delta^{(i+1)}=2\sum_{j=1}^p \min\{\pi_j^{(i+1)},1-\pi_j^{(i+1)}\}$. This ensures that the algorithm will propose to change at least one variable with high probability. The ASI algorithm is described in Algorithm~\ref{explore_ISA} and we indicate its transition kernel at time $i$ as $P_{\eta^{(i)}}^{\textrm{ASI}}$. We use $\kappa = 0.001$ to avoid the estimated probabilities becoming very small.

\begin{algorithm}[h!]
\label{explore_ISA}
\caption{Adaptively Scaled Individual Adaptation (ASI)}
\vspace*{-12pt}
\begin{tabbing}
\enspace for $i=1$ to $i=M$\\
\qquad sample $\gamma' \sim q_{\eta^{(i)}}(\gamma^{(i)}, \cdot) \;$  \textrm{and} $\; U \sim U(0,1);$\\
\qquad if {$\;U < a_{\eta^{(i)}}(\gamma^{(i)}, \gamma')\;$}  then\
    $\gamma^{(i+1)}:=\gamma'$, else  $ \gamma^{(i+1)}:=\gamma^{(i)}$ \\
\qquad Update $\hat\pi_1^{(i+1)},\dots,\hat\pi_p^{(i+1)}$ as in (\ref{eq:RB}) and
 set $\tilde\pi_j^{(i+1)}=\kappa + (1 - 2\kappa)\, \hat\pi_j^{(i+1)}$\\
 \qquad Update $\zeta^{(i+1)}$ as in \eqref{eqn:adap_C} \\
\qquad Calculate
 $A_j^{(i+1)}=\zeta^{(i+1)}\min\left\{1,\tilde\pi^{(i+1)}_j/\left(1-\tilde\pi^{(i+1)}_j\right)\right\}$ for $j=1,\dots,p$\\
 \qquad  Calculate $D_j^{(i+1)}=\zeta^{(i+1)}\min\left\{1,\left(1-\tilde\pi^{(i+1)}_j\right)/\tilde\pi^{(i+1)}_j\right\}$ for $j=1,\dots,p$
 \\
\enspace endfor
\end{tabbing}
\end{algorithm}
\vspace*{-14pt}

\section{Ergodicity of the Algorithms} \label{sec:ergodcity}

Since adaptive Markov chain Monte Carlo algorithms violate the Markov condition, the standard and well developed Markov chain theory can not be used to establish ergodicity and we need to derive appropriate results for our algorithms. 
We verify validity of our algorithms by establishing conditions introduced in \cite{MR2340211}, namely simultaneous uniform ergodicity and diminishing adaptation.


The target posterior specified in Section \ref{sec:setting} on the model space $\Gamma$ is
\begin{equation}\label{eqn:target}
\pi_p(\gamma) = \pi_p(\gamma\mid y) \propto  p(y|\gamma)p(\gamma)
\end{equation} with $p(y|\gamma)$ available analytically, and the vector of adaptive parameters at time $i$ is  \begin{equation} \label{parameters}
\eta^{(i)} = (A^{(i)},D^{(i)}) \;\; \in \;\; [\epsilon, 1-\epsilon]^{2p}  =:  \Delta_{\epsilon}, \quad \textrm{with} \quad 0 < \epsilon < 1/2, \end{equation}
with the update strategies in Algorithm 1 or 2.
$P_{\eta}(\gamma, \cdot)$ denotes the non-adaptive Markov chain kernel corresponding to a fixed choice of $\eta$.
Under the dynamics of either algorithm, for $S \subseteq \Gamma$ we have
\begin{eqnarray}\nonumber
P_{\eta}(\gamma, S) & = &
\mathbb{P}\Big[\gamma^{(i+1)} \in S \, \Big| \, \gamma^{(i)}= \gamma, \eta^{(i)} =\eta \Big]
 \\ & = & \sum_{\gamma' \in S }q_{\eta}(\gamma, \gamma') a_{\eta}(\gamma, \gamma') + \mathbb{I}{\{\gamma \in S\}} \sum_{\gamma' \in \Gamma}
q_{\eta}(\gamma, \gamma')\big(1- a_{\eta}(\gamma, \gamma') \big). \label{trans_kernel}
\end{eqnarray}
In the case of multiple chain acceleration, where $L$ copies of the chain are run, the model state space becomes the product space and the current state of the algorithm at time $i$  is $\gamma^{\otimes L,\, (i)} = (\gamma^{1,\, (i)},\dots, \gamma^{L,\, (i)})   \in \Gamma^L$. The single chain version corresponds to $L=1$ and all results apply.

To assess ergodicity, we need to define the distribution of the adaptive algorithm at time $i$, and the associated total variation distance: for the $l$-th copy of the chain $\{\gamma^{l, (i)}\}_{i=0}^{\infty}$ and $S \subseteq \Gamma$ define 
\begin{align*}
\mathcal{L}^{l,(i)}\big[(\gamma^{l}, \eta), S \big]  & \; := \; \mathbb{P}\Big[\gamma^{l,\, (i)} \in S \, \Big| \, \gamma^{l,\, (0)} = \gamma^{l}, \eta^{(0)} =\eta \Big], \quad \textrm{and}
\\
T^l(\gamma^{l}, \eta,i) & \; := \;  \| \mathcal{L}^{l,(i)}\big[(\gamma^{l}, \eta), \cdot \big] - \pi_p(\cdot) \|_{TV}
\; = \; \sup_{S \in \Gamma}|\mathcal{L}^{l,(i)}\big[(\gamma^{l}, \eta), S \big]  - \pi_p(S)|.
\end{align*}
We show that all the considered algorithms are ergodic and satisfy a strong law of large numbers (SLLN), {\it i.e.}~for any starting point $\gamma^{\otimes L} \in \Gamma^L$ and any initial parameter value $\eta \in \Delta_{\epsilon}$, we have:
\begin{eqnarray}
  \label{eq_thm:IA_erg}
\textrm{ergodicity:}& & \quad \lim_{i \to \infty} T^l(\gamma^{l}, \eta,i) \;  = \; 0, \quad \textrm{for any} \;  l=1, \dots, L;\qquad \textrm{and}
\\
  \label{eq_thm:IA_WLLN} \textrm{SLLN:}&  & \quad {1 \over L} \sum_{l=1}^L
{1 \over k}\sum_{i=1}^{k} f(\gamma^{l,\, (i)}) \;  \stackrel {k\to\infty}{\longrightarrow} \; \pi_p(f) \quad \textrm{almost surely, for any} \; f:  \Gamma \to \mathbb{R}.
\end{eqnarray}
To this end we first establish the following lemmas.
\begin{lemma}[Simultaneous Uniform Ergodicity]\label{lem:SUE}
The family of Markov chains defined by transition kernels $P_{\eta}$ in \eqref{trans_kernel}, targeting $\pi_p(\gamma)$ in \eqref{eqn:target}, is simultaneously uniformly ergodic for any $\epsilon > 0$ in \eqref{parameters}, and so is its multichain version. That is, for any $\delta > 0$ there exists $N=N(\delta,\epsilon) \in \mathbb{N},$ such that  for any starting point $\gamma^{\otimes L} \in \Gamma^L$ and any parameter value $\eta \in \Delta_{\epsilon}$
\[
\|P_{\eta}^{N}(\gamma^{\otimes L}, \cdot) - \pi_p^{\otimes L}(\cdot)\|_{TV} \leq \delta.
\]
\end{lemma}
\begin{lemma}[Diminishing Adaptation]\label{lemma:diminishing_adaptation}
Recall the  constant $1/2 \leq  \lambda \leq 1$ defining the adaptation rate $\phi_i=O(i^{-\lambda})$  in \eqref{eqn:adap_A}, \eqref{eqn:adap_B}, or
\eqref{eqn:adap_C}, and the parameter $\kappa >0$ in Algorithm 2. Then both algorithms: EIA  and ASI satisfy diminishing adaptation. More precisely, their transition kernels satisfy
\begin{eqnarray}\label{eq:dimini}
\sup_{\gamma \in \Gamma} \|P_{\eta^{(i+1)}}^{\bullet}(\gamma, \cdot) - P_{\eta^{(i)}}^{\bullet}(\gamma, \cdot)\| & \leq & C i^{-\lambda} ,\quad \textrm{for some} \; C<\infty,
\end{eqnarray}
where $\bullet$ stands for EIA or ASI.
\end{lemma}
Simultaneous uniform ergodicity together with diminishing adaptation leads to the following
\begin{theorem}[Ergodicity and SLLN] \label{thm:IA-PT_erg} Consider the target $\pi_p(\gamma)$ of \eqref{eqn:target}, the  constants $1/2 \leq  \lambda \leq 1$ and   $\epsilon > 0$ defining respectively the adaptation rate $\phi_i=O(i^{-\lambda})$   and region in \eqref{eqn:adap_A}, \eqref{eqn:adap_B}, or
\eqref{eqn:adap_C}, and the parameter $\kappa >0$ in Algorithm 2. Then ergodicity \eqref{eq_thm:IA_erg}
 and the strong law of large numbers
  \eqref{eq_thm:IA_WLLN} hold for each of the algorithms: EIA, ASI and their multiple chain acceleration versions.
\end{theorem}

\begin{remark} Lemma \ref{lemma:diminishing_adaptation} and  Theorem \ref{thm:IA-PT_erg} remain true with any $\lambda > 0$, however $\lambda >1$ results in finite adaptation (see e.g.  \cite{MR2340211}), and $\lambda < 1/2$ is rarely used in practice for finite sample stability concerns.
\end{remark}

Proofs can be found in Supplementary Material \ref{Proofs}. A comprehensive analysis of the algorithms for other generalised linear models or for linear models using non-conjugate prior distributions requires a case-by-case treatment, and is beyond the scope of this paper.  However, if the prior distributions of additional parameters are continuous, supported on a compact set and everywhere positive, establishing ergodicity will typically be possible with some technical care. 
%
%
%

\section{Results}

\subsection{Simulated Data} \label{sec_sim_data}

We consider the simulated data example of \cite{YaWaJo16}.
 They assume that there are $n$ observations and $p$ regressors and the data is generated from the model
\[
Y = X\beta^{\star} + e
\]
where $e\sim\N(0,\sigma^2 I)$ for $\sigma^2=1$. The first 10 regression coefficients are non-zero and we use
\[
\beta^{\star}=\mbox{SNR}\sqrt{\frac{\sigma^2\log p}{n}}
(2, -3, 2, 2, -3, 3, -2, 3, -2, 3, 0, \dots, 0)^T\in\mathbb{R}^p.
\]
The $i$-th vector of regressors is generated as  $x_i\sim\N(0,\Sigma)$ where $\Sigma_{jk}=\rho^{\vert j - k\vert}$. In our examples, we use the value $\rho=0.6$ which represents a relative large correlation between the regressors.

We are interested in the performance of the two adaptive algorithms (EIA and ASI) relative to an add-delete-swap algorithm.
We define the ratio of the relative time-standardized effective sample size of algorithm $A$ versus algorithm $B$ to be
$
r_{A,B} = ({\mbox{ESS}_A/t_A})/({\mbox{ESS}_B/t_B})
$
where $\mbox{ESS}_A$ is the effective sample size for algorithm $A$. This is estimated by making 200 runs of each algorithm and calculating
$
\hat{r}_{A,B} =({s^2_Bt_B})/({s^2_At_A}),
$
where $t_A$ and $t_B$ are the median run-times  and $s^2_A$ and $s^2_B$ are the sample variances of the posterior inclusion probabilities for algorithms $A$ and $B$.

We use the prior in (\ref{prior}) with $V_\gamma=9 I$ and $h=10/p$, implying a prior mean model size of 10.
The posterior distribution changes substantially with the SNR and the size of the data set. 
All  ten true non-zero coefficients are given posterior inclusion probabilities greater than 0.9 in the two high SNR scenarios (SNR=2 and SNR=3) for each value of $n$ and $p$ and no true non-zero coefficients are given posterior inclusion probabilities greater than 0.2 in the low SNR scenario (SNR=0.5) for each value of $n$ and $p$. In the intermediate SNR scenario (SNR=1), the number of true non-zero coefficients given posterior inclusion probabilities greater than 0.9 are 4 and 8 for $p=500$ and 3 and 0 for $p=5000$. 
Generally, the results are consistent with our intuition that true non-zero regression coefficients should be detected with greater posterior probability for larger SNR, larger value of $n$ and smaller value of $p$.


%

Table~\ref{gene_comp_5}  shows the median relative time-standardized effective sample sizes for the EIA and ASI algorithms with 5 or 25 multiple chains for different combinations of $n$, $p$ and SNR. The median is taken across the estimated relative time-standardized effective sample sizes for all posterior inclusion probabilities. 
{\footnotesize
\begin{table}[h!]
\caption{\small Simulated data: median values of $\hat{r}_{A,B}$ for the posterior inclusion probabilities over all variables where $B$ is the standard Metropolis-Hastings algorithm and $A$ is either the exploratory individual adaptation (EIA) or adaptively scaled individual adaptation (ASI) algorithm}\label{gene_comp_5}
\begin{center}
\begin{tabular}{ccrrrrrrrr}
 && \multicolumn{4}{c}{5 chains} & \multicolumn{4}{c}{25 chains}\\[5pt]
  & & \multicolumn{4}{c}{SNR} & \multicolumn{4}{c}{SNR}\\
$(n,p)$  & & 0.5 & 1 & 2 & 3 & 0.5 & 1 & 2 & 3\\[5pt]
(500, 500)         & EIA        & 4.9	&  1.8		&  5.5		& 5.1		& 1.2		& 1.5		& 2.4		& 2.3\\
                          & ASI      & 1.7	&  21.3	&  31.8	& 7.5		& 2.0		& 36.0	& 42.7	&  12.6 \\
(500, 5000)       & EIA        & 8.7	&  2.2		&  718.0	& 81.5	& 7.1		& 2.9		& 2267.2 	& 147.2 \\
                          & ASI      & 29.9	&  126.9	&  2053.1	& 2271.3	& 53.5	& 353.3	& 12319.5	& 7612.3 \\
(1000, 500)       & EIA        & 5.9   &  16.3	&  7.7   	& 4.2		& 1.6		& 80.7	& 4.4		& 1.8\\
                          & ASI      & 41.9	&  2.1		& 16.9    	& 12.0	& 32.8	& 34.0	& 27.9 	& 14.4  \\
(1000, 5000)     & EIA        & 2.2	&  2.2 	& 9167.2	&11.3		&  5.6		& 2.5		& 15960.7 	& 199.8   \\
                          & ASI      & 15.4	&  37.0	& 4423.1	& 30.8	& 54.9	& 53.4	& 11558.2	& 736.4   \\
\end{tabular}
\end{center}
\end{table}}
Clearly, the ASI algorithm outperforms the EIA algorithm  for most settings with either 5 or 25 multiple chains.  The performance of the EIA and, especially, the adaptive scaled individual adaptation algorithm with 25 chains is better than the corresponding performance with 5 chains for most cases. Concentrating on results with the ASI algorithm, the largest increase in performance compared to a simple Metropolis-Hastings algorithm occurs with SNR=2. In this case, there are three or four orders of magnitude improvements when $p=5000$ and several orders of magnitude improvements for other SNR with $p=5000$. In smaller problems with $p=500$, there are still substantial improvements in efficiency over the simpler Metropolis-Hastings sampler.

The superior performance of the ASI algorithm (which has one tuneable parameter) over the EIA algorithm (which has $2p$ tuneable parameters) is due to the substantially faster convergence of the tuning parameters of the ASI algorithm to optimal values. Plotting posterior inclusion probabilities against $A$ and $D$ at the end of a run shows that, in most cases, the values of $A_j$ are close to the corresponding posterior inclusion probabilities for both algorithms. However, the values of $D_j$ are mostly close to 1 for ASI but not for EIA. 
If $D_j$ is close to 1, then variable $j$ is highly likely to be proposed to be removed if already included in the model. This is consistent with the idealized super-efficient setting (ii) in Proposition~\ref{prop_AD_choices} for $\pi_j<0.5$ and leads to improved mixing rates for small $\pi_j$ since it allows that variable to be included more often in a fixed run length. This is hard to learn through individual adaptation (since variables with low posterior inclusion probabilities will be rarely included in the model and so the algorithm learns the $D_j$ slowly for those variables) whereas the Rao-Blackwellized estimates can often quickly determine which variables have low posterior inclusion probabilities.


\subsection{Behaviour of the exploratory individual adaptation algorithm on the Tecator data} \label{sec_tecator}

The Tecator data contains  172  observations and 100 variables. They have been previously analysed using Bayesian linear regression techniques by \cite{gribro10}, who give a description of the data, and
\cite{lamnisos12}. The regressors show a high degree of multi-collinearity and so this is a challenging example for Bayesian variable selection algorithms.
The prior used was (\ref{prior}) with $V_\gamma=100 I$ and $h=5/100$. Even short runs of the EIA algorithm for this data, such as 5 multiple chains with 3000 burn in and 3000 recorded iterations, taking about $5$ seconds on a laptop, show consistent convergence across runs.

Our purpose was to study the adaptive behaviour of the EIA algorithm on this real data example, in particular to compare the idealized values of the $A_j$'s and $D_j$'s with the values attained by the algorithm.

We use multiple chain acceleration with 50 multiple chains over the total of 6000 iterations (without  thinning). The algorithm parameters were set to $\tau_L = 0.01$ and $\tau_U= 0.1$. The resulting mean acceptance rate was approximately $0.2$ indicating close to optimal efficiency. The average number of variables proposed to be changed in a single accepted proposal was $23$, approximately twice the average model size, meaning that in a typical move all of the current variables were deleted from the model, and a set of completely fresh variables was proposed.

Figure \ref{AD_conv}(a) in the Supplementary Material shows how the EIA algorithm approximates
setting~(ii) of Proposition~\ref{prop_AD_choices}, namely the super-efficient sampling from the idealized posterior \eqref{our-target}. 
Figure \ref{AD_conv}(b) illustrates how the attained values of $A_j$'s somewhat overestimate the idealized values $\min\{1, {\pi_j/(1-\pi_j)}\}$ of setting~(ii) in Proposition~\ref{prop_AD_choices}. This indicates that the chosen parameter values $\tau_L=0.01$ and $\tau_U=0.1$ of the algorithm overcompensates for dependence in the posterior, which is not very pronounced for this dataset. To quantify the performance, we ran both algorithms with adaptation in the burn-in only and calculated the effective sample size. With a burn-in of 10\ 000 iterations and 30\ 000 draws, the effective sample per multiple chain was 4015 with  EIA and 6673 with ASI. This is an impressive performance for both algorithms given the multicollinearity in the regressors. The difference in performance can be explained by the speed of convergence to optimal values for the proposal. To illustrate this, we re-ran the algorithms with the burn-in  extended to 30\ 000 iterations:
the effective sample per multiple chain was now 4503 with EIA but 6533 with ASI, indicating that the first algorithm had caught up somewhat. As a comparison, the effective sample size was 1555 for add-delete-swap and 15039 for the Hamming ball sampler with a burnin of 10\ 000 iterations. However, the Hamming ball sampler required 34 times the run time of the  EIA sampler, rendering the latter nine times more efficient in terms of time-standardized effective sample size.


This example and the previous one show that the simplified posterior \eqref{our-target} is a good fit with many datasets and can indeed be used to guide and design algorithms.

\subsection{Performance on problems with moderate $p$}\label{sec_Shafer}
We consider three more data sets with relatively small values for $p$ (around 100) and high dependencies between the covariates, used to showcase the sequential Monte Carlo method proposed in \cite{ScCh11}. They are the Boston Housing data ($n=506$, $p=104$), the concrete data ($n=1030$, $p=79$) and the protein data ($n=96$, $p=88$), which were constructed by \cite{ScCh11} to lead to challenging, multi-modal posterior mass functions.
Further details about the data can be found in \cite{ScCh11}. We focus here on the comparison of the ASI and the EIA algorithms with the add-delete-swap algorithm and the sequential Monte Carlo algorithm of \cite{ScCh11}, also considering parallel tempering versions of the first three algorithms. In addition, we consider two recently proposed methods for high-dimensional variable selection: the Hamming Ball sampler \citep{TitsiasYau} and the Ji-Schmidler adaptive sampler \citep{JiSchmidler}.
Unlike \cite{ScCh11}, we adopt the prior (\ref{prior}) with $V_\gamma=100 I$ and $h=5/100$, while allowing for any combination of main effects and interactions.
We use the method of \cite{ScCh11} for data visualization of the variation in the posterior marginal inclusion probabilities using boxplots. All algorithms were run for the same amount of time and we run them 200 times for each data set. For each variable, the white box contains the central 80\% of results and the black boxes show the upper and lower 10\% most extreme values. The coloured bars cover 0 up to the smallest recorded posterior inclusion probability across all runs. The results (also including other algorithms) are shown in the Supplementary Figure~\ref{results:boston} (Boston housing), Figure~\ref{results:concrete} (concrete) and Figure~\ref{results:protein} (protein). In the Supplementary material (Figure \ref{results:tecator}), we also include results for the Tecator data.

There are clear variations across the data sets  with
the protein and Tecator data leading to very consistent results whereas there were much greater variations in the inclusion probabilities for the Boston Housing and concrete data sets. Parallel tempering is most helpful for the ASI, a bit less so for the EIA algorithms while the add-delete-swap sampler only benefits somewhat from this modification.

The ASI, EIA, Sch\"afer-Chopin, Hamming Ball and add-delete-swap algorithms all provide similar levels of accuracy, whereas the parallel tempering versions of the ASI and add-delete-swap algorithms provide the most accurate results. This is likely due to the multi-modality in the posterior distribution which is better addressed by parallel tempering than the annealing in the Sch\"afer-Chopin algorithm. For all cases, the Ji-Schmidler sampler performs the worst by some margin.

\subsection{Performance on problems with very large $p$} \label{sec_PCR}

\cite{BoRe12} described a variable selection problem with 22\ 576 variables and 60 observations on two inbred mouse populations. The covariates are gender and gene expression  measurements for 22\ 575 genes. Using quantitative real-time polymerase chain reaction (PCR) three physiological phenotypes are recorded, and used as the response variable in the three data sets called PCR$i, i=1,\dots,3$. 
We use prior (\ref{prior}) with $V_{\gamma}=g I$ where $g$ is given a half-Cauchy hyper-prior distribution and a hierarchical prior was used for $\gamma$ by assuming that $h\sim\Be(1, (p-5)/5)$ which implies that the prior mean number of included variables is 5.

A fourth data set (SNP data) relates to genome-wide mapping of a complex trait \citep{CaZhSt17}. The data are body and weight measurements for 993 outbred mice and 79\ 748 single nucleotide polymorphisms (SNPs) recorded for each mouse.  The testis weight is the response, the body weight is a regressor which is always included in the model and variable selection is performed on the 79\ 748 SNPs. The high dimensionality makes this a difficult problem and \cite{CaZhSt17} use a variational inference algorithm (varbvs) for their analysis.  We have used various prior specifications in (\ref{prior}), and present results for a half-Cauchy hyper-prior on $g$ and $h=5/p$.

For all four datasets, the individual adaptation algorithms were run with $\tau_L=0.05$ and $\tau_U=0.23$, and $\tau=0.234$.
The EIA algorithm had a burn-in of 2\ 150 iterations and 10\ 750 subsequent iterations and no thinning, and the ASI had 500 burn-in and 2\ 500 recorded iterations and no thinning (giving very similar run times). Rao-Blackwellised updates of $\pi^{(i)}$ were only used in the burn-in and posterior inclusion probability for the $j$-th variable was estimated by the mean of the posterior sample of $\gamma_j$.

In addition, we use the add-delete-swap algorithm, with starting models chosen from the prior as well as a version of this algorithm started in the model suggested by the least absolute shrinkage and selection operator, the Hamming Ball sampler with radius 1 and the Sch\"afer-Chopin algorithm. Three independent runs of all algorithms were executed to gauge the degree of agreement across runs. Using MATLAB and an Intel i7 @ 3.60 GHz processor, each algorithm took approximately 25 minutes to run for the PCR data and around 2.5 hours for the SNP data.



Figures~\ref{pcr11:comp_rand_g}-\ref{mice:comp_rand_g_fixed_h} in Supplementary Material \ref{add_fig_real} show the pairwise comparisons between the different runs for all data sets. The estimates from each independent chain for the ASI algorithm and for its parallel tempered version are very similar and indicate that the sampler is able to accurately represent the posterior distribution. The EIA algorithm does not seem to converge rapidly enough to effectively deal with these very high-dimensional model spaces in the relatively modest running time allocated.
 Clearly, the other samplers are not able to adequately characterise the posterior model distribution with runs leading to dramatically different results, especially for the PCR data. For the SNP data, the add-delete-swap method does not do too badly, but provides substantially more variable estimates of the posterior inclusion probabilities than the ASI method. Starting the add-delete-swap algorithm in the model selected by the least absolute shrinkage and selection operator never helps, and can actually harm the performance.



\section{Conclusion}

This paper introduces two adaptive Markov chain Monte Carlo algorithms for variable selection problems with very large $p$ and small $n$. We recommend the adaptively scaled individual adaptation proposal, which is able to quickly find good proposals. This method uses a Rao-Blackwellised estimate of the posterior inclusion probability for each variable in an independent proposal.
On simulated data this algorithm shows orders of magnitude improvements in effective sample size compared to the standard Metropolis-Hastings algorithm. The method is also applied to genetic data with 22\ 576 and 79\ 748 variables and shows excellent agreement in the posterior inclusion probabilities across independent runs of the algorithm, unlike the existing methods we have tried. 
We find that multiple independent chains with a shared proposal lead to better convergence to the optimal parameter values and parallel tempering helps to deal with multimodal posteriors. For smaller data sets (say $p<500$), the exploratory individual adaptation algorithm also performs very well. Code to run both algorithms is available from\par 
\noindent{\smaller\url{https://warwick.ac.uk/go/msteel/steel_homepage/software/version3.0.zip}}.

There are a number of possible directions for future research. We have only considered serial implementations of our algorithms in this paper. However, the algorithms are naturally parallelizable across the multiple chains but work is needed on efficient updating of the shared adaptive parameters.
Finally, it will be interesting to apply these algorithms to more complicated data which may have a non-Gaussian likelihood or a more complicated
 prior distribution. 

\section*{Acknowledgements}
K{\L} acknowledges support of the Royal Society through the Royal Society University Research Fellowship and of EPSRC. The authors thank two anonymous referees and an associate editor for their insightful comments that helped improve the paper.

\bibliographystyle{Chicago}
\bibliography{References2}

\begin{thebibliography}{}

\bibitem[\protect\citeauthoryear{Andrieu and Thoms}{Andrieu and
  Thoms}{2008}]{andrieuthoms08}
Andrieu, C. and J.~Thoms (2008).
\newblock A tutorial on adaptive {MCMC}.
\newblock {\em Statistics and Computing\/}~{\em 18}, 343--373.

\bibitem[\protect\citeauthoryear{Bhattacharya, Chakraborty, and
  Mallick}{Bhattacharya et~al.}{2016}]{ABACBM16}
Bhattacharya, A., A.~Chakraborty, and B.~K. Mallick (2016).
\newblock Fast sampling with {G}aussian scale mixture priors in
  high-dimensional regression.
\newblock {\em Biometrika\/}~{\em 4}, 985--991.

\bibitem[\protect\citeauthoryear{Bondell and Reich}{Bondell and
  Reich}{2012}]{BoRe12}
Bondell, H.~D. and B.~J. Reich (2012).
\newblock Consistent high-dimensional variable selection via penalized credible
  regions.
\newblock {\em Journal of the American Statistical Association\/}~{\em 107},
  1610--1624.

\bibitem[\protect\citeauthoryear{Bornn, Jacob, Del~Moral, and Doucet}{Bornn
  et~al.}{2013}]{BJDD12}
Bornn, L., P.~E. Jacob, P.~Del~Moral, and A.~Doucet (2013).
\newblock An adaptive interacting {W}ang-{L}andau algorithm for automatic
  density exploration.
\newblock {\em Journal of Computational and Graphical Statistics\/}~{\em 22},
  749--773.

\bibitem[\protect\citeauthoryear{Brown, Vannucci, and Fearn}{Brown
  et~al.}{1998}]{BVF98}
Brown, P.~J., M.~Vannucci, and T.~Fearn (1998).
\newblock Multivariate {B}ayesian variable selection and prediction.
\newblock {\em Journal of the Royal Statistical Society, B\/}~{\em 60},
  627--641.

\bibitem[\protect\citeauthoryear{Carbonetto, Zhou, and Stephens}{Carbonetto
  et~al.}{2017}]{CaZhSt17}
Carbonetto, P., X.~Zhou, and M.~Stephens (2017).
\newblock varbvs: Fast variable selection for large-scale regression.
\newblock Technical report.

\bibitem[\protect\citeauthoryear{Castillo, Schmidt-Hieber, and van~der
  Vaart}{Castillo et~al.}{2015}]{CaSHVDV15}
Castillo, I., J.~Schmidt-Hieber, and A.~van~der Vaart (2015).
\newblock Bayesian linear regression with sparse priors.
\newblock {\em The Annals of Statistics\/}~{\em 43}, 1986--2018.

\bibitem[\protect\citeauthoryear{Chipman, George, and McCulloch}{Chipman
  et~al.}{2001}]{Chip01}
Chipman, H., E.~I. George, and R.~E. McCulloch (2001).
\newblock The practical implementation of {B}ayesian model selection.
\newblock In P.~Lahiri (Ed.), {\em Model Selection}. Hayward.

\bibitem[\protect\citeauthoryear{Clyde, Ghosh, and Littman}{Clyde
  et~al.}{2011}]{ClGhLi11}
Clyde, M.~A., J.~Ghosh, and M.~L. Littman (2011).
\newblock Bayesian adaptive sampling for variable selection and model
  averaging.
\newblock {\em Journal of Computational and Graphical Statistics\/}~{\em 20},
  80--101.

\bibitem[\protect\citeauthoryear{Craiu, Rosenthal, and Yang}{Craiu
  et~al.}{2009}]{craiu2009learn}
Craiu, R.~V., J.~Rosenthal, and C.~Yang (2009).
\newblock Learn from thy neighbor: Parallel-chain and regional adaptive {MCMC}.
\newblock {\em Journal of the American Statistical Association\/}~{\em 104},
  1454--1466.

\bibitem[\protect\citeauthoryear{Dezeure, Buehlmann, Meier, and
  Meinshausen}{Dezeure et~al.}{2015}]{DeBuMeMe15}
Dezeure, R., P.~Buehlmann, L.~Meier, and N.~Meinshausen (2015).
\newblock High-dimensional inference: {C}onfidence intervals, p-values and
  {R}-{S}oftware hdi.
\newblock {\em Statistical Science\/}~{\em 30}, 533--558.

\bibitem[\protect\citeauthoryear{Fort, Moulines, and Priouret}{Fort
  et~al.}{2011}]{MR3012408}
Fort, G., E.~Moulines, and P.~Priouret (2011).
\newblock Convergence of adaptive and interacting {M}arkov chain {M}onte
  {C}arlo algorithms.
\newblock {\em Ann. Statist.\/}~{\em 39}, 3262--3289.

\bibitem[\protect\citeauthoryear{Garc\'{\i}a-Donato and
  Mart\'{\i}nez-Beneito}{Garc\'{\i}a-Donato and
  Mart\'{\i}nez-Beneito}{2013}]{GDMB13}
Garc\'{\i}a-Donato, G. and M.~A. Mart\'{\i}nez-Beneito (2013).
\newblock On sampling strategies for {B}ayesian variable selection problems
  with large model spaces.
\newblock {\em Journal of the American Statistical Association\/}~{\em 108},
  340--352.

\bibitem[\protect\citeauthoryear{Gelman, Roberts, and Gilks}{Gelman
  et~al.}{1996}]{MR1425429}
Gelman, A., G.~O. Roberts, and W.~R. Gilks (1996).
\newblock Efficient {M}etropolis jumping rules.
\newblock In {\em Bayesian statistics, 5 ({A}licante, 1994)}, Oxford Sci.
  Publ., pp.\  599--607. Oxford Univ. Press, New York.

\bibitem[\protect\citeauthoryear{George and McCulloch}{George and
  McCulloch}{1997}]{george1997approaches}
George, E.~I. and R.~E. McCulloch (1997).
\newblock Approaches for {B}ayesian variable selection.
\newblock {\em Statistica Sinica\/}~{\em 7}, 339--373.

\bibitem[\protect\citeauthoryear{Ghosh and Clyde}{Ghosh and
  Clyde}{2011}]{GhCl11}
Ghosh, J. and M.~A. Clyde (2011).
\newblock Rao-{B}lackwellisation for {B}ayesian variable selection and model
  averaging in linear and binary regression: A novel data augmentation
  approach.
\newblock {\em Journal of the American Statistical Association\/}~{\em 106},
  1041--1052.

\bibitem[\protect\citeauthoryear{Green, {\L}atuszy\'nski, Pereyra, and
  Robert}{Green et~al.}{2015}]{MR3360496}
Green, P.~J., K.~{\L}atuszy\'nski, M.~Pereyra, and C.~P. Robert (2015).
\newblock Bayesian computation: a summary of the current state, and samples
  backwards and forwards.
\newblock {\em Stat. Comput.\/}~{\em 25\/}(4), 835--862.

\bibitem[\protect\citeauthoryear{Griffin and Brown}{Griffin and
  Brown}{2010}]{gribro10}
Griffin, J.~E. and P.~J. Brown (2010).
\newblock Inference with normal-gamma prior distributions in regression
  problems.
\newblock {\em Bayesian Analysis\/}~{\em 5}, 171--188.

\bibitem[\protect\citeauthoryear{Guan and Stephens}{Guan and
  Stephens}{2011}]{GuSt11}
Guan, Y. and M.~Stephens (2011).
\newblock Bayesian variable selection regression for genome-wide association
  studies and other large-scale problems.
\newblock {\em The Annals of Applied Statistics\/}~{\em 5}, 1780--1815.

\bibitem[\protect\citeauthoryear{Haario, Saksman, and Tamminen}{Haario
  et~al.}{2001}]{haario2001adaptive}
Haario, H., E.~Saksman, and J.~Tamminen (2001).
\newblock An adaptive {M}etropolis algorithm.
\newblock {\em Bernoulli\/}~{\em 7}, 223--242.

\bibitem[\protect\citeauthoryear{Hahn and Carvalho}{Hahn and
  Carvalho}{2015}]{HaCa15}
Hahn, P.~R. and C.~M. Carvalho (2015).
\newblock Decoupling shrinkage and selection in {B}ayesian linear models: a
  posterior summary perspective.
\newblock {\em Journal of the American Statistical Association\/}~{\em 110},
  435--448.

\bibitem[\protect\citeauthoryear{Hans, Dobra, and West}{Hans
  et~al.}{2007}]{HaDoWe07}
Hans, C., A.~Dobra, and M.~West (2007).
\newblock Shotgun stochastic search for ``large p'' regression.
\newblock {\em Journal of the American Statistical Association\/}~{\em 102},
  507--516.

\bibitem[\protect\citeauthoryear{Hastie, Tibshirani, and Wainwright}{Hastie
  et~al.}{2015}]{HaTiWa15}
Hastie, T., R.~Tibshirani, and M.~Wainwright (2015).
\newblock {\em Statistical Learning with Sparsity: The Lasso and
  Generalizations}.
\newblock Chapman \& Hall / CRC.

\bibitem[\protect\citeauthoryear{Ji and Schmidler}{Ji and
  Schmidler}{2013}]{JiSchmidler}
Ji, C. and S.~C. Schmidler (2013).
\newblock Adaptive {M}arkov chain {M}onte {C}arlo for {B}ayesian variable
  selection.
\newblock {\em Journal of Computational and Graphical Statistics\/}~{\em 22},
  708--728.

\bibitem[\protect\citeauthoryear{Johnson and Rossell}{Johnson and
  Rossell}{2012}]{johnson2012bayesian}
Johnson, V.~E. and D.~Rossell (2012).
\newblock Bayesian model selection in high-dimensional settings.
\newblock {\em Journal of the American Statistical Association\/}~{\em 107},
  649--660.

\bibitem[\protect\citeauthoryear{Lamnisos, Griffin, and Steel}{Lamnisos
  et~al.}{2013}]{lamnisos12}
Lamnisos, D.~S., J.~E. Griffin, and M.~F.~J. Steel (2013).
\newblock Adaptive {M}onte {C}arlo for {B}ayesian variable selection in
  regression models.
\newblock {\em Journal of Computational and Graphical Statistics\/}~{\em 22},
  729--748.

\bibitem[\protect\citeauthoryear{{\L}atuszy\'nski and Roberts}{{\L}atuszy\'nski
  and Roberts}{2013}]{MR3030220}
{\L}atuszy\'nski, K. and G.~O. Roberts (2013).
\newblock C{LT}s and asymptotic variance of time-sampled {M}arkov chains.
\newblock {\em Methodol. Comput. Appl. Probab.\/}~{\em 15\/}(1), 237--247.

\bibitem[\protect\citeauthoryear{{\L}atuszy{\'n}ski, Roberts, and
  Rosenthal}{{\L}atuszy{\'n}ski et~al.}{2013}]{latuszynski2013adaptive}
{\L}atuszy{\'n}ski, K., G.~O. Roberts, and J.~S. Rosenthal (2013).
\newblock Adaptive {G}ibbs samplers and related {MCMC} methods.
\newblock {\em The Annals of Applied Probability\/}~{\em 23}, 66--98.

\bibitem[\protect\citeauthoryear{Lee and Neal}{Lee and
  Neal}{2018}]{neal2016optimal}
Lee, C. and P.~J. Neal (2018).
\newblock Optimal scaling of the independence sampler: theory and practice.
\newblock {\em Bernoulli\/}~{\em 24}, 1636--1652.

\bibitem[\protect\citeauthoryear{Liang, Liu, and Carroll}{Liang
  et~al.}{2007}]{LiLiCa07}
Liang, F., C.~Liu, and R.~J. Carroll (2007).
\newblock Stochastic approximation in {M}onte {C}arlo computation.
\newblock {\em Journal of the American Statistical Assocation\/}~{\em 102},
  305--320.

\bibitem[\protect\citeauthoryear{Liang, Song, and Yu}{Liang
  et~al.}{2013}]{LiSoYu13}
Liang, F., Q.~Song, and K.~Yu (2013).
\newblock Bayesian subset modeling for high-dimensional generalized linear
  models.
\newblock {\em Journal of the American Statistical Association\/}~{\em 108},
  589--606.

\bibitem[\protect\citeauthoryear{Miasojedow, Moulines, and Vihola}{Miasojedow
  et~al.}{2013}]{MiMoVi12}
Miasojedow, B., E.~Moulines, and M.~Vihola (2013).
\newblock An adaptive parallel tempering algorithm.
\newblock {\em Journal of Computational and Graphical Statistics\/}~{\em 22},
  649--664.

\bibitem[\protect\citeauthoryear{Neal, Roberts, and Yuen}{Neal
  et~al.}{2012}]{MR3025684}
Neal, P., G.~Roberts, and W.~K. Yuen (2012).
\newblock Optimal scaling of random walk {M}etropolis algorithms with
  discontinuous target densities.
\newblock {\em Ann. Appl. Probab.\/}~{\em 22\/}(5), 1880--1927.

\bibitem[\protect\citeauthoryear{Nikooienejad, Wang, and Johnson}{Nikooienejad
  et~al.}{2016}]{NiWaJo16}
Nikooienejad, A., W.~Wang, and V.~E. Johnson (2016).
\newblock Bayesian variable selection for binary outcomes in high-dimensional
  genomic studies using non-local priors.
\newblock {\em Bioinformatics\/}~{\em 32}, 1338--1345.

\bibitem[\protect\citeauthoryear{Nott and Kohn}{Nott and
  Kohn}{2005}]{nottkohn05}
Nott, D.~J. and R.~Kohn (2005).
\newblock Adaptive sampling for {B}ayesian variable selection.
\newblock {\em Biometrika\/}~{\em 92}, 747--763.

\bibitem[\protect\citeauthoryear{O'Hara and Sillanp{\"a}{\"a}}{O'Hara and
  Sillanp{\"a}{\"a}}{2009}]{o2009review}
O'Hara, R.~B. and M.~J. Sillanp{\"a}{\"a} (2009).
\newblock A review of {B}ayesian variable selection methods: what, how and
  which.
\newblock {\em Bayesian Analysis\/}~{\em 4}, 85--117.

\bibitem[\protect\citeauthoryear{Papaspiliopoulos and Rossell}{Papaspiliopoulos
  and Rossell}{2017}]{papaspiliopoulos2016scalable}
Papaspiliopoulos, O. and D.~Rossell (2017).
\newblock Bayesian block-diagonal variable selection and model averaging.
\newblock {\em Biometrika\/}~{\em 104}, 343--359.

\bibitem[\protect\citeauthoryear{Peskun}{Peskun}{1973}]{MR0362823}
Peskun, P.~H. (1973).
\newblock Optimum {M}onte-{C}arlo sampling using {M}arkov chains.
\newblock {\em Biometrika\/}~{\em 60}, 607--612.

\bibitem[\protect\citeauthoryear{Richardson, Bottolo, and Rosenthal}{Richardson
  et~al.}{2010}]{richardson2010bayesian}
Richardson, S., L.~Bottolo, and J.~S. Rosenthal (2010).
\newblock Bayesian models for sparse regression analysis of high dimensional
  data.
\newblock {\em Bayesian Statistics\/}~{\em 9}, 539--568.

\bibitem[\protect\citeauthoryear{Roberts}{Roberts}{1998}]{MR1613256}
Roberts, G.~O. (1998).
\newblock Optimal {M}etropolis algorithms for product measures on the vertices
  of a hypercube.
\newblock {\em Stochastics Stochastic Rep.\/}~{\em 62\/}(3-4), 275--283.

\bibitem[\protect\citeauthoryear{Roberts, Gelman, and Gilks}{Roberts
  et~al.}{1997}]{RGG97}
Roberts, G.~O., A.~Gelman, and W.~R. Gilks (1997).
\newblock Weak convergence and optimal scaling of random walk {M}etropolis
  algorithms.
\newblock {\em Annals of Applied Probability\/}~{\em 7}, 110--120.

\bibitem[\protect\citeauthoryear{Roberts and Rosenthal}{Roberts and
  Rosenthal}{2004}]{roberts2004general}
Roberts, G.~O. and J.~S. Rosenthal (2004).
\newblock General state space {M}arkov chains and {MCMC} algorithms.
\newblock {\em Probability Surveys\/}~{\em 1}, 20--71.

\bibitem[\protect\citeauthoryear{Roberts and Rosenthal}{Roberts and
  Rosenthal}{2007}]{MR2340211}
Roberts, G.~O. and J.~S. Rosenthal (2007).
\newblock Coupling and ergodicity of adaptive {M}arkov chain {M}onte {C}arlo
  algorithms.
\newblock {\em Journal of Applied Probability\/}~{\em 44}, 458--475.

\bibitem[\protect\citeauthoryear{Roberts and Rosenthal}{Roberts and
  Rosenthal}{2009}]{roberts2009examples}
Roberts, G.~O. and J.~S. Rosenthal (2009).
\newblock Examples of adaptive {MCMC}.
\newblock {\em Journal of Computational and Graphical Statistics\/}~{\em 18},
  349--367.

\bibitem[\protect\citeauthoryear{Rockova and George}{Rockova and
  George}{2014}]{rovckova2014emvs}
Rockova, V. and E.~I. George (2014).
\newblock {EMVS}: The {EM} approach to {B}ayesian variable selection.
\newblock {\em Journal of the American Statistical Association\/}~{\em
  109\/}(506), 828--846.

\bibitem[\protect\citeauthoryear{Sch\"afer and Chopin}{Sch\"afer and
  Chopin}{2013}]{ScCh11}
Sch\"afer, C. and N.~Chopin (2013).
\newblock Sequential {M}onte {C}arlo on large binary sampling spaces.
\newblock {\em Statistics and Computing\/}~{\em 23}, 163--184.

\bibitem[\protect\citeauthoryear{Shah and Samworth}{Shah and
  Samworth}{2013}]{ShSa13}
Shah, R.~D. and R.~J. Samworth (2013).
\newblock Variable selection with error control: Another look at stability
  selection.
\newblock {\em Journal of the Royal Statistical Society, Series B\/}~{\em 75},
  55--80.

\bibitem[\protect\citeauthoryear{Shin, Bhattacharya, and Johnson}{Shin
  et~al.}{2018}]{Shin_etal_18}
Shin, M., A.~Bhattacharya, and V.~E. Johnson (2018).
\newblock Scalable {B}ayesian variable selection using nonlocal prior densities
  in ultrahigh-dimensional settings.
\newblock {\em Statistica Sinica\/}~{\em 28}, 1053--1078.

\bibitem[\protect\citeauthoryear{Titsias and Yau}{Titsias and
  Yau}{2017}]{TitsiasYau}
Titsias, M.~K. and C.~Yau (2017).
\newblock The {H}amming ball sampler.
\newblock {\em Journal of the American Statistical Association\/}~{\em
  112\/}(520), 1598--1611.

\bibitem[\protect\citeauthoryear{Yang, Wainwright, and Jordan}{Yang
  et~al.}{2016}]{YaWaJo16}
Yang, Y., M.~Wainwright, and M.~I. Jordan (2016).
\newblock On the computational complexity of high-dimensional {B}ayesian
  variable selection.
\newblock {\em Annals of Statistics\/}~{\em 44}, 2497--2532.

\bibitem[\protect\citeauthoryear{Zanella and Roberts}{Zanella and
  Roberts}{2019}]{RobertsZanella_19}
Zanella, G. and G.~O. Roberts (2019).
\newblock Scalable importance tempering and {B}ayesian variable selection.
\newblock {\em Journal of the Royal Statistical Society, Series B\/}~{\em 81},
  forthcoming.

\end{thebibliography}

\appendix

\section{Proofs}\label{Proofs}

\subsection{Proof of Proposition 1}

For both $(i)$ and $(ii)$ notice that since the proposal is of product form and probability of acceptance is one, each sequence of individual indicator variables $\{\gamma_j^{(t)}\}_{t=0,1,...}$ evolves independently of other coordinates, and is a Markov chain on $\{0,1\}$ governed by, say, transition kernel $P_j$, with stationary distribution $\pi^{(j)}=\{1-\pi_j, \pi_j\}.$  \\
Part $(i)$: \\
$(a)$ and the first part of $(c)$ are immediate because the proposal samples from the stationary distribution and is accepted with probability 1.\\
To verify the second part of $(c)$, use that individual coordinates are Markovian, and for $f \in \mathbb{L}(\Gamma)$ compute:
\begin{align} \label{as_var_computation}
\sigma^2_{P,f} &=  \sum_{j=1}^pa_j^2 \sigma^2_{P,f_j} =  \sum_{j=1}^pa_j^2 \sigma^2_{P_j,f_j}.
\end{align}
Now recall that $P_j$ in $(i)$ is independent sampling from $\pi^{(j)},$ i.e. $P_j = \Pi_j:= \left[\begin{smallmatrix} 1-\pi_j & \pi_j \\ 1-\pi_j & \pi_j  \end{smallmatrix}\right]$, hence $  \sigma^2_{P_j,f_j} = Var_{\pi^{(j)}} f_j$.
\\
To verify $(b)$, note that
\begin{align} \label{ESJD_computation}
E_{\pi}[\Delta^2] &= E_{\pi}\left[\sum_{j=1}^p|\gamma^{(0)}_j - \gamma^{(1)}_j|^2\right] = \sum_{j=1}^p E_{\pi}|\gamma^{(0)}_j - \gamma^{(1)}_j| ,
\end{align}
and for the independent sampling Markov chain $E_{\pi}|\gamma^{(0)}_j - \gamma^{(1)}_j|=  2 \pi_j (1-\pi_j)$. \\
Part $(ii)$:\\
For maximality in $(a)$,  recall \eqref{ESJD_computation}, and it is enough to check (by simple algebra) that the transition kernel $P_j := \left[ \begin{smallmatrix} 1-A_j & A_j \\ D_j & 1-D_j \end{smallmatrix}\right] $, resulting from this choice of $(A,D),$ maximises $E_{1- \pi_j,\pi_j}|\gamma_j-\gamma'_j|$ over all possible Markov chains on $\{0,1\}$ with stationary distribution $\{1- \pi_j, \pi_j\}.$   \\
For $(a)$, recall \eqref{ESJD_computation}, and note that $E_{\pi}|\gamma^{(0)}_j - \gamma^{(1)}_j|=  2 \min\{1-\pi_j, \pi_j\}$.\\
For Peskun optimality in $(b),$ recall formula \eqref{as_var_computation} and consider $\sigma_{P_j, f_j}$ in this setting. It is enough to verify that for each $j$, the kernel $P_j= \left[ \begin{smallmatrix} 1-A_j & A_j \\ D_j & 1-D_j \end{smallmatrix}\right] $ is optimal with respect to Peskun ordering among all Markov chains on $\{0,1\}$ with stationary distribution $\pi^{(j)}$. Indeed, by simple algebra,  $P_j$ maximises off-diagonal elements among all stochastic matrices with stationary distribution $\pi^{(j)}$ and by Theorem 2.1.1 of \cite{MR0362823}, is optimal.\\
To recover $(b)$, recall \eqref{as_var_computation}, and consider asymptotic variance terms of individual coordinates  $\sigma^2_{P_j,f_j}$ for this case. These can be computed directly, but we take a shortcut noting that
\begin{align*}
P_j & = \left[\begin{smallmatrix} 1- \min\{1, \frac{\pi_j}{1-\pi_j}\}\;  & \; \min\{1, \frac{\pi_j}{1-\pi_j}\} \\  \min\{1, \frac{1-\pi_j}{\pi_j}\} \; & \; 1- \min\{1, \frac{1-\pi_j}{\pi_j}\} \end{smallmatrix}\right] \qquad \textrm{and} \qquad  \Pi_j = \left[\begin{smallmatrix} 1-\pi_j \; & \; \pi_j \\ 1-\pi_j \; & \; \pi_j  \end{smallmatrix}\right]
\end{align*}
admit the representation
\begin{align*}
\Pi_j &= \max\{1-\pi_j, \pi_j\}P_j  + (1-\max\{1-\pi_j, \pi_j\}) I_2.
\end{align*}
Thus $\Pi_j$ is a lazy version of $P_j$ and, by Corollary 1 of \cite{MR3030220}, their asymptotic variances are related by
\begin{align*}
Var_{\pi^{(j)}}f_j &= \sigma^2_{\Pi_j, f_j} = \frac{1}{ \max\{1-\pi_j, \pi_j\}} \sigma^2_{P_j, f_j} +\frac{1-  \max\{1-\pi_j, \pi_j\}}{ \max\{1-\pi_j, \pi_j\}} Var_{\pi^{(j)}}f_j.
\end{align*}
Putting $\sigma^2_{P_j, f_j} = (2\max\{1-\pi_j, \pi_j\}-1) Var_{\pi^{(j)}}f_j $ into \eqref{as_var_computation} concludes  the proof.

\subsection{Proof of Proposition 2}
\begin{align*}
&p(\gamma_j=1\vert \gamma_k=1,\gamma_0,y)
- p(\gamma_j=1\vert \gamma_k=0,\gamma_0,y)\\
=&
\frac{\mbox{BF}(\gamma_j=1\vert \gamma_k=1,\gamma_0)
h}{
(1-h)+
\mbox{BF}(\gamma_j=1\vert \gamma_k=1,\gamma_0)h}
-
\frac{\mbox{BF}(\gamma_j=1\vert \gamma_k=0,\gamma_0)
h}{
(1-h)+
\mbox{BF}(\gamma_j=1\vert \gamma_k=0,\gamma_0)h}\\
=&
\frac{a\mbox{BF}(\gamma_j=1\vert \gamma_k=0,\gamma_0)
h}{
(1-h)+
a\mbox{BF}(\gamma_j=1\vert \gamma_k=0,\gamma_0)h}
-
\frac{\mbox{BF}(\gamma_j=1\vert \gamma_k=0,\gamma_0)
h}{
(1-h)+
\mbox{BF}(\gamma_j=1\vert \gamma_k=0,\gamma_0)h}\\
=&
\mbox{BF}(\gamma_j=1\vert \gamma_k=0,\gamma_0)
h
\left[
\frac{a}{
(1-h)+
a\mbox{BF}(\gamma_j=1\vert \gamma_k=0,\gamma_0)h}
-
\frac{1}{
(1-h)+
\mbox{BF}(\gamma_j=1\vert \gamma_k=0,\gamma_0)h}\right]
\end{align*}
This converges to 0 if (i) $a \rightarrow 1$ or
if (ii) $a\rightarrow A<\infty$ and $\mbox{BF}(\gamma_j=1\vert \gamma_k=0,\gamma_0)
h\rightarrow 0$.

\subsection{Proof of Lemma \ref{lem:SUE}}

To verify the result  it is enough to check that the whole state space $\Gamma^{ L}$ is $1-$small with the same constant $b>0,$ (c.f. \cite{roberts2004general}), that is check, for example, that there exists $b > 0$ s.t. for every $\eta \in \Delta_{\epsilon}$ and every $\gamma^{\otimes L}, \gamma'^{\otimes L} \in \Gamma^{L}$ we have
\begin{equation}\label{eqn:Doeblin}
P_{\eta}(\gamma^{\otimes L}, \gamma'^{\otimes L}) \geq b
\end{equation}
where $P_{\eta}(\gamma^{\otimes L}, \gamma'^{\otimes L})$ is the transition for the $L$ chains.
Decompose the move into proposal and acceptance
\[
P_{\eta}(\gamma^{\otimes L}, \gamma'^{\otimes L}) \; = \; q_{\eta}(\gamma^{\otimes L}, \gamma'^{\otimes L}) \times a_{\eta}(\gamma^{\otimes L}, \gamma'^{\otimes L}),
\]
and notice that by the proposal construction \begin{eqnarray*}
q_{\eta}(\gamma^{\otimes L}, \gamma'^{\otimes L}) & \geq & \epsilon^{pL} \quad \textrm{for EIA;}\\
q_{\eta}(\gamma^{\otimes L}, \gamma'^{\otimes L}) & \geq & (\kappa\epsilon)^{pL} \quad \textrm{for adaptively scaled individual adaptation,}
\end{eqnarray*}
 since $\textrm{diam}(\Gamma^L) = pL.$ Similarly for the acceptance term, since $ \pi^{\otimes L}(\gamma_1^{\otimes L}) q_{\eta}(\gamma_1^{\otimes L}, \gamma_2^{\otimes L})  \leq 1$ for every $\gamma_1^{\otimes L}, \gamma_2^{\otimes L} \in \Gamma^{ L}$, we can write
\begin{eqnarray*}
 a_{\eta}(\gamma^{\otimes L}, \gamma'^{\otimes L}) & = & \min\left\{1, {\pi_p^{\otimes L}(\gamma'^{\otimes L}) q_{\eta}(\gamma'^{\otimes L}, \gamma^{\otimes L}) \over \pi_p^{\otimes L}(\gamma^{\otimes L}) q_{\eta}(\gamma^{\otimes L}, \gamma'^{\otimes L}) }  \right\} \\ & \geq & \pi_p^{\otimes L}(\gamma'^{\otimes L}) q_{\eta}(\gamma'^{\otimes L}, \gamma^{\otimes L}) \; \geq \; \pi_{m}^L\times (\kappa \epsilon)^{pL},
\end{eqnarray*}
where $\pi_{m} := \min_{\gamma \in \Gamma} \pi_p(\gamma).$ Consequently in \eqref{eqn:Doeblin} we can take \[ b = \pi_{m}^L\times (\kappa \epsilon)^{2pL},\] and we have established simultaneous uniform ergodicity.

\subsection{Proof of Lemma \ref{lemma:diminishing_adaptation}}
First, observe that if we prove that the proposals of EIA or adaptively scaled individual adaptation satisfy the following diminishing property
\begin{equation}\label{prop_dimini}
\sup_{\gamma \in \Gamma}\|q_{\eta^{(i+1)}}( \gamma, \cdot) - q_{\eta^{(i)}}( \gamma, \cdot)\| \leq C a_i \to 0,
\end{equation}
then by Lemma 4.21(ii) of \cite{latuszynski2013adaptive} (and precisely by inspection of their proof), also the transition kernels satisfy
\begin{equation}\label{kern_dimini}
\sup_{\gamma \in \Gamma}\|P_{\eta^{(i+1)}}( \gamma, \cdot) - P_{\eta^{(i)}}( \gamma, \cdot)\| \leq C_1 a_i \to 0,
\end{equation}
for some $C_1$.

To establish \eqref{prop_dimini}, recall the proposal form \eqref{gen_prop}, and compute
\begin{eqnarray}
\sup_{\gamma \in \Gamma}\|q_{\eta^{(i+1)}}( \gamma, \cdot) - q_{\eta^{(i)}}( \gamma, \cdot)\|
& = & {1 \over 2}
\sup_{\gamma \in \Gamma} \left\{
\sum_{\gamma' \in \Gamma}
| q_{\eta^{(i+1)}}( \gamma, \gamma') - q_{\eta^{(i)}}( \gamma, \gamma')|
\right\}  \nonumber \\
& = & {1 \over 2}
\sup_{\gamma \in \Gamma} \left\{
\sum_{\gamma' \in \Gamma}
\left|
\prod_{j=1}^pq_{\eta^{(i+1)}, j}(\gamma_j, \gamma'_j)
- \prod_{j=1}^pq_{\eta^{(i)}, j}(\gamma_j, \gamma'_j)
\right|
\right\} \nonumber \\
& \leq & {1 \over 2}
\sup_{\gamma \in \Gamma} \left\{
\sum_{\gamma' \in \Gamma}
\sum_{j=1}^p
\left|  q_{\eta^{(i+1)}, j}(\gamma_j, \gamma'_j)
- q_{\eta^{(i)}, j}(\gamma_j, \gamma'_j)
\right|
\right\} =: \spadesuit_1, \qquad \label{eq_spadesuit}
\end{eqnarray}
where the last inequality follows from $|\prod_{j=1}^p a_j - \prod_{j=1}^pb_j| \leq \sum_{j=1}^p |a_j - b_j|$ for $a_j, b_j \in [0,1]$. From
\[
q_{\eta^{(i+1)}, j}(\gamma_j, \gamma'_j) = \left(A_j^{(i+1)}\right)^{(1-\gamma_j)\gamma'_j}\left(1-A_j^{(i+1)}\right)^{(1-\gamma_j)(1-\gamma'_j)}
\left(D_j^{(i+1)}\right)^{\gamma_j(1-\gamma'_j)}\left(1-D_j^{(i+1)}\right)^{\gamma_j\gamma'_j},
\]
we obtain
\begin{eqnarray}
\spadesuit_1 & \leq &
{1 \over 2}
\sup_{\gamma \in \Gamma} \left\{
\sum_{\gamma' \in \Gamma}
\sum_{j=1}^p
\max\left\{
|A_j^{(i+1)} - A_j^{(i)}|, |D_j^{(i+1)} - D_j^{(i)}|
\right\}
\right\} \nonumber \\
& \leq &
{1 \over 2}
\left\{
\sum_{\gamma' \in \Gamma}
\sum_{j=1}^p
\max\left\{
\max_j\left\{|A_j^{(i+1)} - A_j^{(i)}|\right\},
\max_j\left\{|D_j^{(i+1)} - D_j^{(i)}|\right\}
\right\}
\right\} \nonumber \\
& \leq &
p2^{p-1}
\max\left\{
\max_j\left\{|A_j^{(i+1)} - A_j^{(i)}|\right\},
\max_j\left\{|D_j^{(i+1)} - D_j^{(i)}|\right\}
\right\} \nonumber \\
& = &
C
\max\left\{
\max_j\left\{|A_j^{(i+1)} - A_j^{(i)}|\right\},
\max_j\left\{|D_j^{(i+1)} - D_j^{(i)}|\right\}
\right\} =: \spadesuit_2. \nonumber
\end{eqnarray}
For EIA the difference $|A_j^{(i+1)} - A_j^{(i)}|$ comes from the $\logit_{\epsilon}$ update \eqref{eqn:adap_A}, and analogously the difference $|D_j^{(i+1)} - D_j^{(i)}|$ from  the $\logit_{\epsilon}$ update \eqref{eqn:adap_B}. Recall that $\logit_{\epsilon}:(\epsilon, 1-\epsilon) \to \mathbb{R}$ and noticing $ {\partial \logit_{\epsilon}(x) \over \partial x} > 4,$ yields  $ {\partial \logit^{-1}_{\epsilon}(y) \over \partial y} < 1/4$. Consequently, updating the $\logit_{\epsilon}$ by $\phi_i=O(i^{-\lambda})$ is equivalent to updating $A_j^{(i)}$ or $D_j^{(i)}$ by a term of at most the same order $O(i^{-\lambda})$. Hence for EIA
\begin{eqnarray}
\spadesuit_2 & = & O(i^{-\lambda}). \label{logit_order}
\end{eqnarray}
For adaptively scaled individual adaptation recall \eqref{ASI_update} and its use in Algorithm \ref{explore_ISA}. As the components are in $[0,1]$, we apply the triangle inequality to obtain:
\begin{eqnarray}
|A_j^{(i+1)} - A_j^{(i)}| & \leq & |\zeta^{(i+1)} - \zeta^{(i)}| +
\left|
\min\bigg\{1, { \tilde\pi^{(i+1)}_j \over 1-\tilde\pi^{(i+1)}_j } \bigg\} - \min\bigg\{1, { \tilde\pi^{(i)}_j \over 1-\tilde\pi^{(i)}_j } \bigg\}  \right|; \label{A_for_ASI} \\
|D_j^{(i+1)} - D_j^{(i)}| & \leq & |\zeta^{(i+1)} - \zeta^{(i)}| +
\left|
\min\bigg\{1, {1-\tilde\pi^{(i+1)}_j \over \tilde\pi^{(i+1)}_j } \bigg\} - \min\bigg\{1, { 1-\tilde\pi^{(i)}_j  \over \tilde\pi^{(i)}_j } \bigg\}  \right|. \label{D_for_ASI}
\end{eqnarray}
The update equation for $\zeta^{(i)}$ is \eqref{eqn:adap_C}, hence
\begin{equation} |\zeta^{(i+1)} - \zeta^{(i)}| =O(i^{-\lambda})
\label{ASI_zetas}
\end{equation}
 by the same argument that led to  \eqref{logit_order}. The term $\tilde\pi^{(i)}$ introduced in Algorithm \ref{explore_ISA} is
\begin{equation} \label{tilde_from_hat}
\tilde\pi_j^{(i)}=\kappa + (1 - 2\kappa) \hat\pi_j^{(i)}
\end{equation}
where $\hat\pi_j^{(i)}$ is the Rao-Blackwellised estimate of $\pi_j=p(\gamma_j=1\vert y)$ defined in \eqref{eq:RB}. It remains to show that the second terms in \eqref{A_for_ASI} and \eqref{D_for_ASI} are at most $O(i^{-\lambda}).$ We shall see that the terms are $O(i^{-1}).$ To this end, first note that $\hat\pi_j^{(i)}$  using posterior samples $\gamma^{(1)},\dots,\gamma^{(i)}$ is
\begin{equation} \nonumber
\hat\pi_j^{(i)} = \frac{1}{i} \sum_{k=1}^i  p\left(\gamma_j=1\left\vert \gamma^{(k)}_{-j}, y\right.\right).
\end{equation}
and therefore
\begin{eqnarray} \nonumber
|\hat\pi_j^{(i+1)} - \hat\pi_j^{(i)}| & = &  \left|{i \over i+1} \hat\pi_j^{(i)} + {1 \over i+1} p\left(\gamma_j=1\left\vert \gamma^{(i+1)}_{-j}, y\right.\right) - \hat\pi_j^{(i)} \right| \\ \nonumber
& \leq &  \left|{i \over i+1} \hat\pi_j^{(i)}  - \hat\pi_j^{(i)} \right|+\left| {1 \over i+1} p\left(\gamma_j=1\left\vert \gamma^{(i+1)}_{-j}, y\right.\right)\right| \\ \label{eq:diff_RB}
& \leq & {2 \over i+1}.
\end{eqnarray}
Next, consider the function $f_{\kappa}(x) = {\kappa + (1-2\kappa)x \over \kappa + (1-2\kappa)(1-x) }$ and compute its derivative to see that it is Lipshitz with constant ${1 \over \kappa}.$ Consequently, so is $g_{\kappa}(x) := \min\{ 1, f_{\kappa}(x)\}$, hence
\begin{eqnarray}
\left|
\min\bigg\{1, { \tilde\pi^{(i+1)}_j \over 1-\tilde\pi^{(i+1)}_j } \bigg\} - \min\bigg\{1, { \tilde\pi^{(i)}_j \over 1-\tilde\pi^{(i)}_j } \bigg\}  \right| & = & |g_{\kappa}(\hat\pi_j^{(i+1)}) - g_{\kappa}(\hat\pi_j^{(i)})|
 \; \leq  \;
{1 \over \kappa} |\hat\pi_j^{(i+1)} - \hat\pi_j^{(i)}|  \nonumber \\
& \leq & {2 \over \kappa} {1\over i+1} \; = \; O(i^{-1}). \label{ASI_pi}\qquad
\end{eqnarray}
Combining \eqref{ASI_zetas} and \eqref{ASI_pi} shows that the right-hand side of \eqref{A_for_ASI} is $O(i^{-\lambda})$ and a symmetric reasoning yields the result for \eqref{D_for_ASI}, which completes the proof.

\subsection{Proof of Theorem \ref{thm:IA-PT_erg}}
We shall use Theorem 1 of  \cite{MR2340211} to obtain ergodicity~\eqref{eq_thm:IA_erg}. This requires simultaneous uniform ergodicity of the transition kernels, established in Lemma \ref{lem:SUE} and diminishing adaptation. Lemma \ref{lemma:diminishing_adaptation} verifies diminishing adaptation in single chain implementations. For multiple chain implementations there will be up to $L$ updates of the adaptive parameter between consecutive updates of chain number $l$, hence applying the triangle inequality and Lemma \ref{lemma:diminishing_adaptation} yields diminishing adaptation with the same rate (and a constant multiplied by $L$).

The strong law of large numbers \eqref{eq_thm:IA_WLLN} will be demonstrated by first establishing it for each chain in the multiple chain implementation separately, i.e.
\begin{eqnarray} \label{SLLN_separate}
{1 \over k}\sum_{i=1}^{k} f(\gamma^{l,\, (i)}) &\stackrel{k\to\infty}{\longrightarrow} & \pi(f) \quad \textrm{almost surely,}
\end{eqnarray}
and then combining it into \eqref{eq_thm:IA_WLLN} by averaging the $L$ chains.
To obtain \eqref{SLLN_separate} we use Corollary 2.8 following Theorem 2.7 in \cite{MR3012408}.
This requires establishing the following conditions:
\begin{enumerate}
\item[C1.] (Drift condition A3 in \cite{MR3012408}) There exists a function $V: \Gamma \to [1, +\infty)$ and for any $\eta$ there exist some constants $b_{\eta} < \infty,$ $\delta_{\eta} \in (0,1), \lambda_{\eta} \in (0,1)$ and a probability measure $\nu_{\eta}$ on $\Gamma$, such that
\begin{eqnarray} P_{\eta}V &\leq & \lambda_{\eta}V + b_{\eta}; \nonumber \\
P_{\eta}(\gamma, \cdot) & \geq & \delta_{\eta} \nu_{\eta}(\cdot)\mathbb{I}{\{V\leq c_{\eta}\}}(\gamma), \qquad c_{\eta} := 2b_{\eta}(1-\lambda_{\eta})^{-1} - 1. \nonumber
\end{eqnarray}
Indeed, it is immediate to check that the above condition is met with $V\equiv 1,$ $\lambda_{\eta} = 1/2,$ $b_{\eta} = 1,$ the measure $\nu_{\eta}$ uniform on $\Gamma$ and, following the proof of Lemma \ref{lem:SUE},  with $\delta_{\eta} = \pi_m ( \kappa \epsilon)^{2p},$ where $\pi_{m} := \min_{\gamma \in \Gamma} \pi_p(\gamma).$
\item[C2.] (Condition A4 in  \cite{MR3012408} after specialising to the parameters  in C1 above)
\begin{eqnarray} \nonumber
\sum_{i=1}^{\infty} i^{-1}  \sup_{\gamma \in \Gamma} \|P_{\eta^{(i+1)}}(\gamma, \cdot) - P_{\eta^{(i)}}(\gamma, \cdot)\| & < & \infty.
\end{eqnarray}
The condition indeed holds for EIA, adaptively scaled individual adaptation, and their multiple chain implementations, by the diminishing adaptation rate established in Lemma \ref{lemma:diminishing_adaptation}.
\item[C3.] Condition A5 in  \cite{MR3012408} which is trivially satisfied with the parameters chosen in C1.
\end{enumerate}
Since we have established C1, C2, and C3,  by Corollary 2.8 in  \cite{MR3012408},
\eqref{SLLN_separate} holds, and so does~\eqref{eq_thm:IA_WLLN}.

\section{Parallel Tempering implementations}\label{PT}

We will consider an adaptive parallel tempering scheme
to avoid the algorithms becoming trapped in very well separated modes. A sequence of distributions $\pi_1,\dots,\pi_m$ are constructed with
\[
\pi_k(\gamma \vert y)\propto p(y\vert \gamma)^{t_k} \pi(\gamma),\qquad  k=1,\dots,m
\]
where the parameters $0<t_1<t_2<\dots<t_m=1$ are referred to as temperatures (with smaller $t_j$ referring to higher temperatures). The density $\pi_m(\gamma\vert y)$ is the posterior density $p(\gamma\vert y)$ of interest. The distribution becomes flatter at higher temperatures.
The sequence of posterior distribution can be sampled using Markov chain Monte Carlo methods by defining the joint target 
for $\gamma^{\star}=(\gamma^{\star}_1,\dots,\gamma^{\star}_m)$,
$
\pi(\gamma^{\star}\vert y)=\prod_{k=1}^m \pi_k(\gamma^{\star}_k\vert y)
$
 where $\pi_k(\gamma^{\star}_k\vert y)\propto
p(y\vert \gamma^{\star}_k)^{t_k} \pi(\gamma^{\star}_k)$.
The Markov chain Monte Carlo algorithm uses two types of moves. Firstly, $\gamma^{\star}_k$ is updated for all values of $k$. Secondly,
 a Metropolis-Hastings update proposes
 to swap $\gamma^{\star}_k$ with $\gamma^{\star}_{k+1}$ for $k$ chosen uniformly at random from $\{1,\dots,m-1\}$.
The temperature schedule is chosen adaptively using the method proposed by   \cite{MiMoVi12}.
As the temperature increases (as $t_k\rightarrow 0$) , the  dependence becomes weaker between $\gamma_i$ and $\gamma_j$ under $\pi_k(\gamma^{\star}_k \vert y)$.

\section{Derivation of Rao-Blackwellised estimate of $\pi_j$}\label{SM:RB}

The Rao-Blackwellised estimate of $\pi_j=p(\gamma_j=1\vert y)$ using posterior samples $\gamma^{(1)},\dots,\gamma^{(N)}$ is
\begin{equation} \nonumber
\hat\pi_j = \frac{1}{N} \sum_{k=1}^N  p\left(\gamma_j=1\left\vert \gamma^{(k)}_{-j}, y\right.\right).
\end{equation}
We know that
\[
p(\gamma\vert y) \propto p(y\vert \gamma)p(\gamma)
\]
where
\[
p(y\vert \gamma)= |Z_{\gamma}^TZ_{\gamma} + \Lambda_{\gamma}|^{-1/2} g^{-p_{\gamma}/2}
\left(
y^Ty - y^T Z_{\gamma} (Z_{\gamma}^T Z_{\gamma} + \Lambda_{\gamma})^{-1}Z_{\gamma}^T y
\right)^{-n/2}\equiv m(\gamma)
\]
and $\Lambda_{\gamma} = \mbox{diag}(0,\overbrace{g^{-1},\dots,g^{-1}}^{p_{\gamma}\mbox{-times}})$. Then, if $h$ is fixed,
\[
p(\gamma_j=1\vert \gamma_{-j}, y)
=
\frac{h m(\gamma_j^{\uparrow})}
{(1-h) m(\gamma_j^{\downarrow})+h m(\gamma_j^{\uparrow})}=
\frac{h\mbox{BF}_j(\gamma_{-j})}
{1-h+h\mbox{BF}_j(\gamma_{-j})}
\]
where $\gamma_j^{\uparrow}$ represents $\gamma_{-j}$ and $\gamma_j=1$,
$\gamma_j^{\downarrow}$ represents $\gamma_{-j}$ and $\gamma_j=0$, and
$\mbox{BF}_j(\gamma_{-j})={m(\gamma_j^{\uparrow})}/{m(\gamma_j^{\downarrow})}$ is the Bayes factor in favour of including the $j$-th variable given $\gamma_{-j}$.
If $h\sim\Be(a, b)$,
\[
p(\gamma_j=1\vert \gamma_{-j}, y)
=\frac{h^{\star} m(\gamma_j^{\uparrow})}
{(1-h^{\star}) m(\gamma_j^{\downarrow})+h^{\star} m(\gamma_j^{\uparrow})}=
\frac{h^{\star}\mbox{BF}_j(\gamma_{-j})}
{1-h^{\star}+h^{\star}\mbox{BF}_j(\gamma_{-j})}
\]
where $h^{\star} = \frac{\#\gamma_{-j}+1+a}{p+a+b}$.

Therefore, calculating the Rao-Blackwellised estimates reduces to calculating the Bayes factors $\mbox{BF}_j(\gamma_{-j})$ using values calculated in the MCMC chain. 
We distinguish the two cases, $\gamma_j=0$ and $\gamma_j=1$ in the current state of the chain.

If $\gamma_j=0$,  we define
 $Z_{\gamma_j^\uparrow} = [Z_{\gamma}\ x_j]$ and the Bayes factor is
\begin{align*}
\mbox{BF}_j(\gamma_{-j}) &=
\frac{ |Z_{\gamma_j^\uparrow}^TZ_{\gamma_j^\uparrow} + \Lambda_{\gamma_j^\uparrow}|^{-1/2} }{ |Z_{\gamma}^TZ_{\gamma} + \Lambda_{\gamma}|^{-1/2}}
\,g^{-1/2}
\left(\frac{
y^Ty - y^T Z_{\gamma_j^\uparrow} (Z_{\gamma_j^\uparrow}^T Z_{\gamma_j^\uparrow} + \Lambda_{\gamma_j^\uparrow})^{-1}Z_{\gamma_j^\uparrow}^T y
}
{
y^Ty - y^T Z_{\gamma} (Z_{\gamma}^T Z_{\gamma} + \Lambda_{\gamma})^{-1}Z_{\gamma}^T y
}\right)^{-n/2}\\
&=
\frac{ |Z_{\gamma_j^\uparrow}^TZ_{\gamma_j^\uparrow} + \Lambda_{\gamma_j^\uparrow}|^{-1/2} }{ |Z_{\gamma}^TZ_{\gamma} + \Lambda_{\gamma}|^{-1/2}}
\,g^{-1/2}
\left(\frac{
y^Ty - y^T Z_{\gamma_j^\uparrow} \tilde{F} Z_{\gamma_j^\uparrow}^T y
}
{
y^Ty - y^T Z_{\gamma} F Z_{\gamma}^T y
}\right)^{-n/2}
\end{align*}
where
$F = (Z_{\gamma}^TZ_{\gamma} + \Lambda_{\gamma})^{-1}$
and $\tilde{F}=\left(Z_{\gamma^\uparrow_j}^TZ_{\gamma^\uparrow_j}
+\Lambda_{\gamma_j^\uparrow}\right)^{-1}$.
The standard formulae for the Schur complement show that
\begin{equation}
\left|Z_{\gamma_j^\uparrow}^TZ_{\gamma_j^\uparrow} + \Lambda_{\gamma_j^\uparrow}\right|
= \left|Z_{\gamma}^TZ_{\gamma} + \Lambda_{\gamma}\right|d_j^\uparrow
\end{equation}
and
\begin{equation}
\tilde{F}=\left(Z_{\gamma^\uparrow_j}^TZ_{\gamma^\uparrow_j}
+\Lambda_{\gamma_j^\uparrow}\right)^{-1}=
\left(
\begin{array}{cc}
F + F (Z_{\gamma}^T x_j)(x_j^T Z_{\gamma})F/d_j^\uparrow &
- F(Z_{\gamma}^Tx_j)/d_j^\uparrow\\
-x_j^T Z_{\gamma} F/d_j^\uparrow  & 1/d_j^\uparrow
\end{array}
\right)
\label{schur}
\end{equation}
where
$
d_j^\uparrow = x_j^T x_j + g^{-1} - (x_j^T Z_{\gamma})F(Z_{\gamma}^T x_j).
$
The result follows from the application of (C1) and (C2) to the expression for $\mbox{BF}_j(
\gamma_{-j})$ and noticing that
\begin{align*}
y^T Z_{\gamma_j^\uparrow}
\tilde{F} Z_{\gamma_j^\uparrow}^T y
=y^T Z_{\gamma}
F Z_{\gamma}^T y
+ \frac{1}{d^\uparrow_j}\left( y^TZ_{\gamma} F (Z_{\gamma}^T x_j)(x_j^T Z_{\gamma}) F Z_{\gamma}^Ty
-2y^T Z_{\gamma}  F Z_{\gamma}^T x_j x_j^T y
+ y^T x_j  x_j^T y\right).
\end{align*}

%
%

Similarly,
if $\gamma_j=1$, we can write $Z_{\gamma} = [Z_{\gamma^\downarrow_j} \ x_j]$
\[
\mbox{BF}_j(\gamma_{-j}) =
\frac{ |Z_{\gamma}^TZ_{\gamma} + \Lambda_{\gamma}|^{-1/2} }{ |Z_{\gamma_j^\downarrow}^TZ_{\gamma_j^\downarrow} + \Lambda_{\gamma_j^\downarrow}|^{-1/2}}
\,g^{-1/2}
\left(\frac{
y^Ty - y^T Z_{\gamma} (Z_{\gamma}^T Z_{\gamma} + \Lambda_{\gamma})^{-1}Z_{\gamma}^T y
}
{
y^Ty - y^T Z_{\gamma_j^\downarrow} (Z_{\gamma_j^\downarrow}^T Z_{\gamma_j^\downarrow} + \Lambda_{\gamma_j^\downarrow})^{-1}Z_{\gamma_j^\downarrow}^T y
}\right)^{-n/2}.
\]
 Again,
the standard formulae for the Schur complement show that
\begin{equation}
\left|Z_{\gamma}^TZ_{\gamma} + \Lambda_{\gamma}\right|
= \left|Z_{\gamma_j^\downarrow}^TZ_{\gamma_j^\downarrow} + \Lambda_{\gamma_j^\downarrow}\right|d_j^\downarrow
\end{equation}
and
\begin{equation}
F=\left(Z_{\gamma}^TZ_{\gamma}
+\Lambda_{\gamma}\right)^{-1}=
\left(
\begin{array}{cc}
\tilde{F} + \tilde{F} (Z_{\gamma_j^\downarrow}^T x_j)(x_j^T Z_{\gamma_j^\downarrow})\tilde{F}/d^\downarrow_j &
- \tilde{F}(Z_{\gamma_j^\downarrow}^Tx_j)/d_j^\downarrow\\
-x_j^T Z_{\gamma_j^\downarrow} \tilde{F}/d_j^\downarrow  & 1/d_j^\downarrow
\end{array}
\right)
\label{schur}
\end{equation}
where
$\tilde{F} = (Z_{\gamma_j^\downarrow}^TZ_{\gamma_j^\downarrow} + \Lambda_{\gamma_j^\downarrow})^{-1}$
and
$
d_j^\downarrow = x_j^T x_j + g^{-1} - (x_j^T Z_{\gamma_j^\downarrow})\tilde{F}(Z_{\gamma_j^\downarrow}^T x_j).
$
We have that
$\tilde{F} = F_{1:p_\gamma,1:p_\gamma} - F_{1:\gamma, p_{\gamma}+1} F_{p_{\gamma}+1, 1:p_\gamma} \frac{1}{F_{p_\gamma+1, p_\gamma+1}}
$ and $d^\downarrow_j = \frac{1}{F_{p_\gamma+1, p_\gamma+1}}$.
Using
\[
y^T Z_{\gamma_j^\downarrow} F_{1:p_\gamma,1:p_\gamma} Z_{\gamma_j^\downarrow}y=
y^T Z_{\gamma} F Z_{\gamma}^T y
 - y^T x_j F_{p_\gamma+1, p_\gamma+1} x_j y - 2\,y^T Z_{\gamma^\downarrow_j} F_{1:p_{\gamma}, p_\gamma+1} x_j^T y
\]
we can write
\begin{align*}
 y^T Z_{\gamma_j^\downarrow} (Z_{\gamma_j^\downarrow}^T Z_{\gamma_j^\downarrow} + \Lambda_{\gamma_j^\downarrow})^{-1}Z_{\gamma_j^\downarrow}^T y
=\,&
y^T Z_{\gamma_j^\downarrow} \tilde{F} Z_{\gamma_j^\downarrow}y\\
=\,&
y^T Z_{\gamma} F Z_{\gamma}^T y
 - y^T x_j \frac{1}{d_j^\downarrow} x_j y - 2\,y^T Z_{\gamma^\downarrow_j} F_{1:p_\gamma, p_\gamma+1} x_j^T y \\
 &- y^T Z_{\gamma^\downarrow_j} F_{1:p_\gamma, p_\gamma+1} F_{p_\gamma+1, 1:p_\gamma} d_j^\downarrow Z_{\gamma^\downarrow_j} y^T\\
=\,&
y^T Z_{\gamma} F Z_{\gamma}^T y
- (a_j + b_j)^2
\end{align*}
where
\[
a_j =
y^T x_j
({d_j^\downarrow})^{-1/2}
\]
\[
b_j =
y^T Z_{\gamma_j^\downarrow}
{d_j^\downarrow}^{1/2}F_{1:p_\gamma,p_\gamma+1}
=
{d_j^\downarrow}^{1/2}
\left(y^T Z_{\gamma}
F_{\cdot, p_\gamma+1}
-y^T x_j F_{p_\gamma+1, p_\gamma+1}
\right)
=
{d_j^\downarrow}^{1/2}
y^T Z_{\gamma}
F_{\cdot, p_\gamma+1}
-y^T x_j ({d_j^\downarrow})^{-1/2}.
\]
This sum simplifies to
\[
a_j + b_j
=y^T x_j
({d_j^\downarrow})^{-1/2}
+{d_j^\downarrow}^{1/2}
y^T Z_{\gamma}
F_{\cdot, p_\gamma+1}
-y^T x_j
({d_j^\downarrow})^{-1/2}
={d_j^\downarrow}^{1/2}
y^T Z_{\gamma}
F_{\cdot, p_\gamma+1}.
\]
Applying this with (C3) to $\mbox{BF}_j(\gamma_{-j})$ leads to the result.
\newpage

\section{Additional graphical material for Section 2}\label{add_fig}

\newpage

\section{Graphical results for the examples with real data}\label{add_fig_real}

 \begin{figure}[!h]
\subfigure[Limiting values of the $(A_j,D_j)$ pairs align at the top ends of the segments of Figure~\ref{fig_segment}, with $D_j$'s close to $1$, corresponding to the super-efficient setting~(ii) of Proposition~\ref{prop_AD_choices}. ]{\includegraphics[width=.45\linewidth]{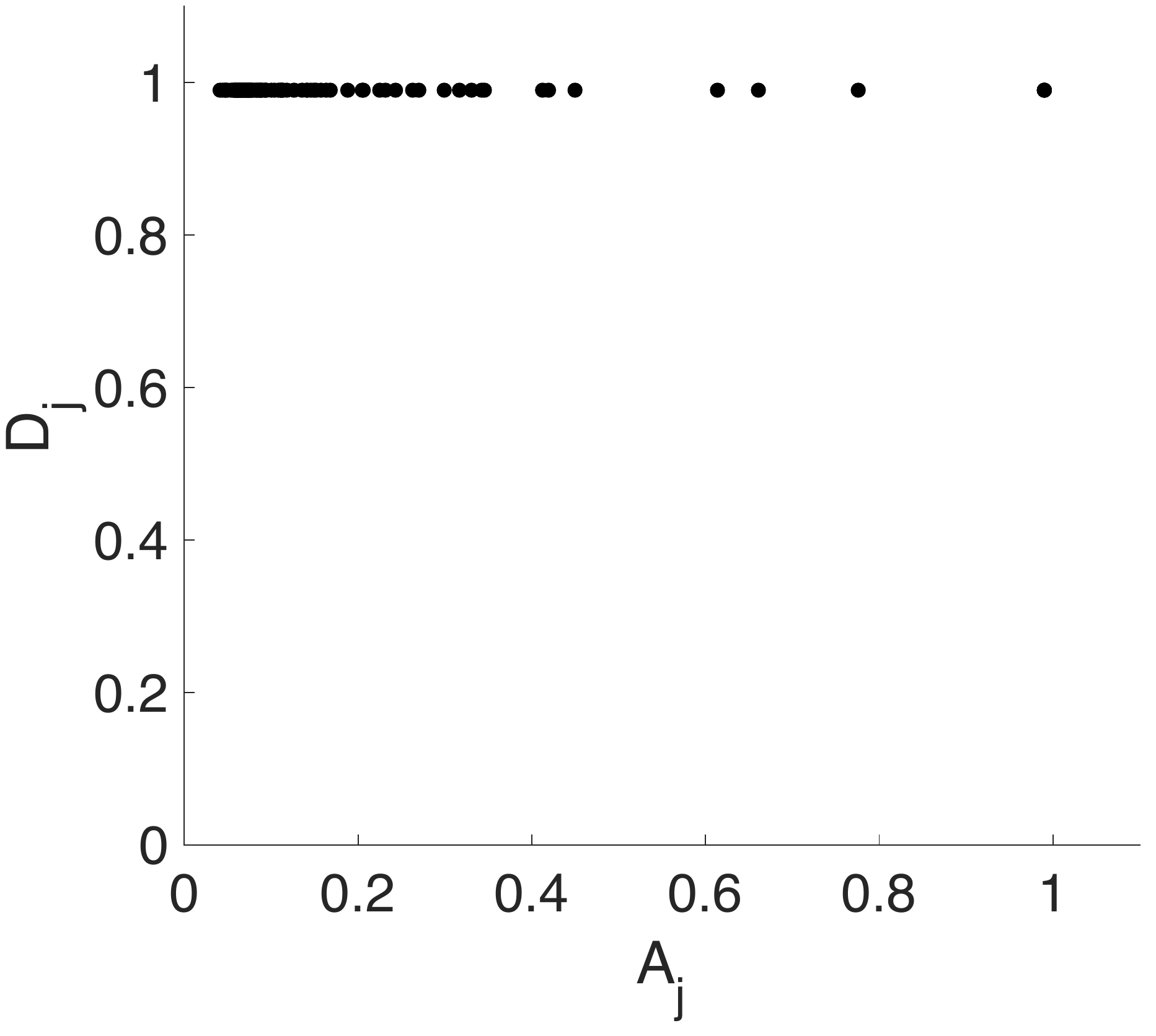}}
\hfil
\subfigure[The attained values of $A_j$'s overestimate the idealized values $\min\{1, {\pi_j \over 1-\pi_j}\}$ of setting~(ii) in Proposition~\ref{prop_AD_choices}, indicating low dependence in the posterior. ]{\includegraphics[width=.45\linewidth]{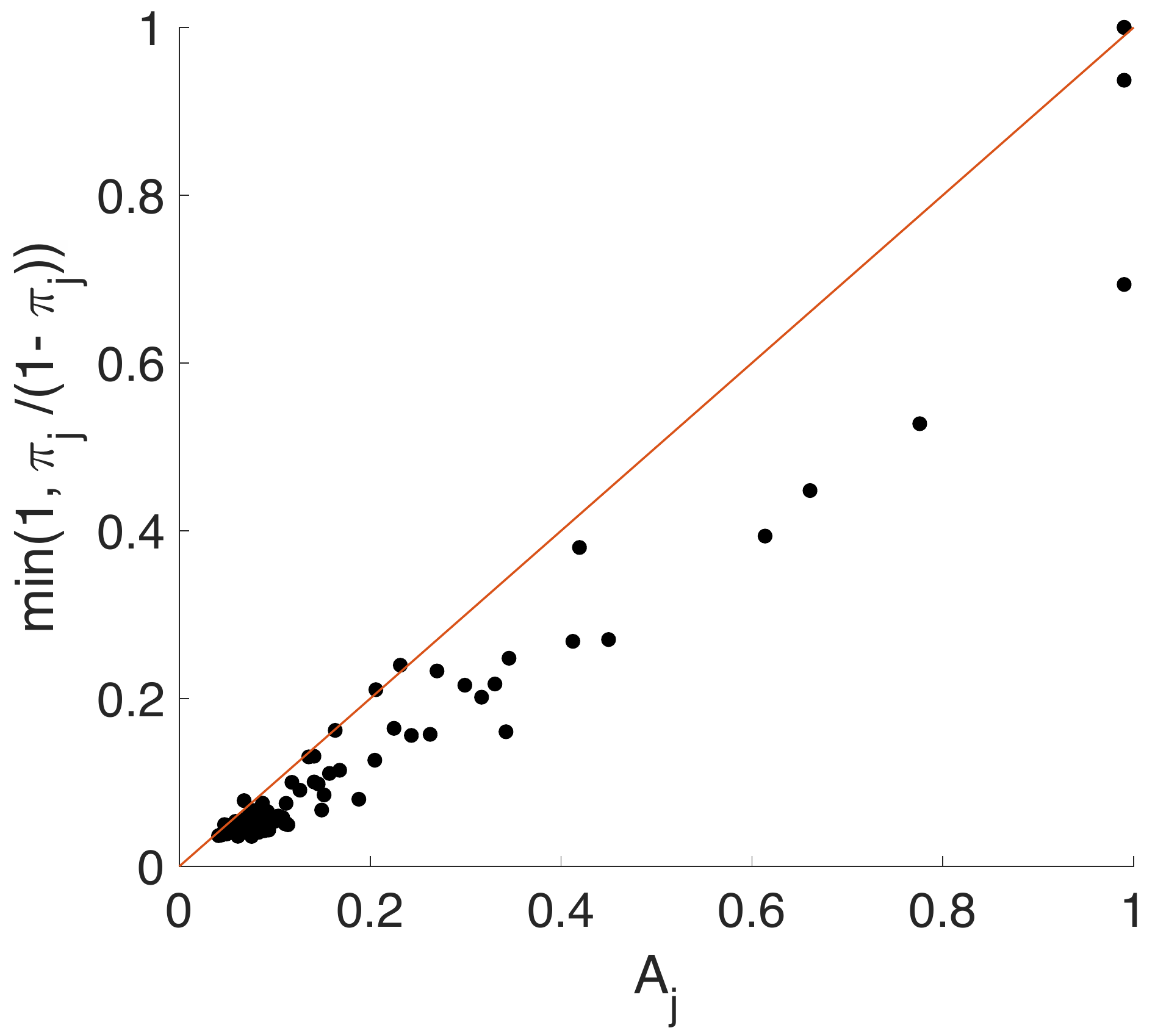}}
\caption{\label{AD_conv} \small Tecator data: the adaptive parameter $\eta = (A,D)$ for the exploratory individual adaptation algorithm.}
\end{figure}

\begin{figure}[h!]
\begin{center}
\includegraphics[trim=40mm 100mm 40mm 90mm, clip, scale=0.6]{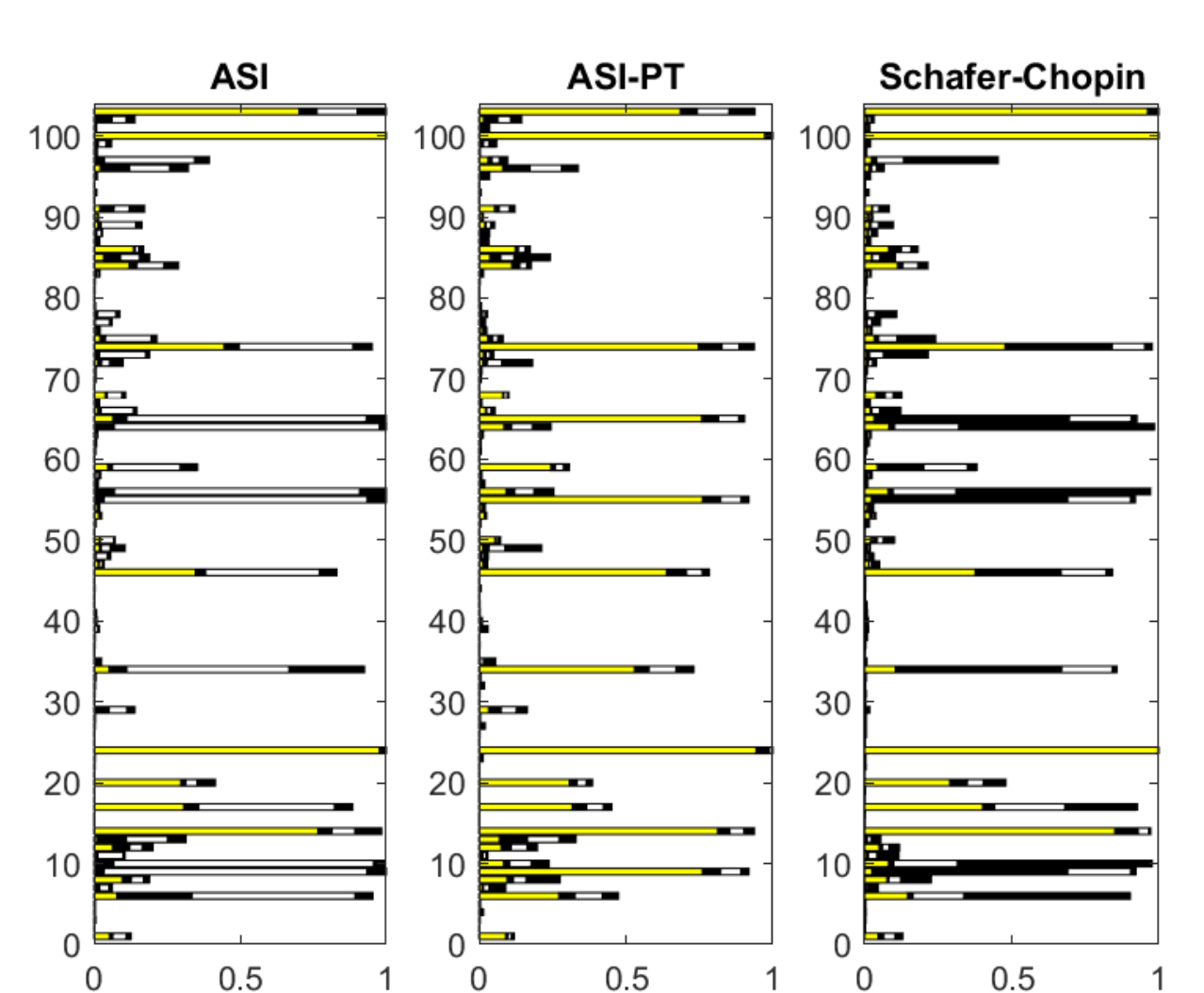}\\
\includegraphics[trim=40mm 100mm 40mm 90mm, clip, scale=0.6]{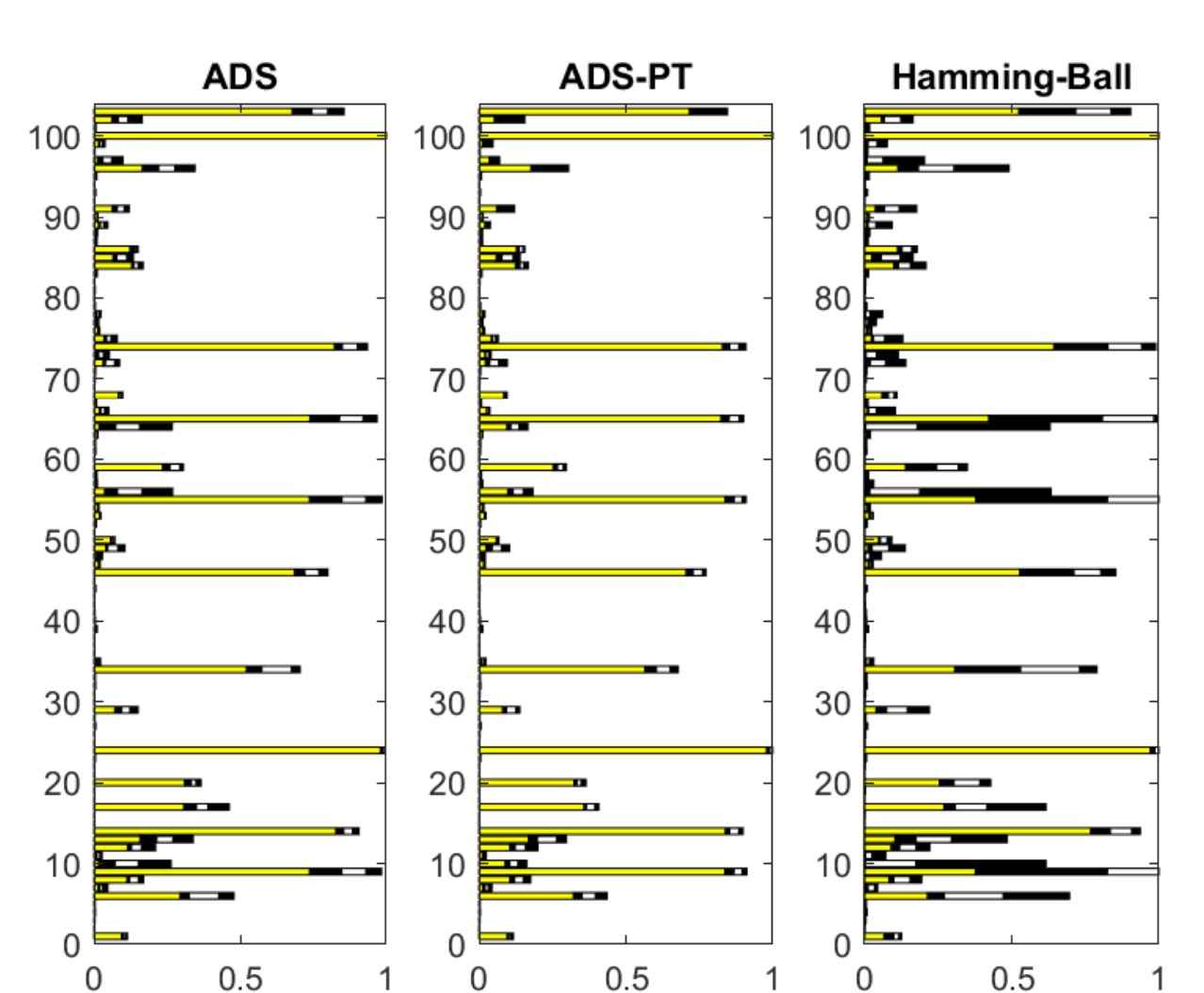}\\
\includegraphics[trim=40mm 100mm 40mm 90mm, clip, scale=0.6]{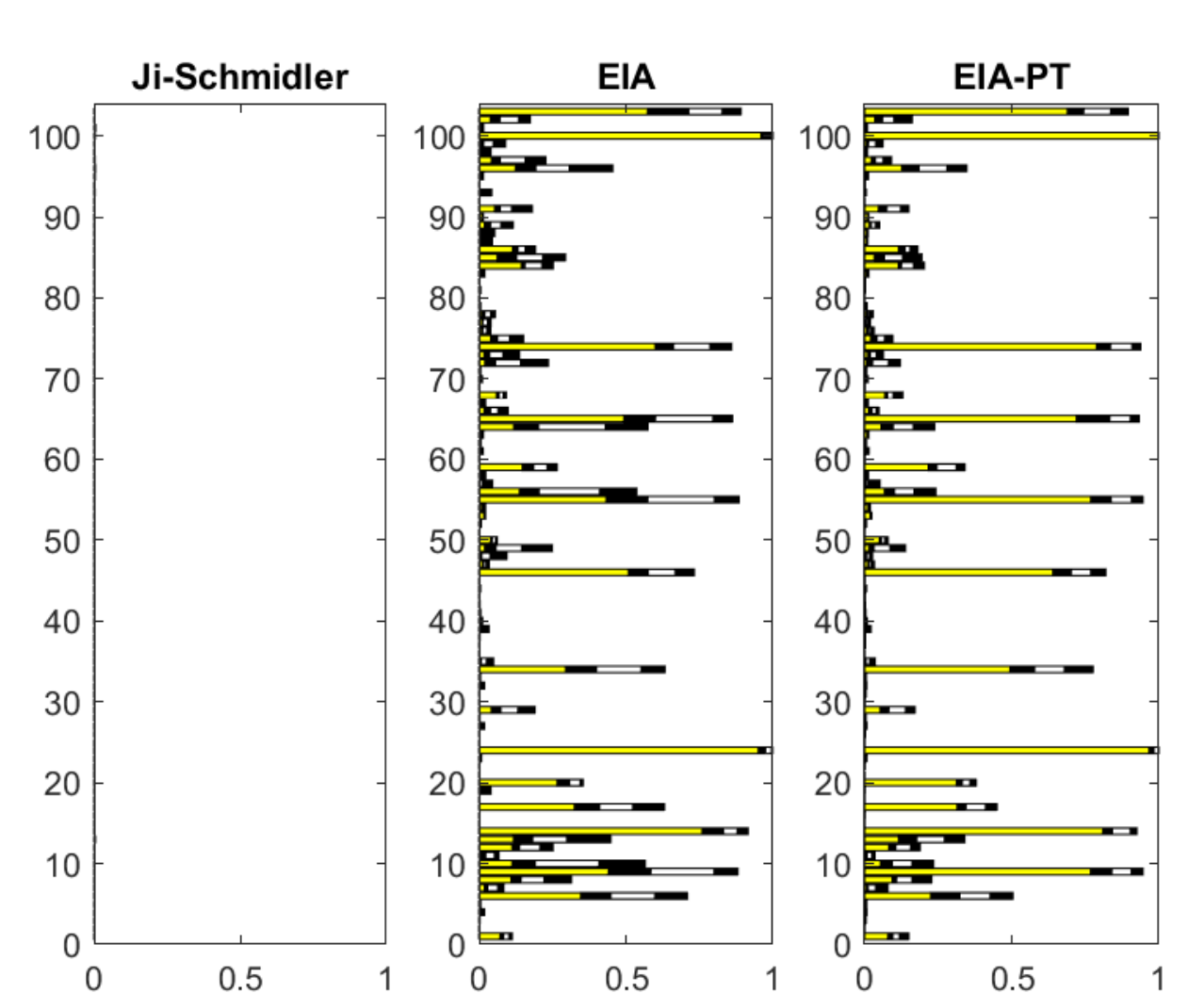}
\end{center}
\caption{\small Boston housing data: Inclusion probabilities boxplots using adaptively scaled individual adaptation (ASI), exploratory individual adaptation (EIA), add-delete-swap (ADS) and the sequential Monte Carlo algorithm of \cite{ScCh11}, with parallel tempering (PT) versions of the first three algorithms. We also consider the Hamming Ball and the Ji-Schmidler samplers}
\label{results:boston}
\end{figure}

\begin{figure}[h!]
\begin{center}
\begin{tabular}{c}
\includegraphics[trim=40mm 100mm 40mm 90mm, clip, scale=0.6]{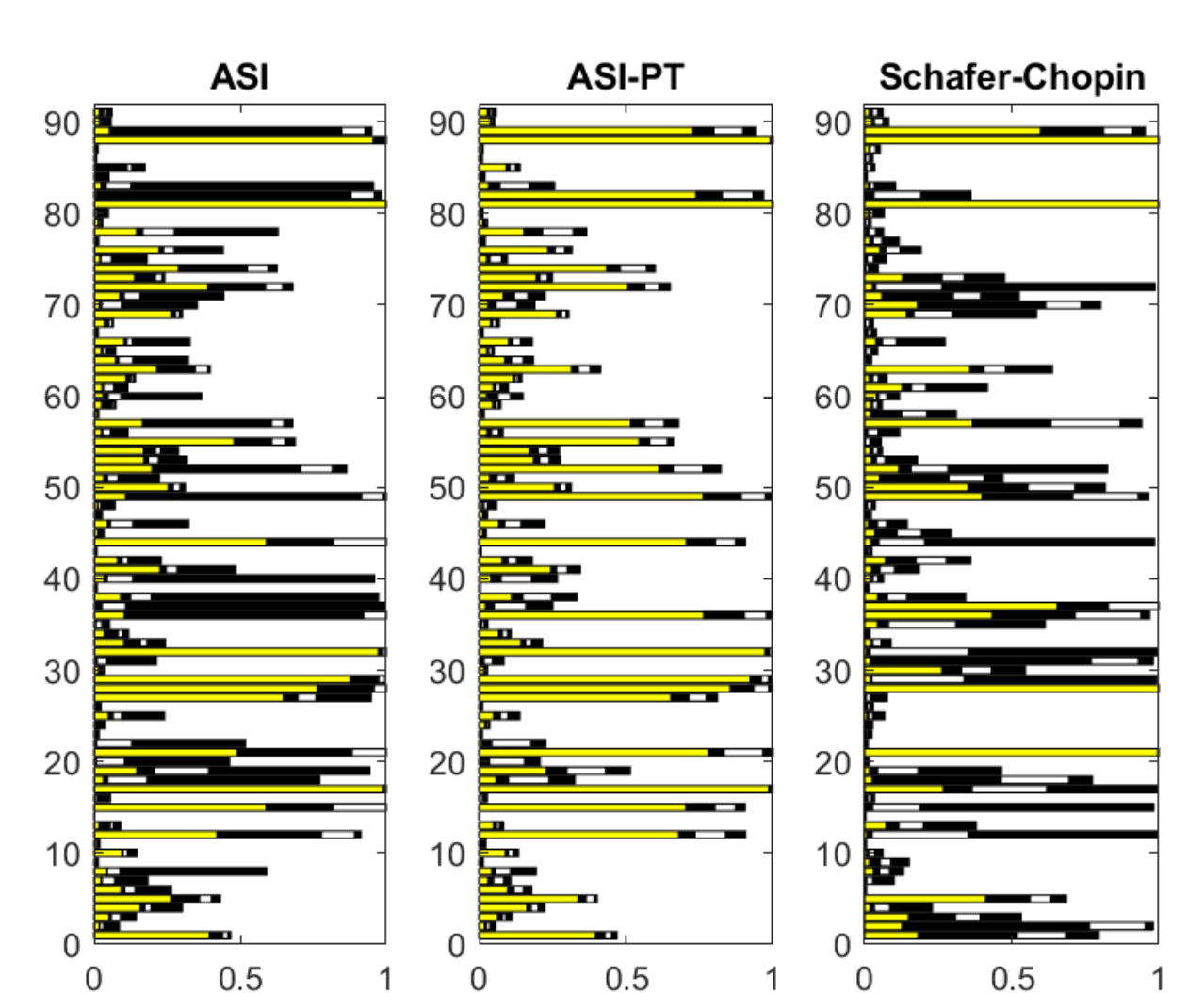} \\
\includegraphics[trim=40mm 100mm 40mm 90mm, clip, scale=0.6]{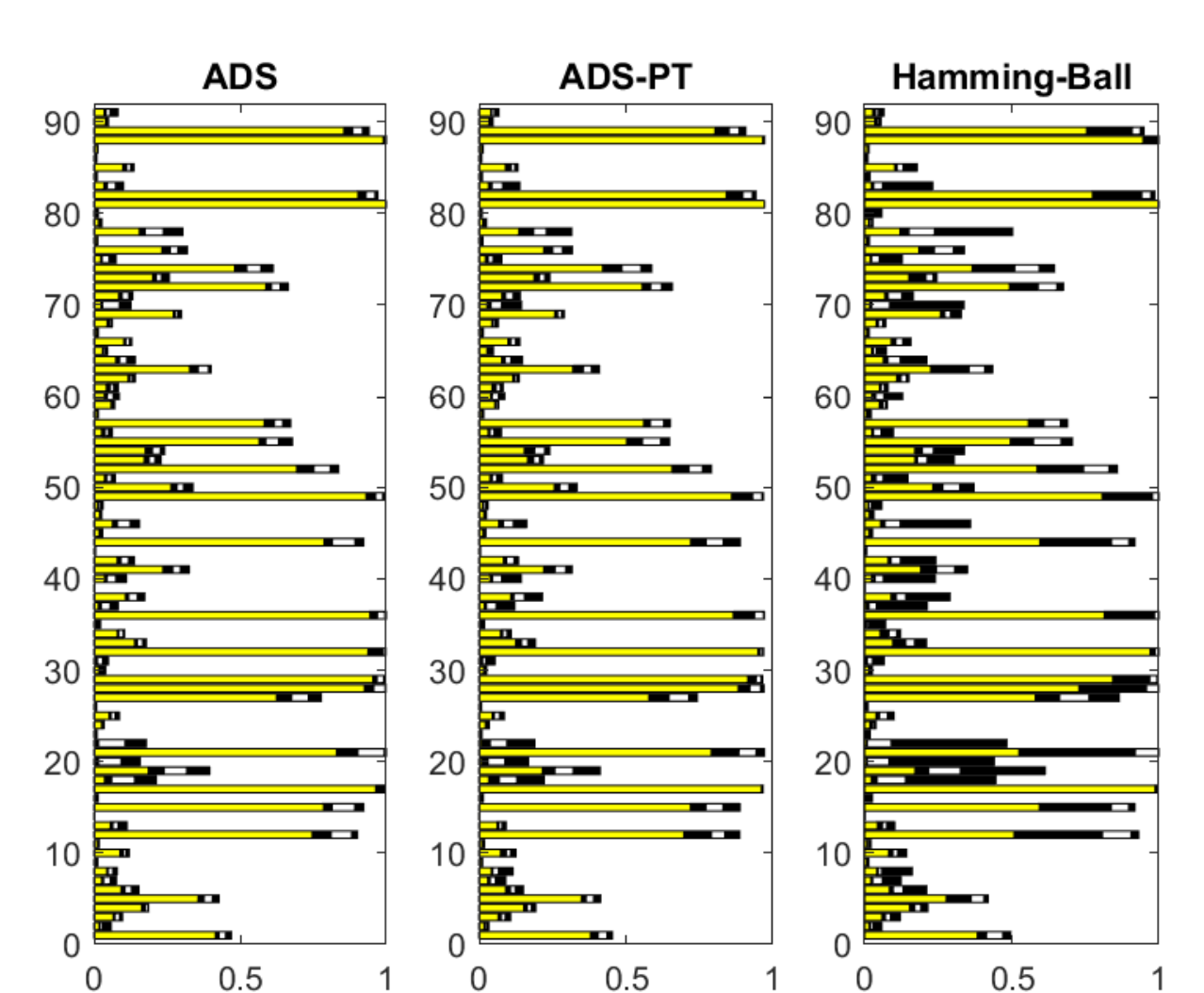} \\
\includegraphics[trim=40mm 100mm 40mm 90mm, clip, scale=0.6]{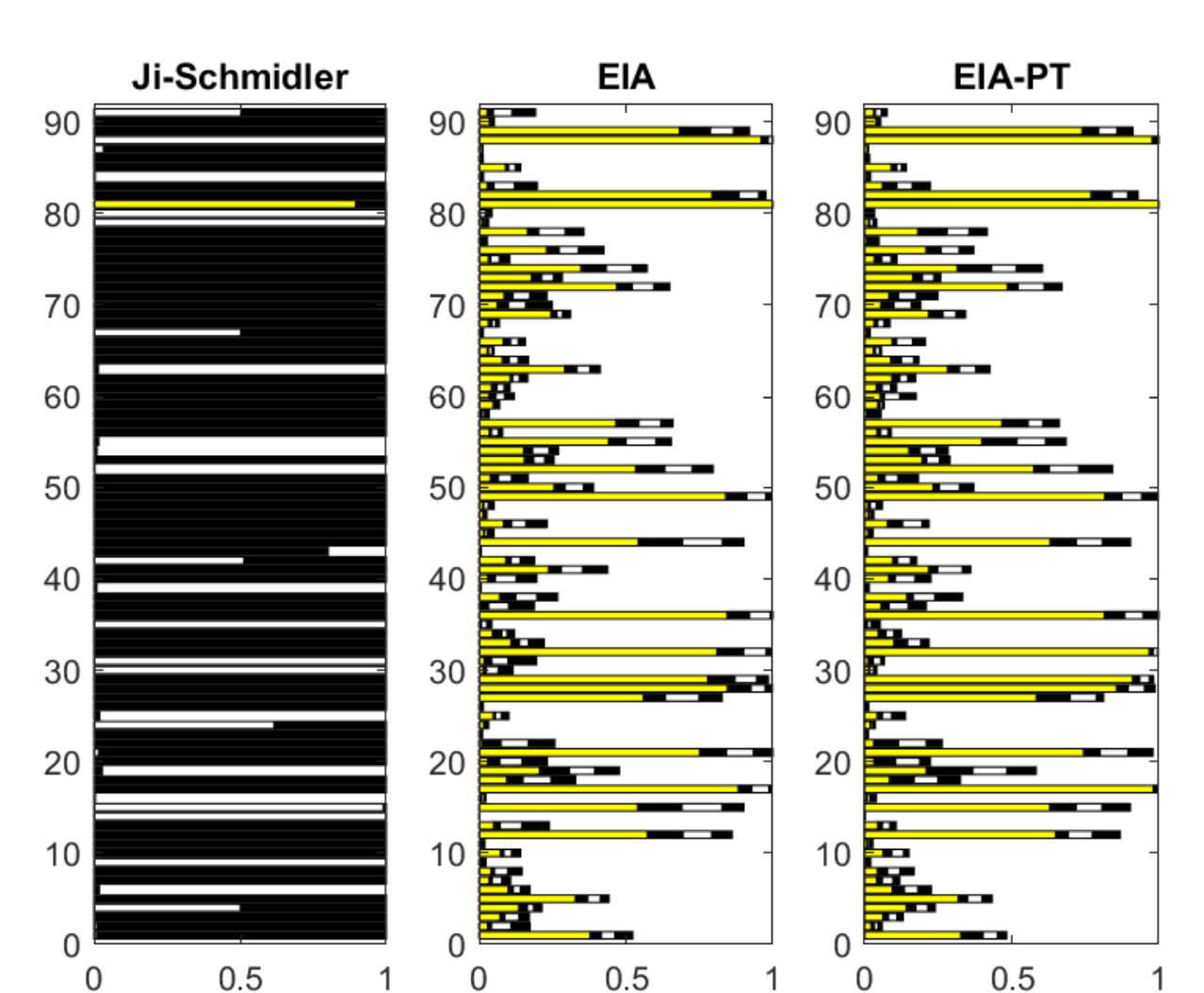}
\end{tabular}
\end{center}
\caption{\small Concrete data: Inclusion probabilities boxplots using adaptively scaled individual adaptation (ASI), exploratory individual adaptation (EIA), add-delete-swap (ADS) and the sequential Monte Carlo algorithm of \cite{ScCh11}, with parallel tempering (PT) versions of the first three algorithms. We also consider the Hamming Ball and the Ji-Schmidler samplers}
\label{results:concrete}
\end{figure}

\begin{figure}[h!]
\begin{center}
\begin{tabular}{c}
\includegraphics[trim=40mm 100mm 40mm 90mm, clip, scale=0.6]{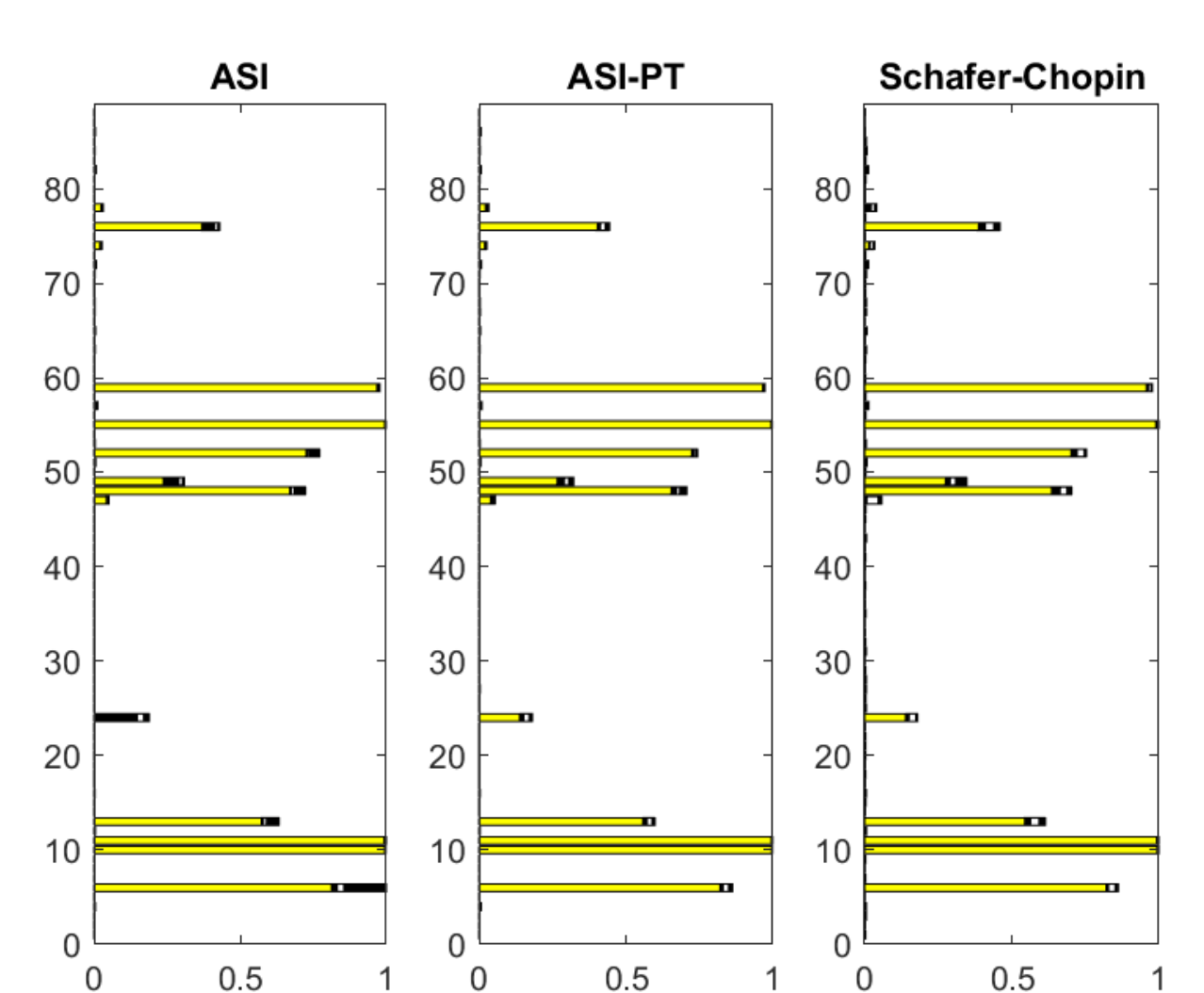}\\
\includegraphics[trim=40mm 100mm 40mm 90mm, clip, scale=0.6]{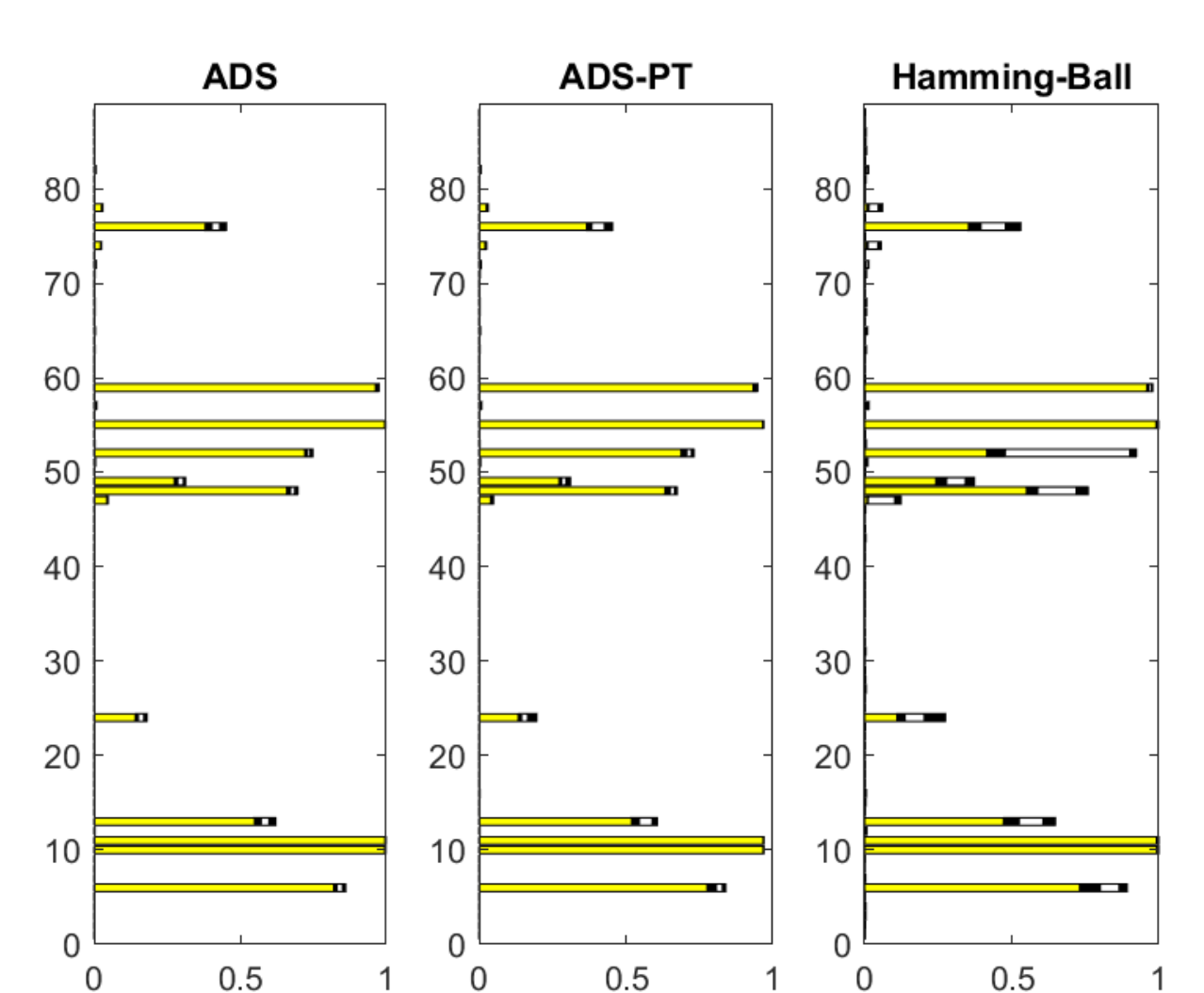}\\
\includegraphics[trim=40mm 100mm 40mm 90mm, clip, scale=0.6]{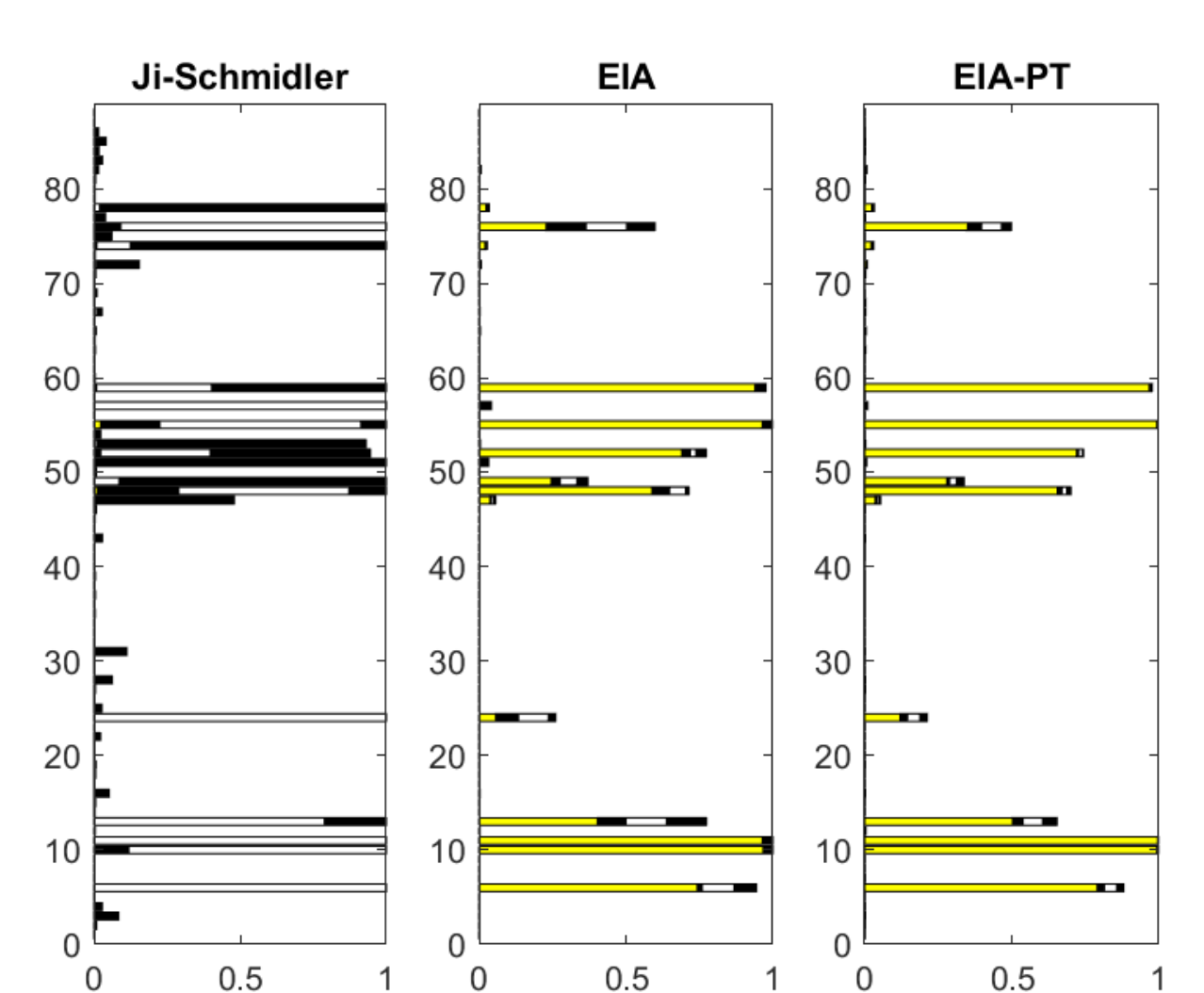}
\end{tabular}
\end{center}
\caption{\small Protein data: Inclusion probabilities boxplots using adaptively scaled individual adaptation (ASI), exploratory individual adaptation (EIA), add-delete-swap (ADS) and the sequential Monte Carlo algorithm of \cite{ScCh11}, with parallel tempering (PT) versions of the first three algorithms. We also consider the Hamming Ball  and the Ji-Schmidler samplers}
\label{results:protein}
\end{figure}

\begin{figure}[h!]
\begin{center}
\begin{tabular}{c}
\includegraphics[trim=40mm 100mm 40mm 90mm, clip, scale=0.6]{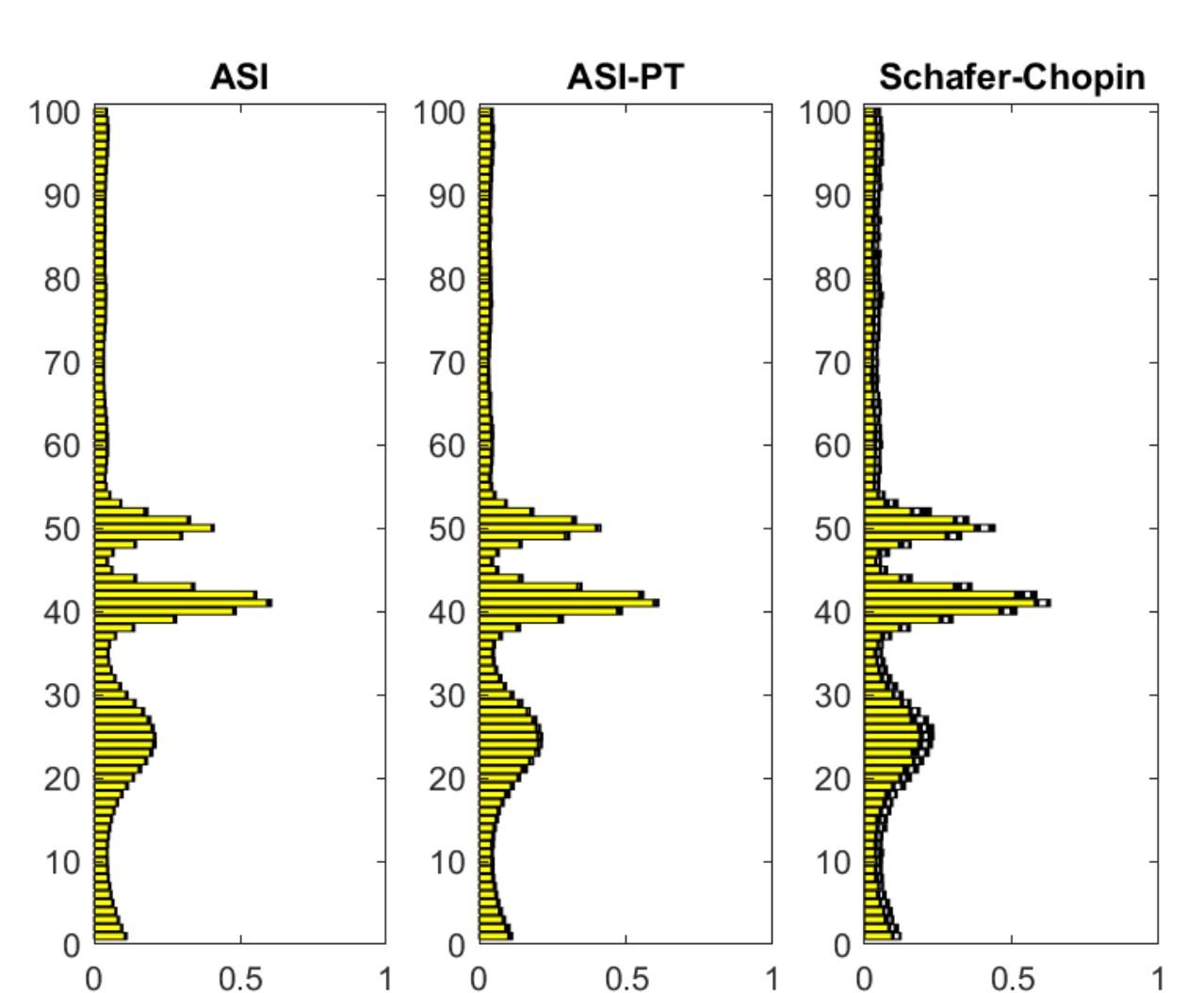}\\
\includegraphics[trim=40mm 100mm 40mm 90mm, clip, scale=0.6]{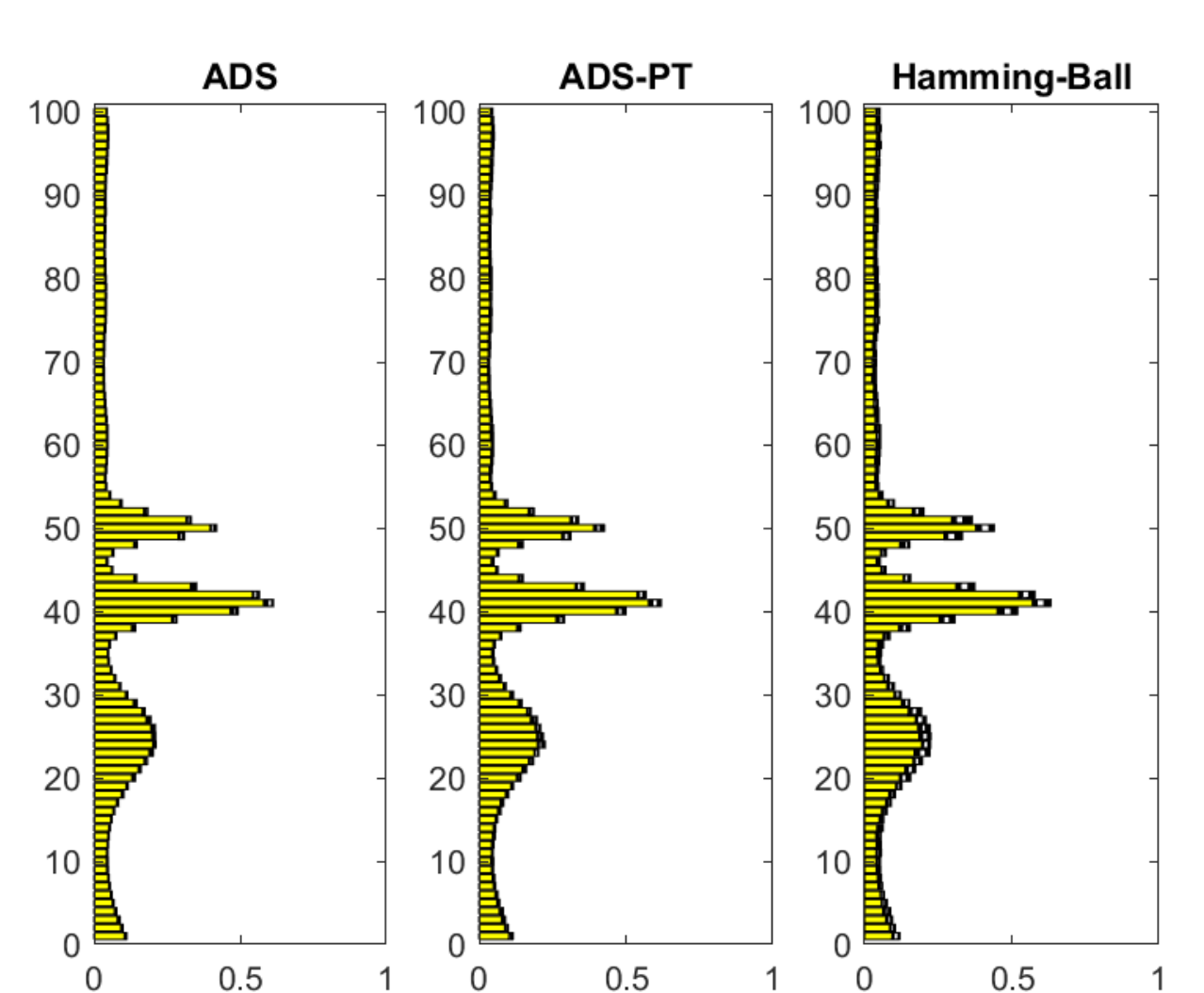}\\
\includegraphics[trim=40mm 100mm 40mm 90mm, clip, scale=0.6]{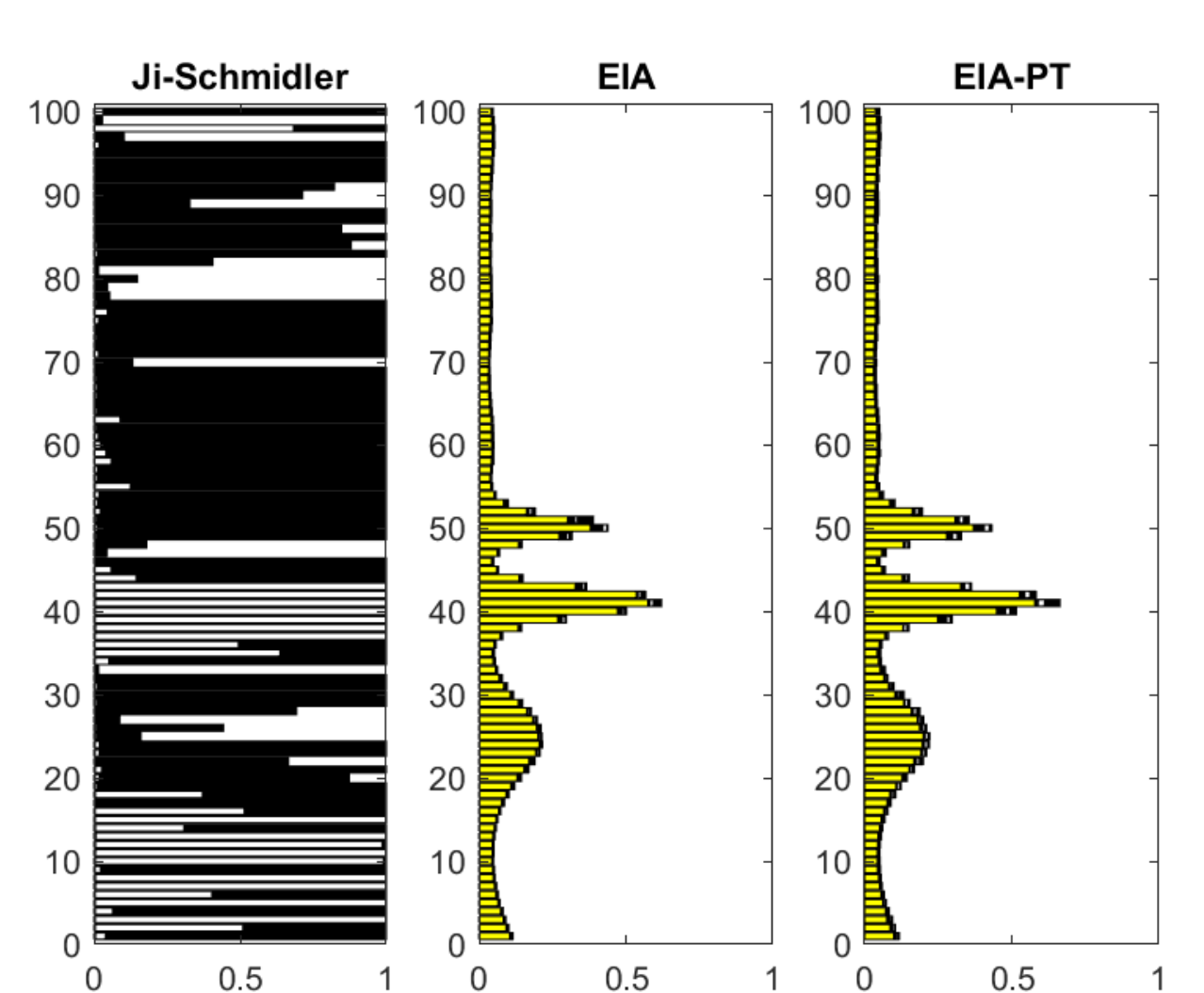}
\end{tabular}
\end{center}
\caption\small {Tecator data: Inclusion probabilities boxplots using adaptively scaled individual adaptation (ASI), exploratory individual adaptation (EIA), add-delete-swap (ADS) and the sequential Monte Carlo algorithm of \cite{ScCh11}, with parallel tempering (PT) versions of the first three algorithms. We also consider the Hamming Ball and the Ji-Schmidler samplers}
\label{results:tecator}
\end{figure}

\begin{figure}[h!]
\begin{center}
\includegraphics[trim=40mm 100mm 40mm 100mm, clip]{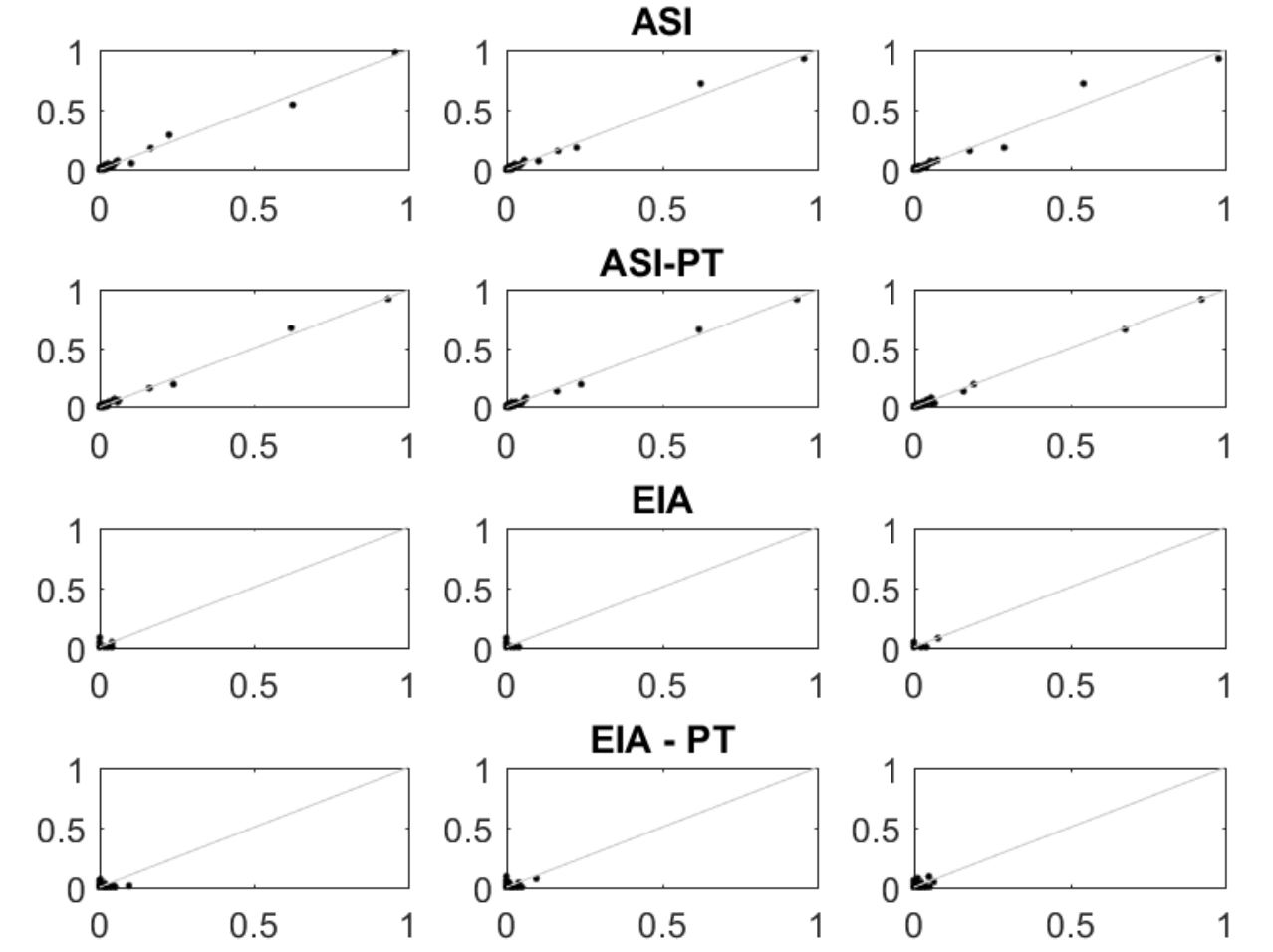}
\end{center}
\caption{\small PCR$1$ data: comparisons of pairs of runs with random $g$ and $h$ using adaptively scaled individual adaptation (ASI), adaptively scaled individual adaptation with parallel tempering (ASI-PT), exploratory individual adaptation (EIA) and exploratory individual adaptation with parallel tempering (EIA-PT)}\label{pcr11:comp_rand_g}
\end{figure}

\begin{figure}[h!]
\begin{center}
\includegraphics[trim=40mm 100mm 40mm 100mm, clip]{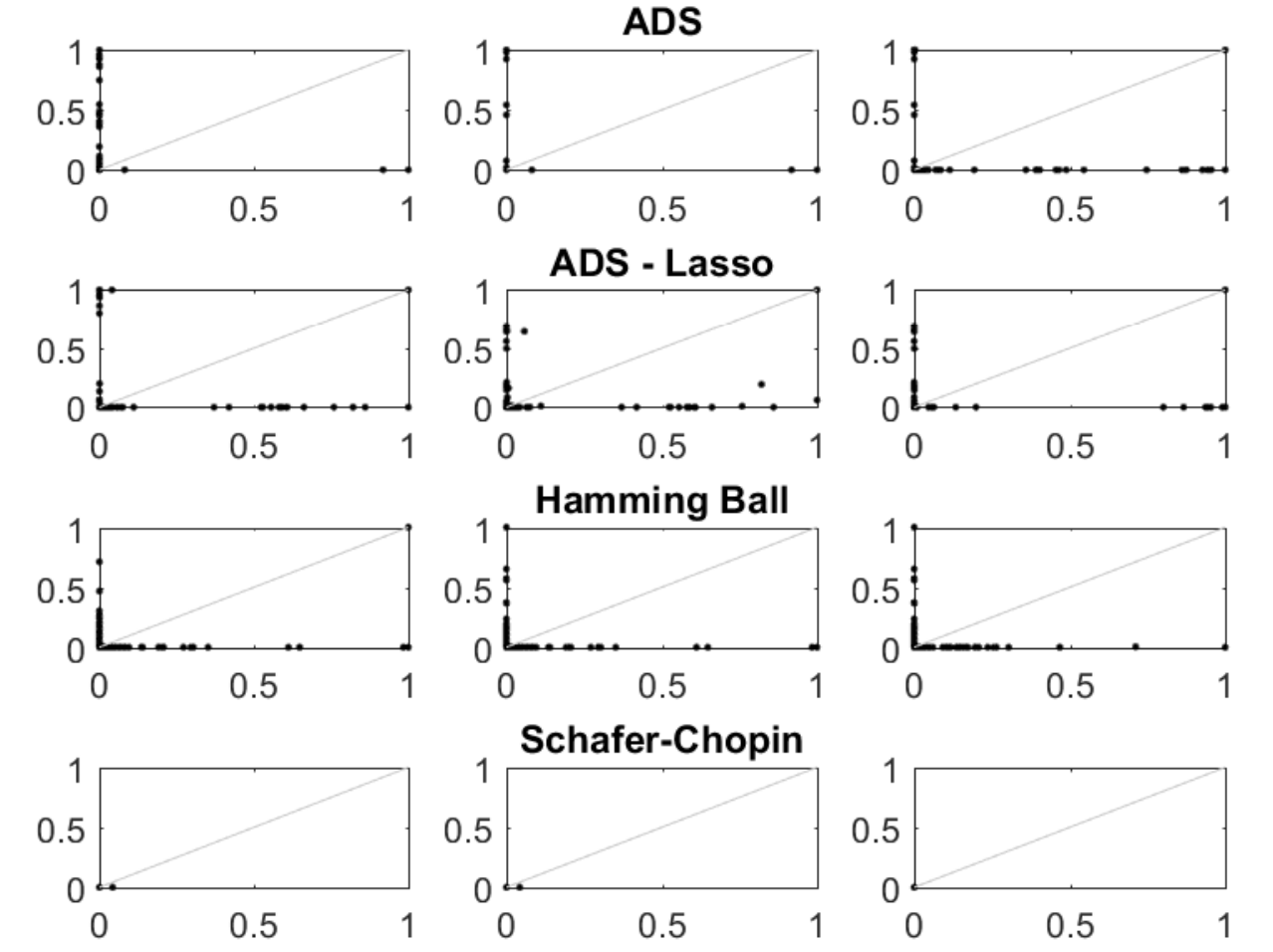}
\end{center}
\caption{\small PCR$1$ data: comparisons of pairs of runs with random $g$ and $h$ using add-delete-swap (ADS), add-delete-swap with lasso start (ADS-Lasso), Hamming Ball and Sch\"afer-Chopin samplers}\label{pcr12:comp_rand_g}
\end{figure}

\begin{figure}[h!]
\begin{center}
\includegraphics[trim=40mm 100mm 40mm 100mm, clip]{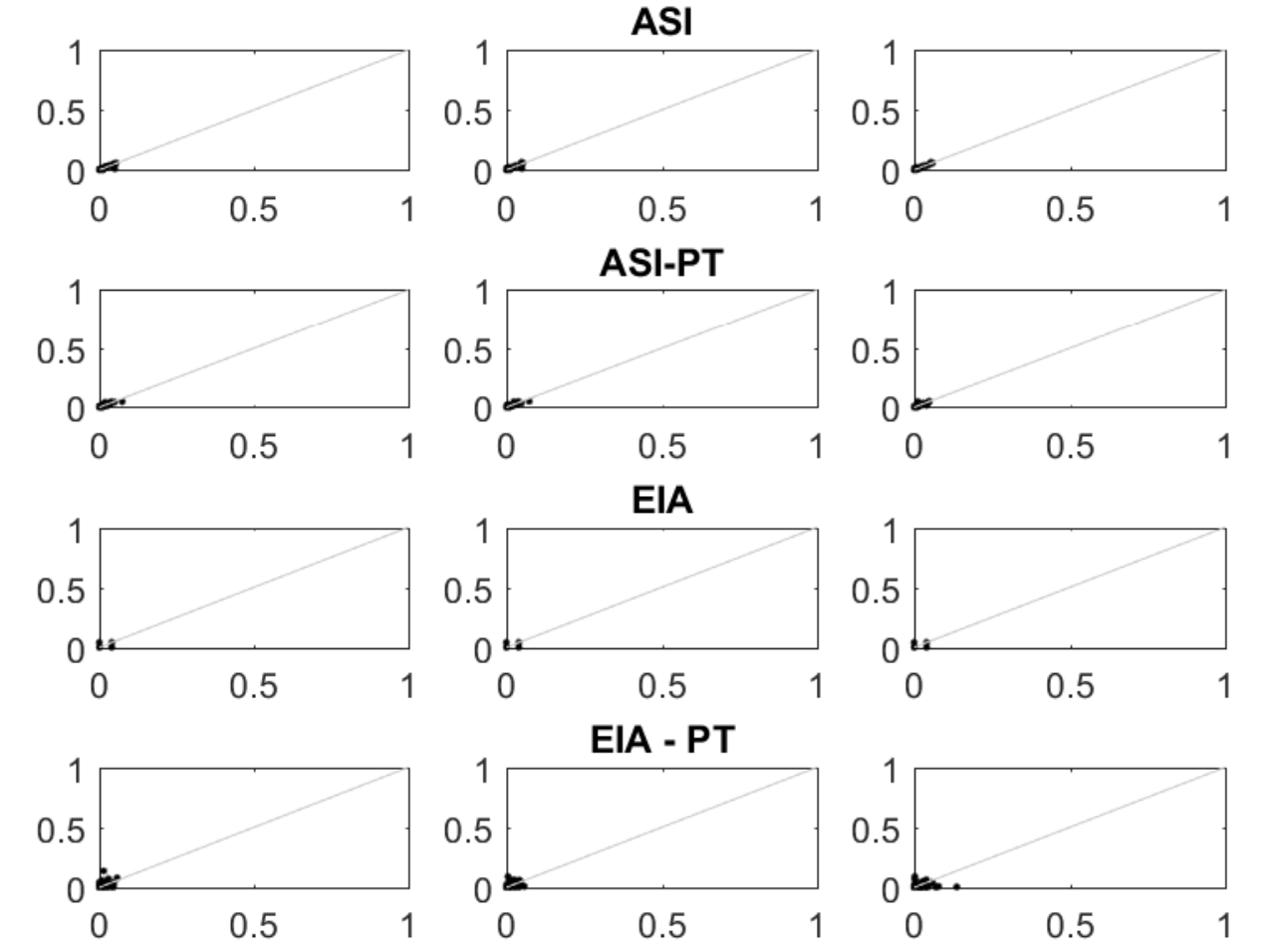}
\end{center}
\caption{\small PCR$2$ data: comparisons of pairs of runs with random $g$ and $h$ using adaptively scaled individual adaptation (ASI), adaptively scaled individual adaptation with parallel tempering (ASI-PT), exploratory individual adaptation (EIA) and exploratory individual adaptation with parallel tempering (EIA-PT)}\label{pcr21:comp_rand_g}
\end{figure}

\begin{figure}[h!]
\begin{center}
\includegraphics[trim=40mm 100mm 40mm 100mm, clip, scale=0.9]{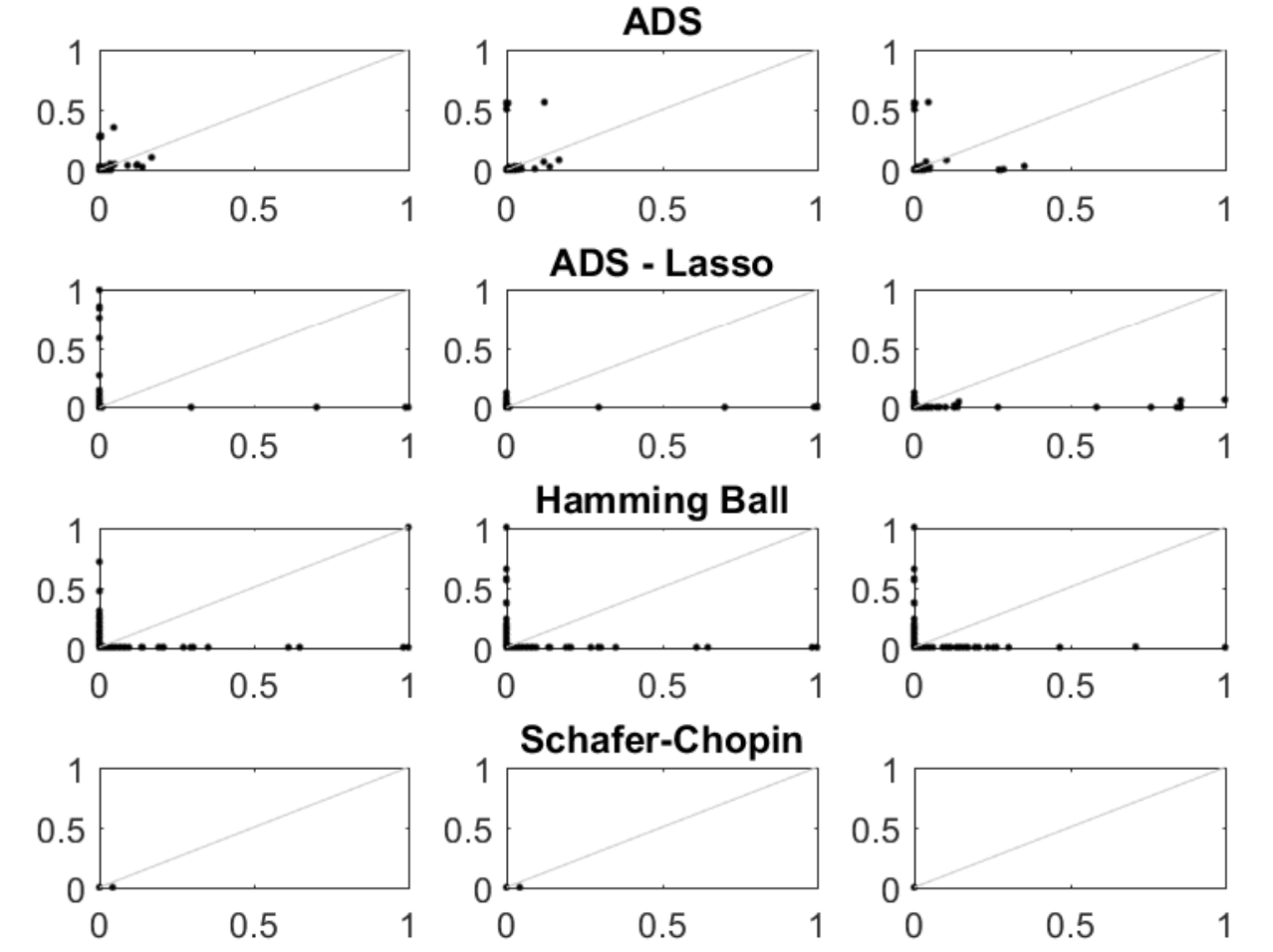}
\end{center}
\caption{\small PCR$2$ data: comparisons of pairs of runs with random $g$ and $h$ using add-delete-swap (ADS), add-delete-swap with lasso start (ADS-Lasso), Hamming Ball and Sch\"afer-Chopin samplers}\label{pcr22:comp_rand_g}
\end{figure}

 \begin{figure}[h!]
\begin{center}
\includegraphics[trim=40mm 100mm 40mm 100mm, clip, scale=0.9]{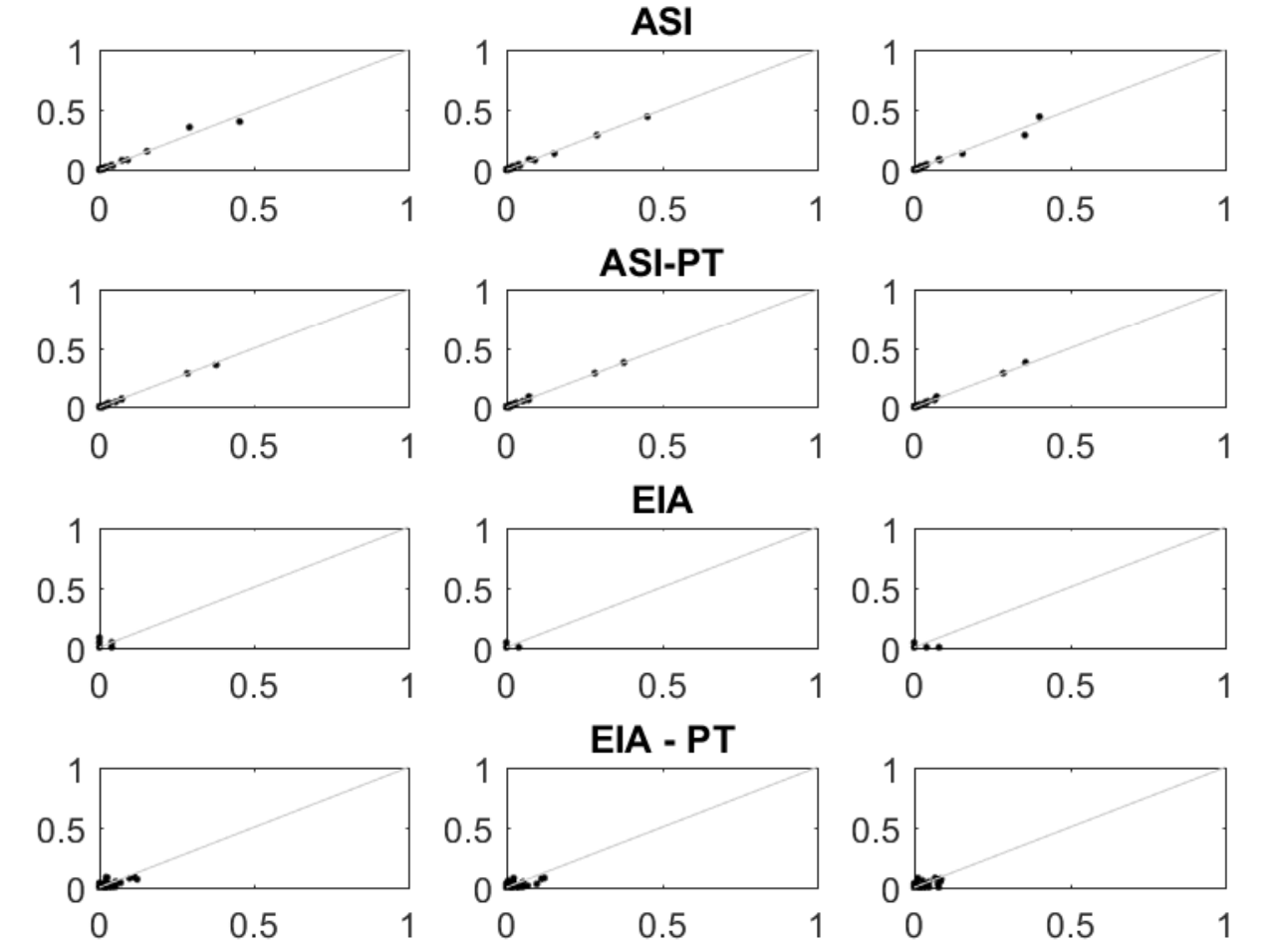}
\end{center}
\caption{\small PCR$3$ data: comparisons of pairs of runs with random $g$ and $h$ using adaptively scaled individual adaptation (ASI), adaptively scaled individual adaptation with parallel tempering (ASI-PT), exploratory individual adaptation (EIA) and exploratory individual adaptation with parallel tempering (EIA-PT)}\label{pcr31:comp_rand_g}
\end{figure}

\begin{figure}[h!]
\begin{center}
\includegraphics[trim=40mm 100mm 40mm 100mm, clip, scale=0.9]{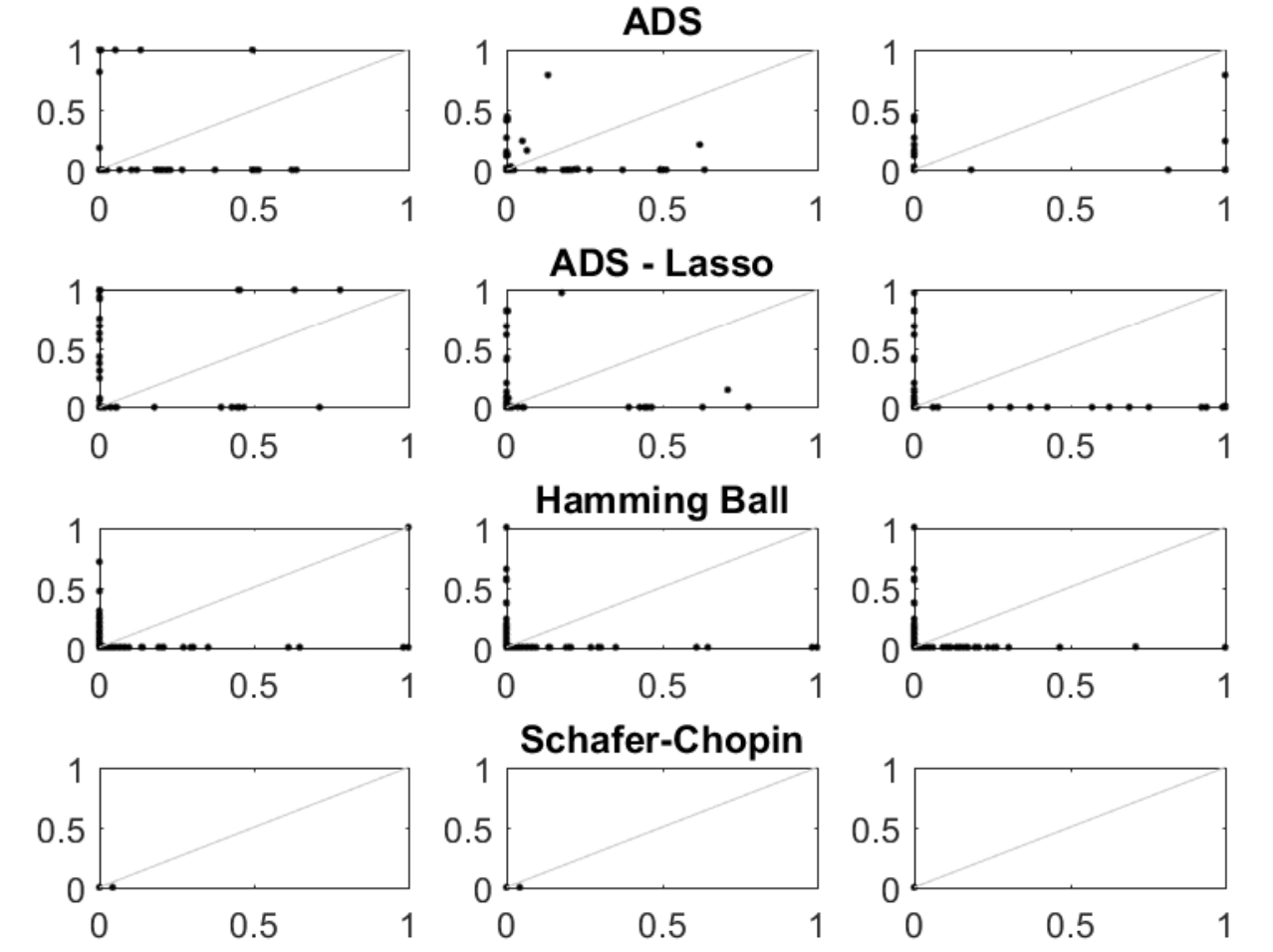}
\end{center}
\caption{\small PCR$3$ data: comparisons of pairs of runs with random $g$ and $h$ using add-delete-swap (ADS), add-delete-swap with lasso start (ADS-Lasso), Hamming Ball and Sch\"afer-Chopin samplers}\label{pcr32:comp_rand_g}
\end{figure}

\begin{figure}[h!]
\begin{center}
\includegraphics[trim=40mm 100mm 40mm 100mm, clip, scale=0.9]{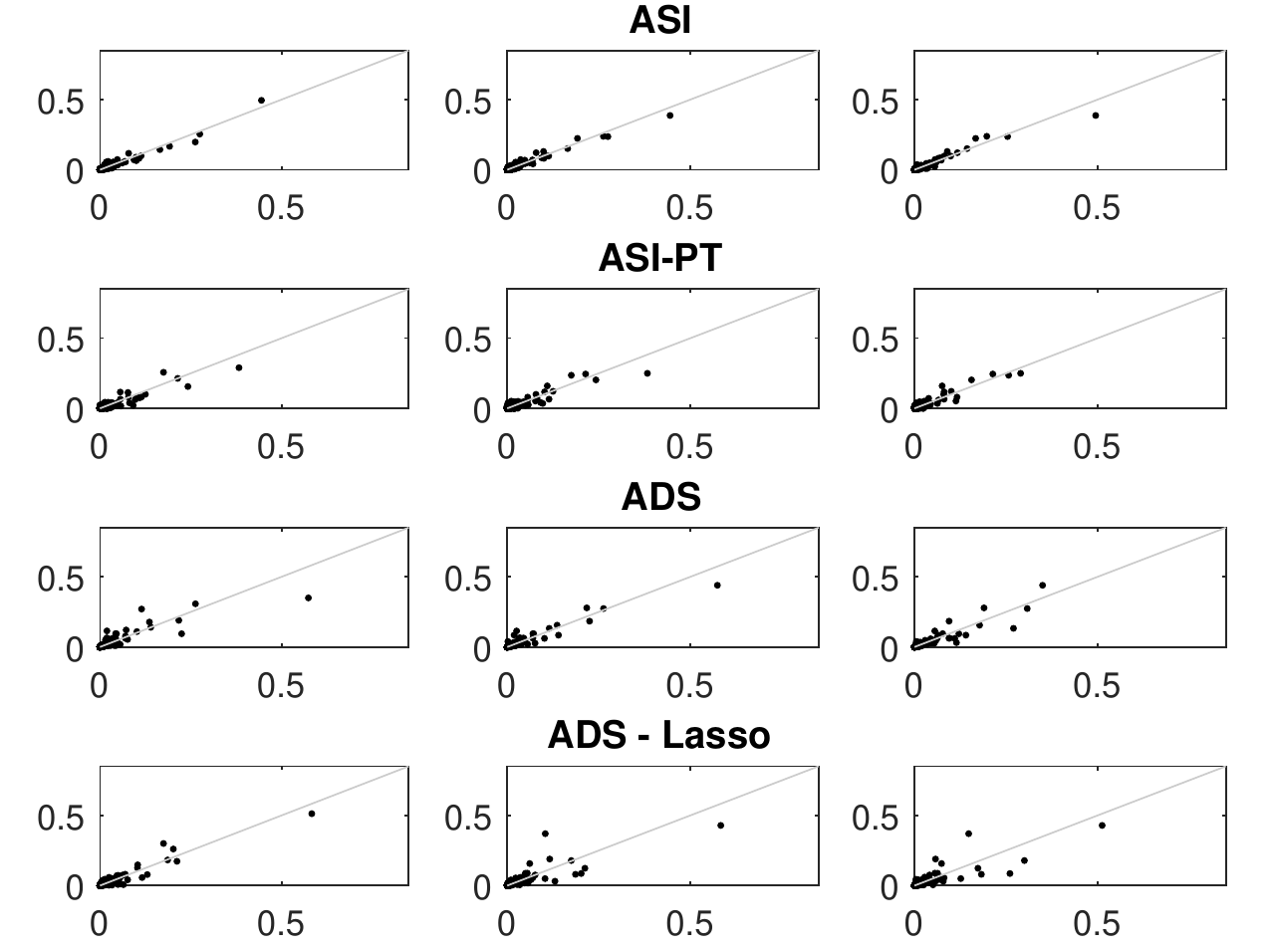}
\end{center}
\caption{\small SNP data: comparisons of pairs of runs with random $g$ and fixed $h$ using adaptively scaled individual adaptation (ASI), adaptively scaled individual adaptation with parallel tempering (ASI-PT), add-delete-swap (ADS) and add-delete-swap with lasso start (ADS-L)}\label{mice:comp_rand_g_fixed_h}
\end{figure}

\end{document}